\documentclass[useAMS,usenatbib]{mn2e}

\usepackage{aas_macros}
\usepackage{dcolumn}
\usepackage{amsmath}
\usepackage{amssymb}
\usepackage{graphicx}

\begin{document}

\title[Compact Stellar Systems in Coma]{The HST/ACS Coma Cluster Survey: V - Compact Stellar Systems in the Coma Cluster}
\author[J. Price et al.]
{J. Price$^{1}$\thanks{james.price@bristol.ac.uk},
S. Phillipps$^{1}$, 
A. Huxor$^{1}$,
N. Trentham$^{2}$, H.C. Ferguson$^{3}$,   
\newauthor R.O. Marzke$^4$, A. Hornschemeier$^5$, P. Goudfrooij$^{3}$, D. Hammer$^{6}$, R.B. Tully$^7$, 
\newauthor K. Chiboucas$^7$, R.J. Smith$^8$, D. Carter$^{9}$, D. Merritt$^{10}$, M. Balcells$^{11}$,
\newauthor P. Erwin$^{12,13}$, T.H. Puzia$^{14}$ \\
$^{1}$Astrophysics Group, H.H. Wills Physics Laboratory, University of Bristol, Tyndall Avenue,  Bristol BS8 1TL, UK \\
$^{2}$Institute of Astronomy, University of Cambridge, Madingley Road, Cambridge, CB3 0HA, UK \\
$^{3}$Space Telescope Science Institute, 3700 San Martin Drive, Baltimore, MD 21218, USA \\
$^{4}$Department of Physics and Astronomy, San Franciso State University, San Francisco, CA 94132, USA \\
$^{5}$NASA Goddard Space Flight Centre, Code 662, Greenbelt, MD 20771, USA \\
$^{6}$Department of Physics and Astronomy, Johns Hopkins University, 3400 North Charles Street, Baltimore, \\
MD 21218, USA \\
$^{7}$Institute for Astronomy, University of Hawaii, 2680 Woodlawn Drive, Honolulu, HI 96822, USA \\
$^{8}$Department of Physics, University of Durham, Durham, DH1 3LE, UK \\
$^{9}$Astrophysics Research Institute, Liverpool John Moores University, Twelve Quays House, Egerton Wharf, \\ Birkenhead, CH41 1LD, UK \\
$^{10}$Department of Physics, Rochester Institute of Technology, 85 Lomb Memorial Drive, Rochester, NY \\
14623, USA \\
$^{11}$Instituto de Astrof\'isica de Canarias, C/V\'ia Lactea s/n, 38200 La Laguna, Tenerife, Spain \\
$^{12}$Max-Planck-Institute for Extraterrestrial Physics, Giessenbachstrasse, 85748 Garching, Germany \\
$^{13}$Universit\"{a}tssternwarte, Scheinerstrasse 1, 81679 M\"{u}nchen, Germany \\
$^{14}$Herzberg Institute of Astrophysics, 5071 West Saanich Road, Victoria, BC V9E 2E7, Canada}
\date{MNRAS accepted}

\pagerange{\pageref{firstpage}--\pageref{lastpage}} \pubyear{2009}

\maketitle

\label{firstpage}

\begin{abstract}
The HST ACS Coma Cluster Treasury Survey is a deep two passband imaging survey
of the nearest very rich cluster of galaxies, covering a range of galaxy density environments. The imaging is complemented by a recent wide field redshift survey of the cluster conducted with Hectospec on the 6.5m MMT. Among the many scientific applications for this data are the search for compact galaxies. In this paper, we present the discovery of seven compact 
(but quite luminous) stellar systems, ranging from M32-like galaxies down to ultra-compact dwarfs 
(UCDs)/dwarf to globular transition objects (DGTOs).

We find that all seven compact galaxies require a two-component fit to their light profile and have measured velocity dispersions that exceed those expected for typical early-type galaxies at their luminosity. From our structural parameter analysis we conclude that three of the sample should be classified as compact ellipticals or M32-like galaxies, the remaining four being less extreme systems. The three compact ellipticals are all found to have old luminosity weighted ages ($\gtrsim$ 12 Gyr), intermediate metallicities (-0.6 $<$ [Fe/H] $<$ -0.1) and high [Mg/Fe] ($\gtrsim$ 0.25). 

Our findings support a tidal stripping scenario as the formation mode of compact galaxies covering the luminosity range studied here. We speculate that at least two early-type morphologies may serve as the progenitor of compact galaxies in clusters.
\end{abstract}


\section{Introduction}

The term compact stellar system covers a range of possible object types,
characterised by small physical dimensions and high surface brightness.
The largest are the compact elliptical (cE) galaxies exemplified by M32. As a close companion of M31, its truncated radial profile has long been suspected to
be due to tidal or other interactions with its giant neighbour \citep{king62,faber73}.
M32 itself has $M_V \simeq -16.5$ mag and a half-light radius $R_e \simeq 120$ pc \citep{kent87}. Despite a luminosity in the dwarf galaxy regime, it has 
other properties more akin to those of a giant galaxy or a classical bulge, such as the presence of a massive central black hole \citep{vandermarel97} and a high metallicity \citep{rose05}, reinforcing the suggestion that it has suffered major loss of mass. 

Few other galaxies with similar
properties are known. The other usually quoted examples - somewhat brighter at $M_V \sim -18$ mag, and with $R_e \sim200$-300 pc - are NGC 4486B, a companion of M87 near the centre of the Virgo Cluster \citep{rood65}, NGC 5846A \citep[in the NGC 5846 group;][]{faber73}, two objects
in Abell 1689 \citep{mieske05} and one in Abell 496 \citep{chilingarian07}.

Globular clusters (GCs) obviously also satisfy the requirements for compact
stellar systems and are, of course, extremely numerous. Their typical luminosities (the turn over point in
their luminosity function) are around $M_V = -7.5$ mag \citep{dicriscienzo06} and the brightest examples
around our Galaxy ($\omega$ Cen) and M31 (G1) reach $M_V \simeq -10$ to $-11$ mag;
the very populous GC systems around central cluster galaxies
may extend slightly brighter, to $M_V \simeq -11.5$ mag \citep{harris09}.
  
No compact objects\footnote{We exclude blue compact dwarfs (BCDs) from this list, as they
appear to be relatively diffuse underlying galaxies with very compact
central star forming regions, see e.g. \cite{papaderos06}. We note, however, that some BCDs, such as the object reported by \cite{trentham01}, are analogous to blue cEs.} at luminosities between these two extremes were known until the discovery in the Fornax Cluster \citep{hilker99,drinkwater00} of what are generally referred to
as ultra-compact dwarfs \citep[UCDs;][]{phillipps01}.
These have $M_V$ between $-11.5$ and $-13.5$ mag and $R_e \sim 20$-30 pc. Further similar
examples were found in the
Virgo Cluster \citep{drinkwater04,jones06} and in a couple of large groups \citep{firth06,evstigneeva07a} while \cite{hasegan05} added
the slightly fainter, intermediate objects which they call DGTOs, for dwarf/globular transition objects \citep[see also][]{drinkwater05,gregg09}. \cite{marzke06} and \cite{chilandmamon08} have also suggested the existence, in Sloan Digital Sky Survey\footnote{http://www.sdss.org}(SDSS) data, of compact
systems with luminosities between UCDs and compact ellipticals.
Between them, these new types of system now span essentially all of the luminosity
space from normal globular clusters upwards. The majority appear to reside in reasonably dense environments where interactions can be expected. Indeed even those found in relatively low density environments, such as the recently discovered UCD associated with the Sombrero galaxy \citep[M104;][]{hau09} and M32 itself, have more massive companions.

In the present paper we present the discovery of compact stellar systems
in the very rich and dense Coma Cluster. The sample was observed as part of the Hubble Space Telescope (HST) Treasury Survey of the Coma Cluster, undertaken using the
Advanced Camera for Surveys (ACS). We describe the survey and the
spectroscopic follow-up in section 2 \citep[see also][for further details.]{carter08}. In section 3 we present details of our search methods and
examine a total of seven compact systems which have redshifts proving 
them to be cluster members, three of which are extreme enough to be classed as cEs. Section 4 provides details of our photometric and structural analysis of these objects and section 5 an analysis of their spectroscopic properties, including velocity dispersions and stellar populations. Section 6 summarises our results. A separate paper will explore the range from UCDs to large globular clusters in Coma.

Before continuing it is prudent to comment on the nomenclature used throughout this paper. The term cE is used here in a general sense to refer to those galaxies in our sample which possess comparable high surface brightness to that of M32. It is not used to make assumptions regarding their specific morphology or indeed possible progenitor morphology.

\section{Observations}
\subsection{The HST/ACS Coma Cluster Treasury Survey}

The HST Treasury Survey of the Coma Cluster (HST program GO10861, P.I. D. Carter)
currently covers an area of
just over 200 square arcminutes; unfortunately it was curtailed when only 28\% complete due to the failure of ACS on 2007 January 27. The majority
of the 25 fields surveyed (21 completely, 4 partially)
lie in the central regions of the cluster,
but a few sparsely sample the outer (infall) regions, around 1 degree 
from the core and are centred on known cluster members
\citep[see][for details of the pointings]{carter08}. Note that {\em all}
of the fields observed contain one or more moderately bright galaxies from the catalogue of \cite[GMP]{gmp83}.

Each individual ACS field is 11.3 arcmin$^2$ in area, imaged onto
two $4096 \times 2048$ pixel CCDs \citep[see][for details of the instrument]{acshandbook}. We assume hereafter a luminosity distance
to Coma of 100 Mpc \citep[equivalent to $H_0 \simeq 71$~km~s$^{-1}$, see e.g.][]{kavelaars00}, i.e. a distance modulus $m-M = 35.0$.
At this distance the pixel size of the ACS ($0.05^{\prime\prime}$) corresponds to 23 pc
and the ACS field of view is approximately 100 kpc square. 
Exposure times were $4 \times 640$s in the F475W filter and $4 \times 350$s in F814W, giving
limiting point source magnitudes of $g_{F475W} = 26.7$ mag, $I_{F814W} = 25.9$ mag ($10 \sigma$ detections) in the AB system. Standard data reduction
was carried out at STScI, using PyRAF/STSDAS and MultiDrizzle \citep{koekemoer02}. Additional second-pass processing is described in the main
survey paper \citep{carter08}.  Final data products and object catalogues are discussed fully in Hammer et al. (2009, in preparation), hereafter Paper 2 of this series.

\subsection{MMT / Hectospec Observations}

As part of the ACS Coma Cluster Survey, we are carrying out a  
comprehensive spectroscopic survey of galaxies in the Coma region  
using Hectospec, a 300-fibre spectrograph at the MMT \citep{fabricant05}. Observing time was allocated for this survey by the NOAO  
TAC through the NSF/TSIP program. Hectospec observations were  
obtained during three semesters (2007A - 2009A) in a classical-queue  
mode overseen by Nelson Caldwell at the Harvard-Smithsonian Center  
for Astrophysics. We describe the survey in detail in Marzke et al.  
(2009, in preparation).

Here, we use observations obtained during 2007 and 2008.  For these  
observing runs, we selected targets drawn from SDSS imaging based on  
their apparent magnitude, $g-r$ colour and surface brightness.  In the  
regions observed with HST/ACS, the overall spectroscopic completeness  
is  90\% at $r=19.9$ mag and 50\% at $r=20.8$ mag.  Because the ACS imaging  
was unavailable when the first round of Hectospec targeting was  
completed in 2007, we chose to exclude sources identified as stars by  
the SDSS photometric pipeline. In 2008, we took a similar approach,  
but we added four candidate cEs identified using the procedure  
described in section 3.1.

We used the 270 lines/mm grating, which delivers approximately 4.5 \AA 
\ resolution at a pixel scale of 1.21 \AA/pixel.  The useful spectral  
range is approximately 3700-9000\AA. For the vast majority of  
targets (including those discussed in this paper), we integrated for  
a total of one hour divided into three twenty-minute exposures.   
Seeing ranged from $0.7^{\prime\prime}$ to $1.8^{\prime\prime}$, comparable to the $1.5^{\prime\prime}$ fibre  
diameter, which subtends roughly 730 pc at the Coma distance.

We reduced the data using HSRED, an IDL package written by Richard  
Cool and based on the IDL pipeline developed for the SDSS. HSRED establishes an approximate flux calibration  
using spectra of bright F stars, which have well-measured $ugriz$  
photometry and are observed in each fibre configuration.

The HSRED pipeline produces redshifts computed using a $\chi^2$  
template-fitting algorithm.  We also measured redshifts independently  
by cross-correlating the spectra with SDSS templates using XCSAO  
\citep{kurtz98}. We selected final redshifts and judged their  
quality by examining each of the 6,259 spectra by eye.

\section{The Sample}

\subsection{Selection Criteria}

To begin our search we first made use of the SExtractor \citep[SE]{bertin96} derived catalogues detailed in Paper 2. Based on all the ACS Coma images in both F475W and F814W filters this database provided total magnitudes, matched aperture colours and first pass size and shape parameters for every object in our frames. We next combined the Hectospec redshift survey results with the imaging data and constructed the relevant total magnitude, surface brightness and size diagnostic plots to differentiate objects of interest that have confirmed Coma membership. To refine the search further we impose additional constraints in terms of object colour, size and the level of resolution of the image ('class\_star'). 

To constrain object colour we convert the colours of five of the known cEs to our ACS filter set by de-reddening and K-correcting to rest frame colours where necessary, transforming to the survey's F475W-F814W colour and then applying Coma reddening and K-correction. Following this procedure we obtain a colour cut of 1.1 $<$ F475W-F814W $<$ 1.5 that can be applied directly to our observed SE photometry. The cEs used were M32 (V-I$_{c}$ from HyperLeda\footnote{http://leda.univ-lyon1.fr/}), NGC 4486B and NGC 5846A from the SDSS and the two objects in Abell 1689 from \cite{mieske05}. Colour transformations and K-corrections are derived using the synphot routine in IRAF/STSDAS and spectral energy distributions (SEDs) from BaSTI\footnote{http://albione.oa-teramo.inaf.it/}. For the colour transformations we use SEDs with -1.27 $\leq$ [Fe/H] $\leq$ 0.4 and 3 Gyr $\leq$ Age $\leq$ 14 Gyr and for the K-correction we employ a 10 Gyr SED with [Fe/H] = -0.66. Tests indicate that the K-correction is dependent on metallicity and age of the assumed SED as one would expect, with variations of $\pm$20\%. This only becomes significant when removing the effects of K-correction on the g-i colours of the two cEs in Abell 1689 where it is $\sim\pm$0.1 mag, and even then it is not sufficient to move either galaxy out of the colour cut. Indeed we also find no further likely compact candidates within $\pm$0.1 mag of our selected colour range. Galactic extinction is corrected for according to \cite{schlegel98}. 

\begin{figure}
\flushleft
\scalebox{0.42}[0.42]{\includegraphics{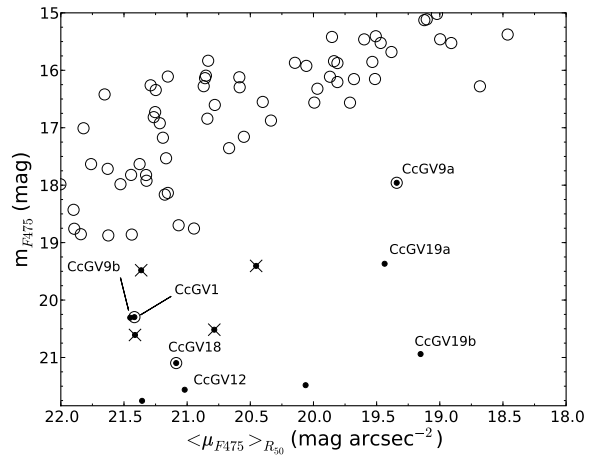}}
\caption{Diagram used to select our initial sample and followup spectroscopic targets. Unfilled circles are confirmed cluster members inclusive of the Hectospec 2007 observations. Crosses are confirmed background objects. Filled circles show those objects that meet our size, colour and star classifier constraints. The data presented here is as observed with no corrections for extinction, K-correction or cosmological dimming.}
\label{mflux50}
\end{figure}

\begin{table}
\caption{Coma compact galaxy (CcG) photometry and adopted redshifts. The full identifier refers to the Survey visit number in which the galaxy is located (see \citet{carter08} Table 2 for details). An additional character, in alphabetical order brightest to faintest, is used to differentiate galaxies when a visit contains more than one object of interest.}
\label{objtab}
\begin{tabular}{l@{\hspace{3mm}}c@{\hspace{3mm}}c@{\hspace{3mm}}c@{\hspace{3mm}}c@{\hspace{2mm}}c}
\hline
ID & Ra$_{2000}$ & Dec$_{2000}$ & m$_{B}$ & $B-I_{c}$ & z \\
\hline
CcGV1 & 13:00:47.67 & +28:05:33.9 & 20.72 & 2.19 & 0.0226 \\
CcGV9a$^{\dagger}$ & 13:00:18.85 & +28:00:33.4 & 18.34 & 2.04 & 0.0207 \\
CcGV9b$^{\star}$ & 13:00:27.31 & +28:00:33.3 & 20.71 & 2.14 & 0.0213 \\
CcGV12$^{\star}$ & 12:59:42.29 & +28:00:55.1 & 21.98 & 2.17 & 0.0261 \\
CcGV18 & 12:59:59.88 & +27:59:21.6 & 21.50 & 2.13 & 0.0218 \\
CcGV19a$^{\star}$ & 12:59:37.18 & +27:58:19.8 & 19.79 & 2.19 & 0.0254 \\
CcGV19b$^{\star}$ & 12:59:39.20 & +27:59:54.4 & 21.33 & 2.07 & 0.0236 \\
\hline
\end{tabular}
$^{\star}$Spectroscopic membership confirmed following photometric selection. $^{\dagger}$GMP 2777
\end{table}

\begin{figure}
\scalebox{0.42}[0.42]{\includegraphics{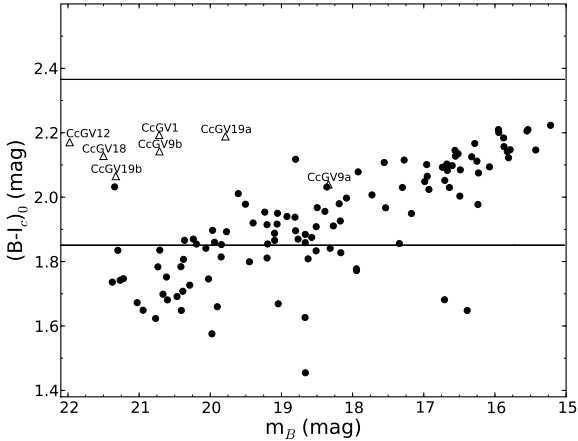}}
\caption{Colour-magnitude relation for all confirmed Coma members imaged by our ACS observations. Compact galaxies are denoted by triangles. Horizontal lines show the upper and lower extent of the colour cut used when selecting the compact galaxy sample.}
\label{cmr}
\end{figure}

Making use of the flux radius parameter provided by SE, we impose an upper limit of 500 pc on the radius that contains 50\% of the flux inside the MAG\_AUTO Kron-like aperture. This value was determined by inspecting the magnitude-size relation for our data set and largely acts to guide the eye to areas of interest in the parameter space.  

Finally we enforce a maximum value for the SE star classifier of 0.5 in order to exclude unresolved objects from our search. In real terms this cut removes objects with SE 50\% flux radius $\lesssim$ 1.5 pixels ($\sim$ 35 pc at Coma) and so excludes stars, GCs and some UCDs.

Fig. \ref{mflux50} shows the $m - <\!\!\mu\!\!>_{R_{50}}$ diagram used to select our initial sample and candidates. Coma membership assignment was based on data from NED\footnote{http://nedwww.ipac.caltech.edu/} and our 2007 spectroscopic data. The mean surface brightness used here is computed as,
\vspace{2ex}
\begin{center}
$<\!\!\mu\!\!>_{R_{50}} = m_{F475W} + 2.5$log$[2\pi R_{50}^{2}]$
\end{center}
\vspace{2ex}

\noindent where $m_{F475W}$ is the SE Kron-like magnitude and $R_{50}$ the radius of a circular aperture which contains 50\% of the total flux within the Kron-like aperture in arcseconds.

This process revealed three confirmed Coma compact galaxies (CcG) with a mean surface brightness well above the global Coma trend (CcGV1, 9a and 18 in Fig. 1). In addition it highlighted a number of other candidate objects without redshifts which were in turn eyeballed and a qualitative probability of cluster membership assigned. Five were deemed to have a high probability of membership based on their near circular appearance and lack of visible substructure. No new high probability candidates were found when the colour cut was extended 0.1 mag blue or redward. Of the five probable members four were on the footprint for the Hectospec followup run in May 2008 and subsequently all four were confirmed as cluster members. Unfortunately the fifth could not be observed due to fibre deployment issues.

Table \ref{objtab} lists our sample with SE magnitudes and matched MAG\_AUTO aperture colours corrected for extinction, K-correction and converted to B and I$_{c}$ bandpasses on the Vega system via,
\vspace{1.5ex}
\begin{center}
B = F475W$_{AB}$ + 0.329(F475W-F814W)$_{AB}$ + 0.097 \\
\vspace{1ex}
(B - I$_{c}$) = 1.287(F475W-F814W)$_{AB}$ + 0.538
\end{center}
\vspace{1.5ex}
\noindent which are derived using the methods described previously. The SE Kron-like magnitudes are near total being measured within an aperture that contains greater than 90\% of the light for typical galaxy profiles. Matched aperture photometry is achieved by running SE in dual image mode.

\begin{figure*}
\centering
\begin{minipage}{120mm}
\scalebox{0.56}[0.56]{\includegraphics{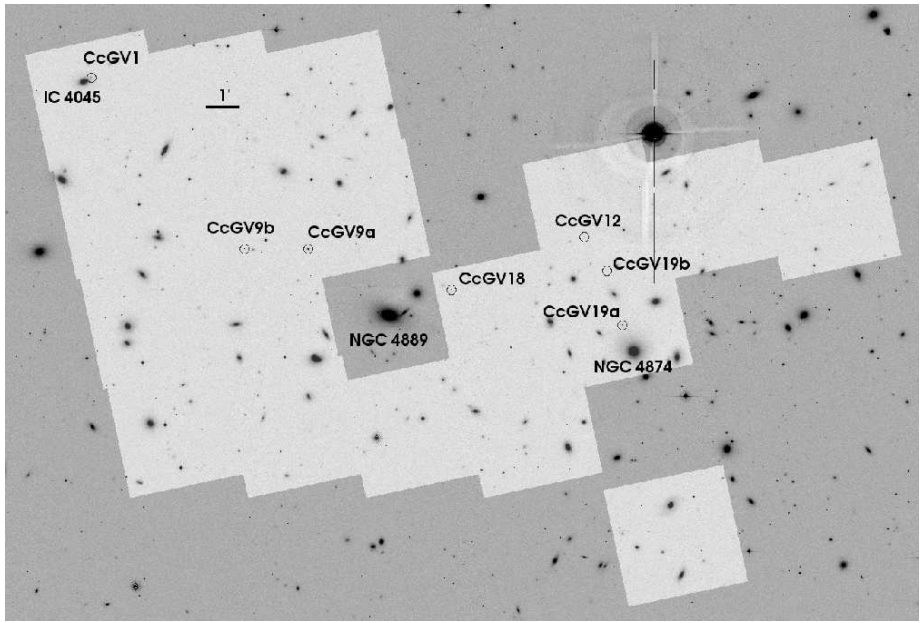}}
\caption{Portion from a CFHT MegaCam g-band image of the central part of the
Coma cluster. The locations of our compact galaxies are indicated, as
well as the three giant galaxies with which some of them may be
associated. The darkened regions show areas outside of the HST/ACS
survey footprint.}
\label{mosaic}
\end{minipage}
\end{figure*}

In Fig. \ref{cmr} we present the colour-magnitude relation for all confirmed Coma members imaged by the survey. Of our compact sample, six of the galaxies sit well above the sequence defined by normal cluster galaxies at their respective luminosities with only CcGV9a being close to the general trend. It is noted that there is some scatter in the correlation but the systematic deviation of our sample toward redder colours by 0.2 to 0.4 mag is clearly apparent. Put another way, almost all the CcGs are $\sim$ 3 magnitudes, or 20 times, fainter than other cluster galaxies of the same colour.

\subsection{Location in the Cluster}

In Fig. \ref{mosaic} we present a CFHT g band mosaic of the inner regions of the Coma cluster overlayed with the positions of our sample and relevant giant galaxies for reference. We note that throughout this subsection the redshifts and their respective errors for the bright neighbouring galaxies are taken from the SDSS as they were not observed by Hectospec.

Both CcGV1 and CcGV19a are seen in projection close to significantly larger galaxies. CcGV1 is exceptionally close to IC 4045 at a projected distance of $\sim$6.7 kpc and given a difference in their radial velocities of $\Delta v_{r}$ = 117$\pm$53 km s$^{-1}$ it can be assumed they are gravitationally interacting. CcGV19a is less immediately located near its neighbour at $\sim$23~kpc from NGC 4874, yet still sits in the cDs diffuse halo. In this case $\Delta v_{r}$ = 438$\pm$65 km s$^{-1}$.

CcGV19b and CcGV12 are located $\sim$68~kpc and 101~kpc from NGC 4874 respectively. With the former having $\Delta v_{r}$ = 99$\pm$65 km s$^{-1}$, it is likely to be associated with the cD while CcGV12, having $\Delta v_{r}$ = 650$\pm$66 km s$^{-1}$ with respect to it, would appear not to be.

CcGV18 also has a prominent near neighbour in the form of NGC 4886 which is located at a projected distance of $\sim$28 kpc. Comparing their radial velocities we find $\Delta v_{r}$ = 138$\pm$54 km s$^{-1}$ which again is possible evidence for their interaction. Additionally we note that NGC 4889, the brightest galaxy in the cluster, is a projected distance of $\sim$54 kpc away with $\Delta v_{r}$ = 35$\pm$82 km s$^{-1}$.

Finally, both CcGV9a and CcGV9b are relatively isolated spatially. CcGV9a has a projected distance of $\sim$84 kpc to NGC 4889 and few major galaxies within this radius. Similarly CcGV9b has no major galaxies within 55 kpc.

\section{Photometric Properties}

\subsection{Structural Parameters}

We have analysed our sample using the two-dimensional fitting algorithm Galfit \citep{peng02} to recover their structural parameters. Each galaxy's light profile is modelled via axially symmetric ellipses and fit with one or more user specified analytic functions. The free parameters of each function are adjusted until the $\chi^{2}$ residual between the real and model image (summed over all pixels) is minimised. As well as the imaging data Galfit has a further five important inputs which are discussed in detail below. These inputs are the uncertainty map, the instrumental point spread function (PSF), image masks, the choice of parametric light profile model and finally the sky level.

\subsubsection{Uncertainty Maps}

To compute a meaningful $\chi^{2}$ for a given model and set of parameters Galfit requires a per pixel uncertainty or sigma image. Such maps may be created by the program itself or otherwise supplied by the user. Here we opt to take advantage of the 'ERR' inverse variance map produced by Multidrizzle which includes poisson noise from both sky and object together with instrument dependent contributions due to dark current and read noise. The maps are scaled by $\sim$0.77 to correct for correlated pixel noise (D. Hammer 2008, private communcation) and converted to 1$\sigma$ per pixel images ready for input into Galfit.

\subsubsection{PSF}

\begin{figure*}
\begin{minipage}{135mm}
\centering
\scalebox{0.19}[0.19]{\includegraphics{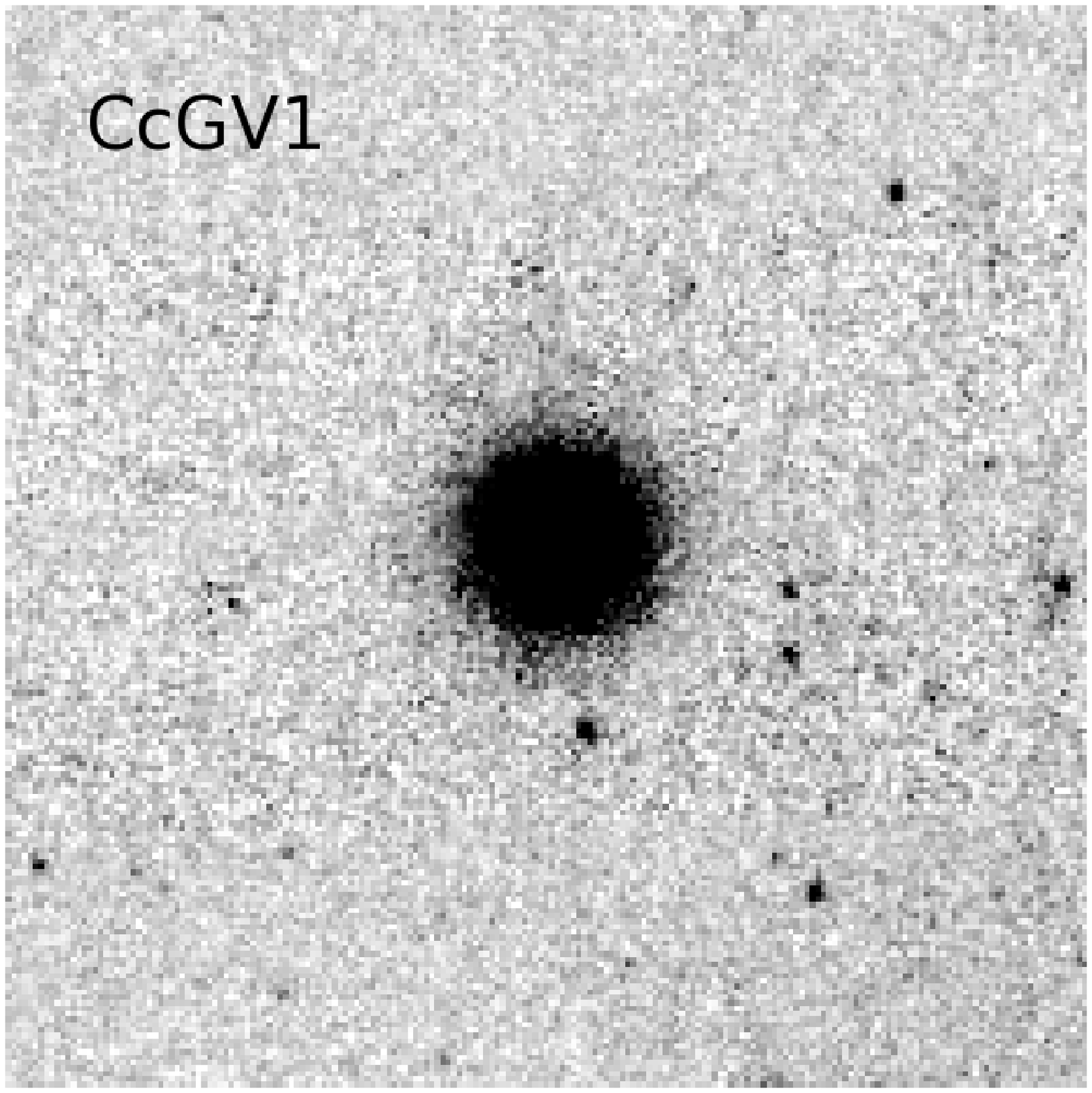}} 
\scalebox{0.19}[0.19]{\includegraphics{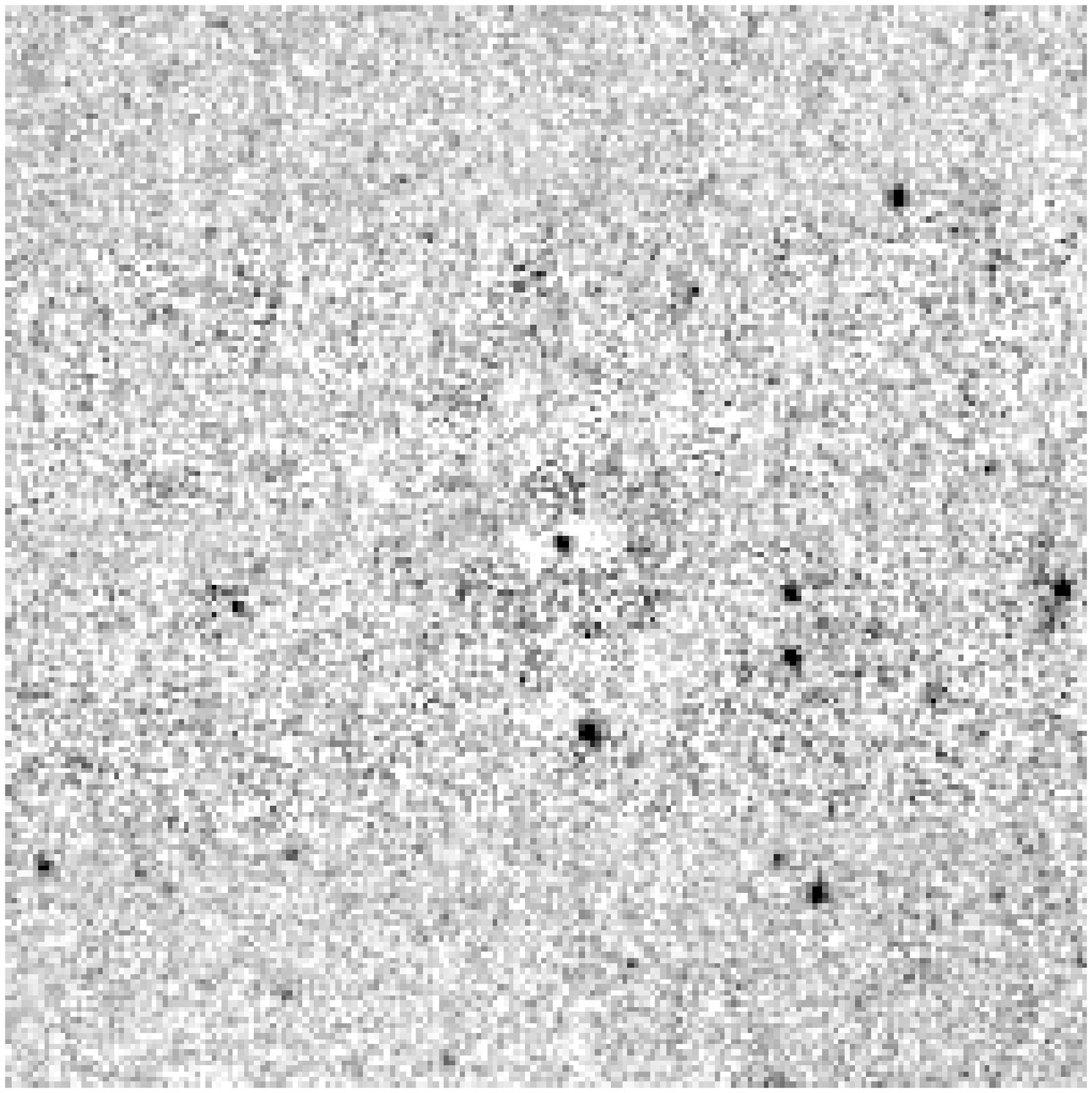}}
\scalebox{0.19}[0.19]{\includegraphics{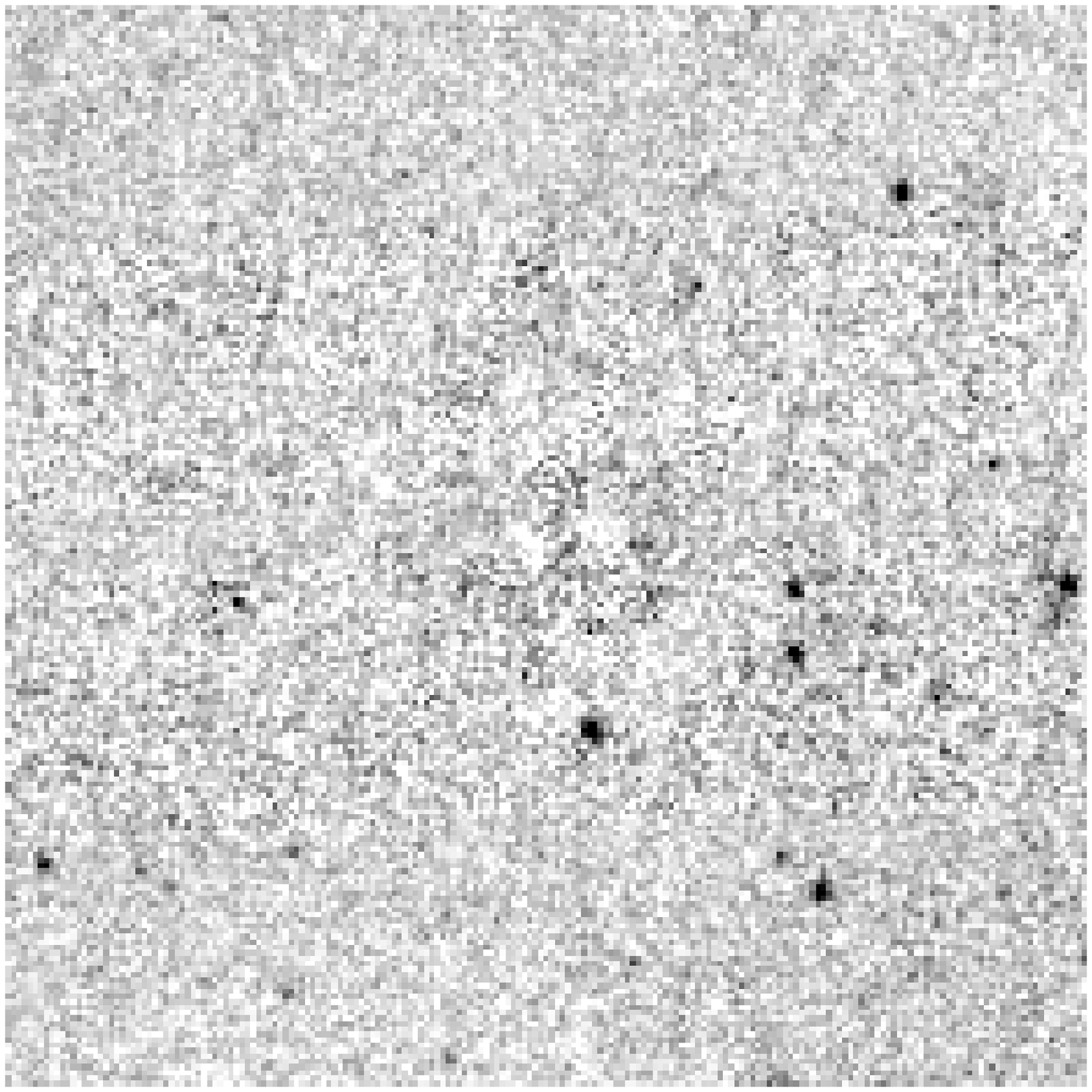}} 
\\
\scalebox{0.19}[0.19]{\includegraphics{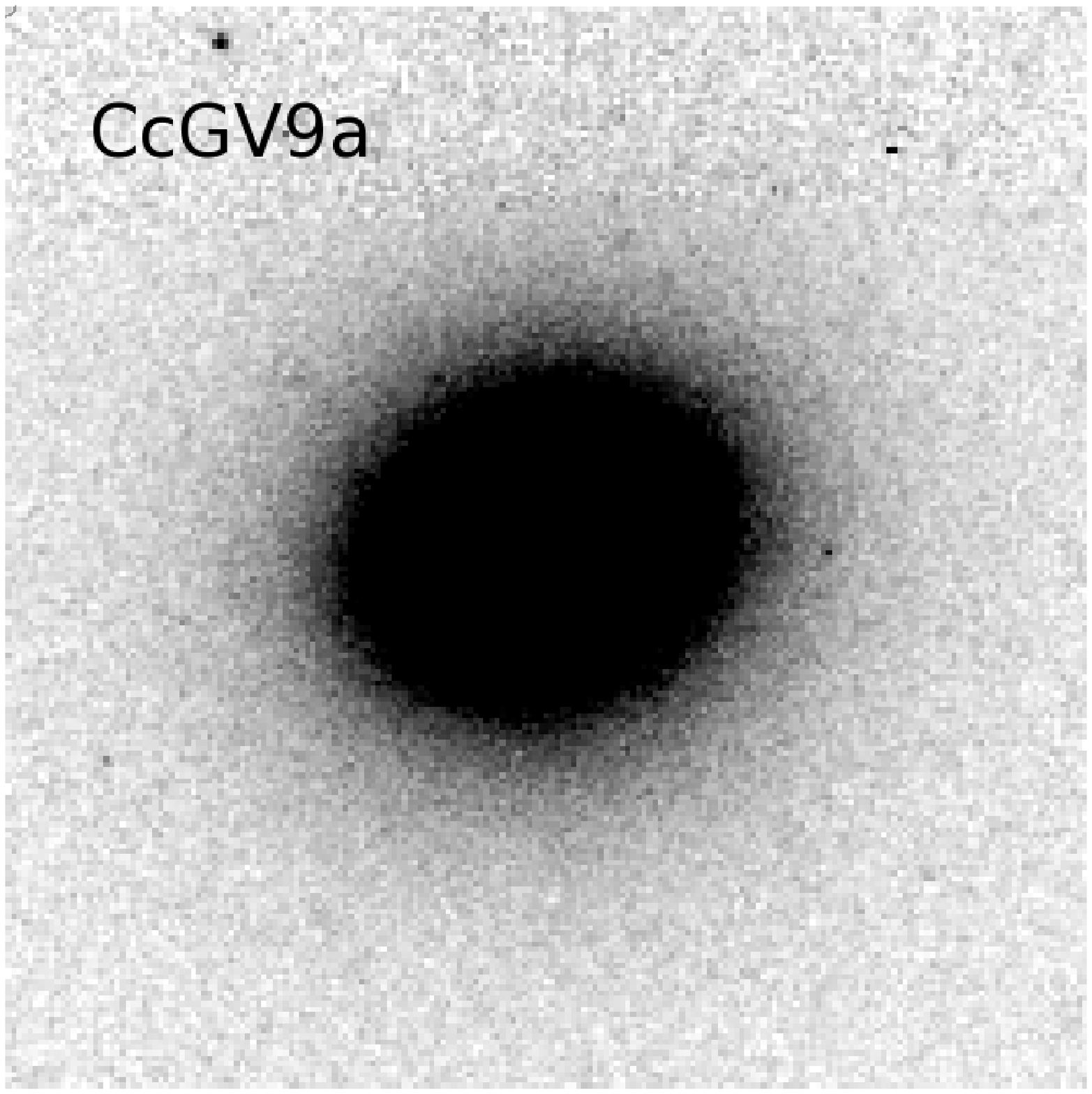}} 
\scalebox{0.19}[0.19]{\includegraphics{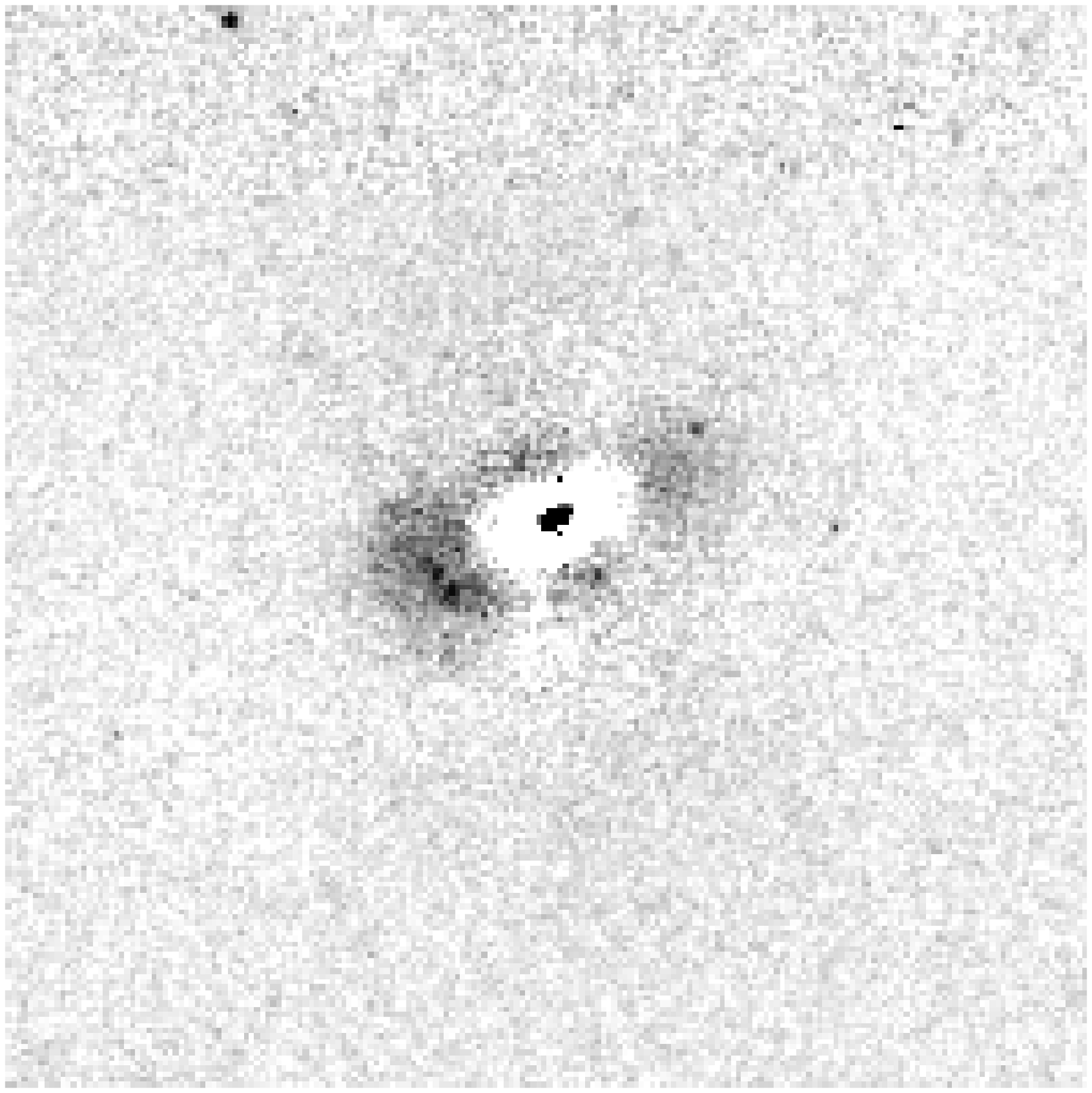}}
\scalebox{0.19}[0.19]{\includegraphics{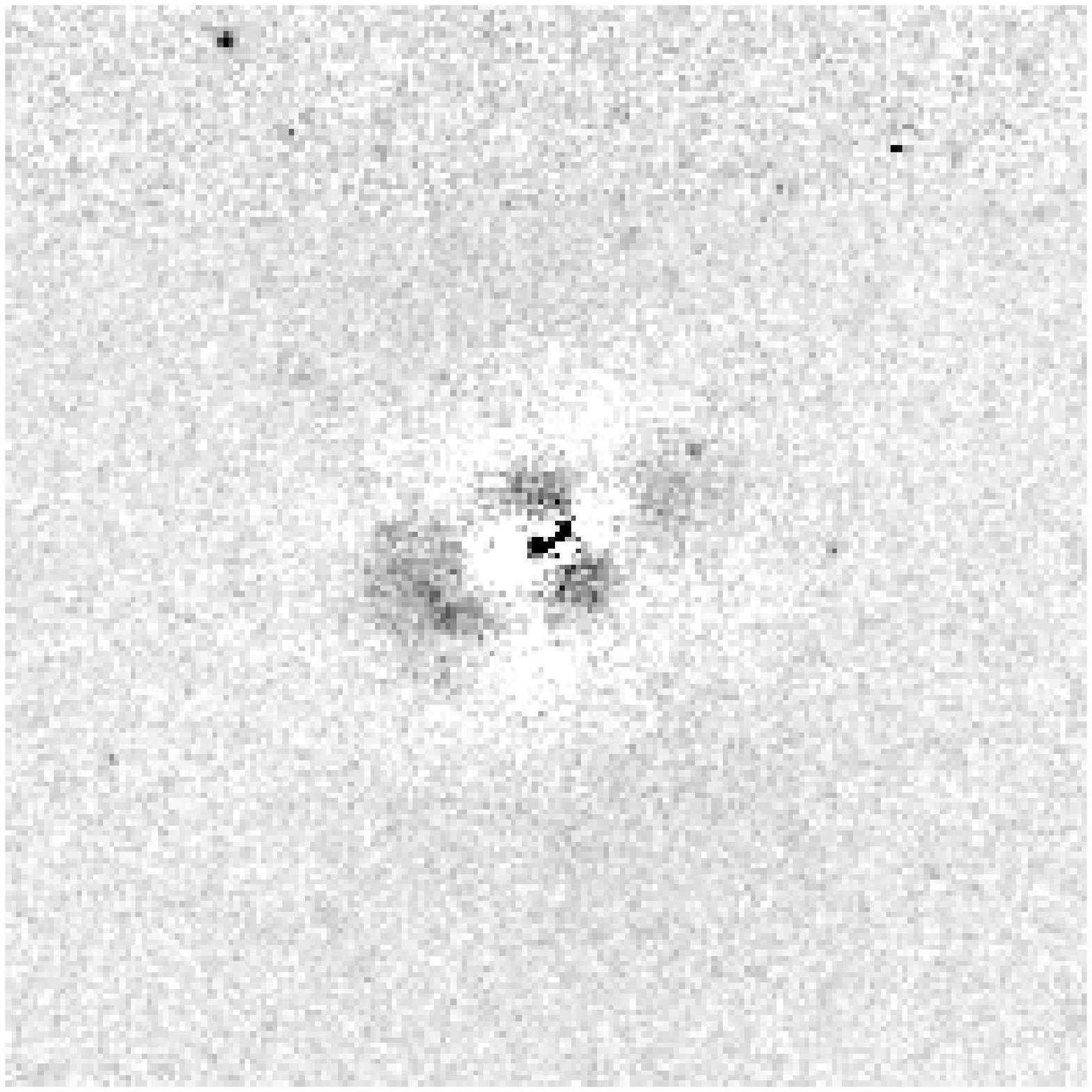}} 
\\
\scalebox{0.19}[0.19]{\includegraphics{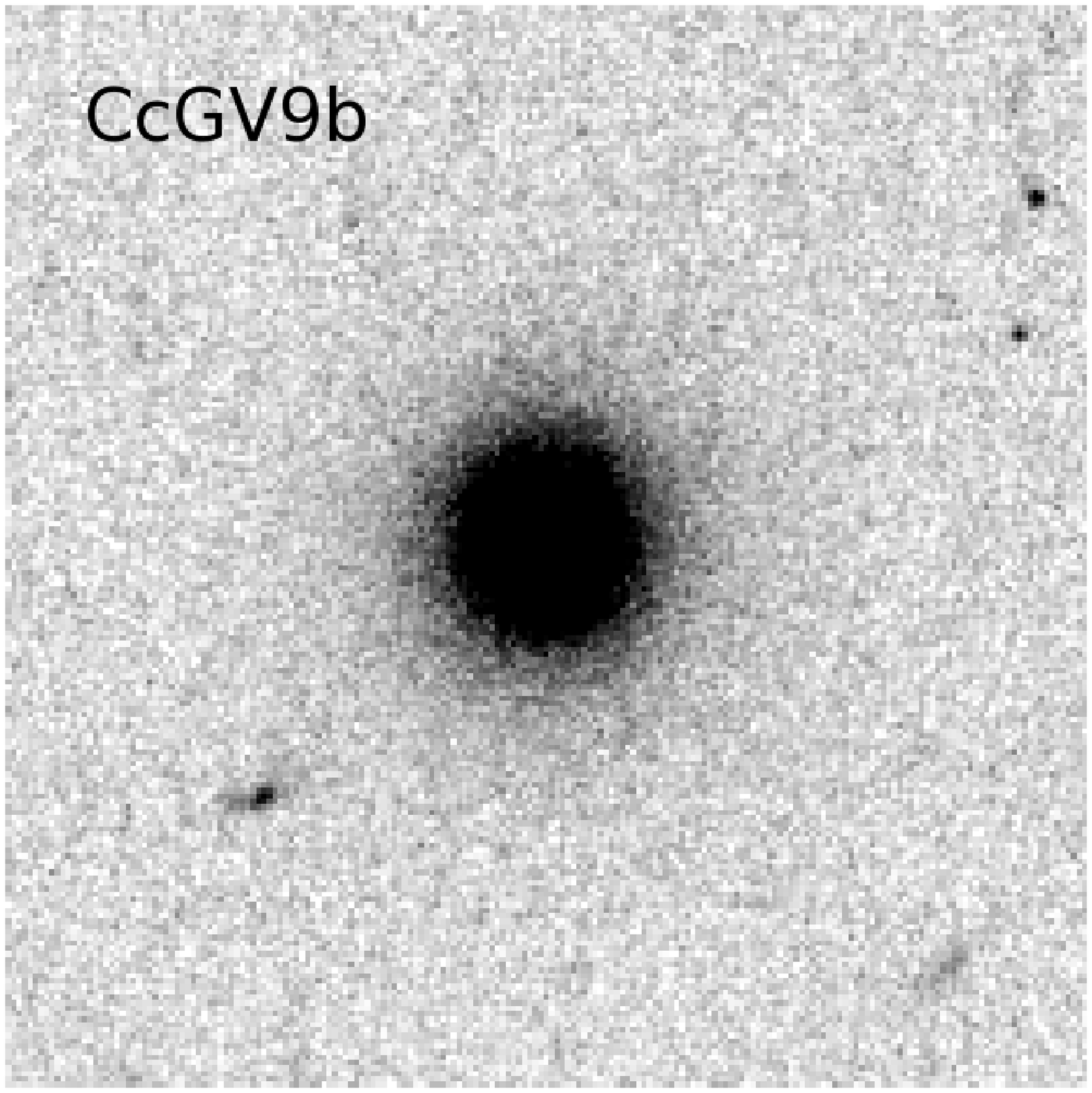}}
\scalebox{0.19}[0.19]{\includegraphics{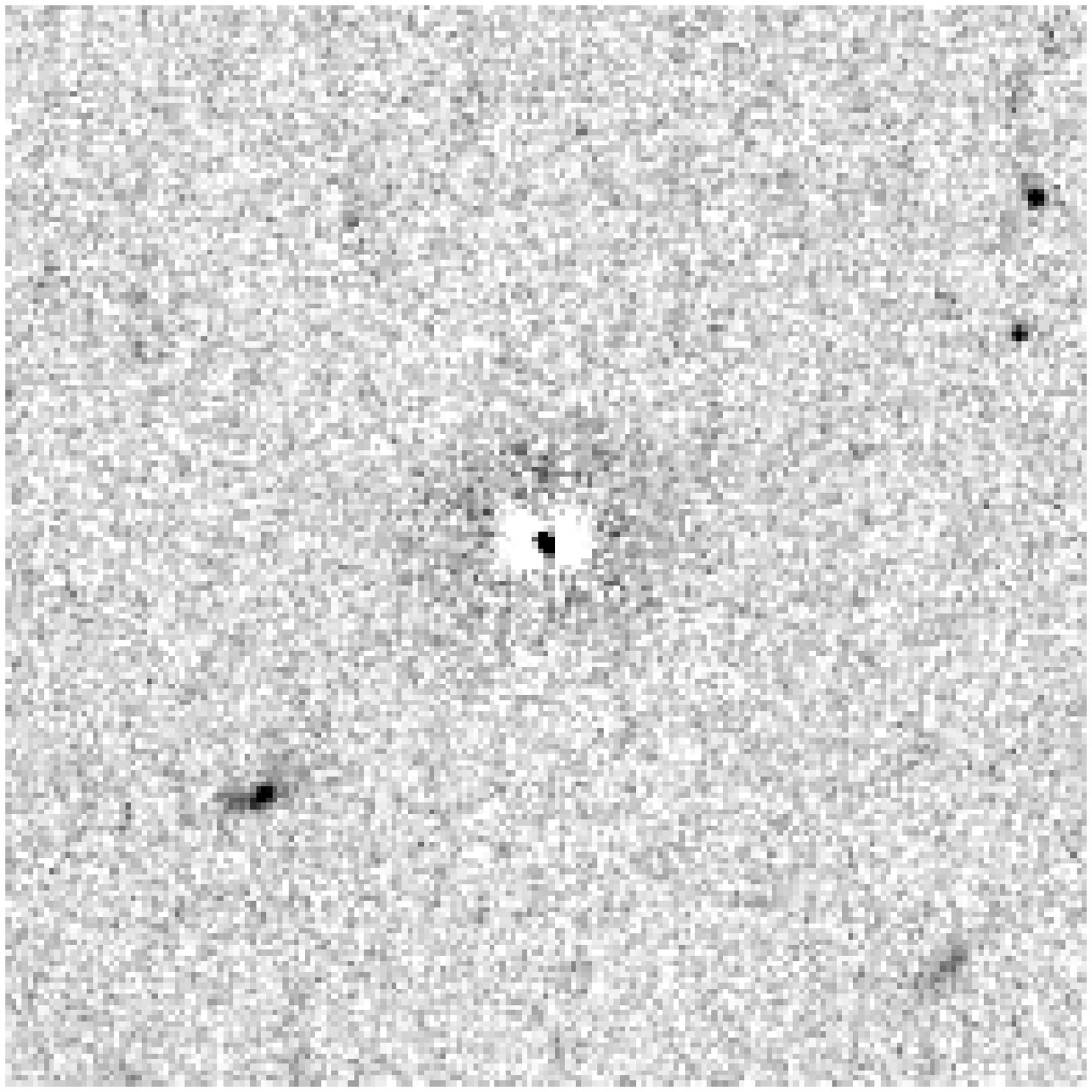}} 
\scalebox{0.19}[0.19]{\includegraphics{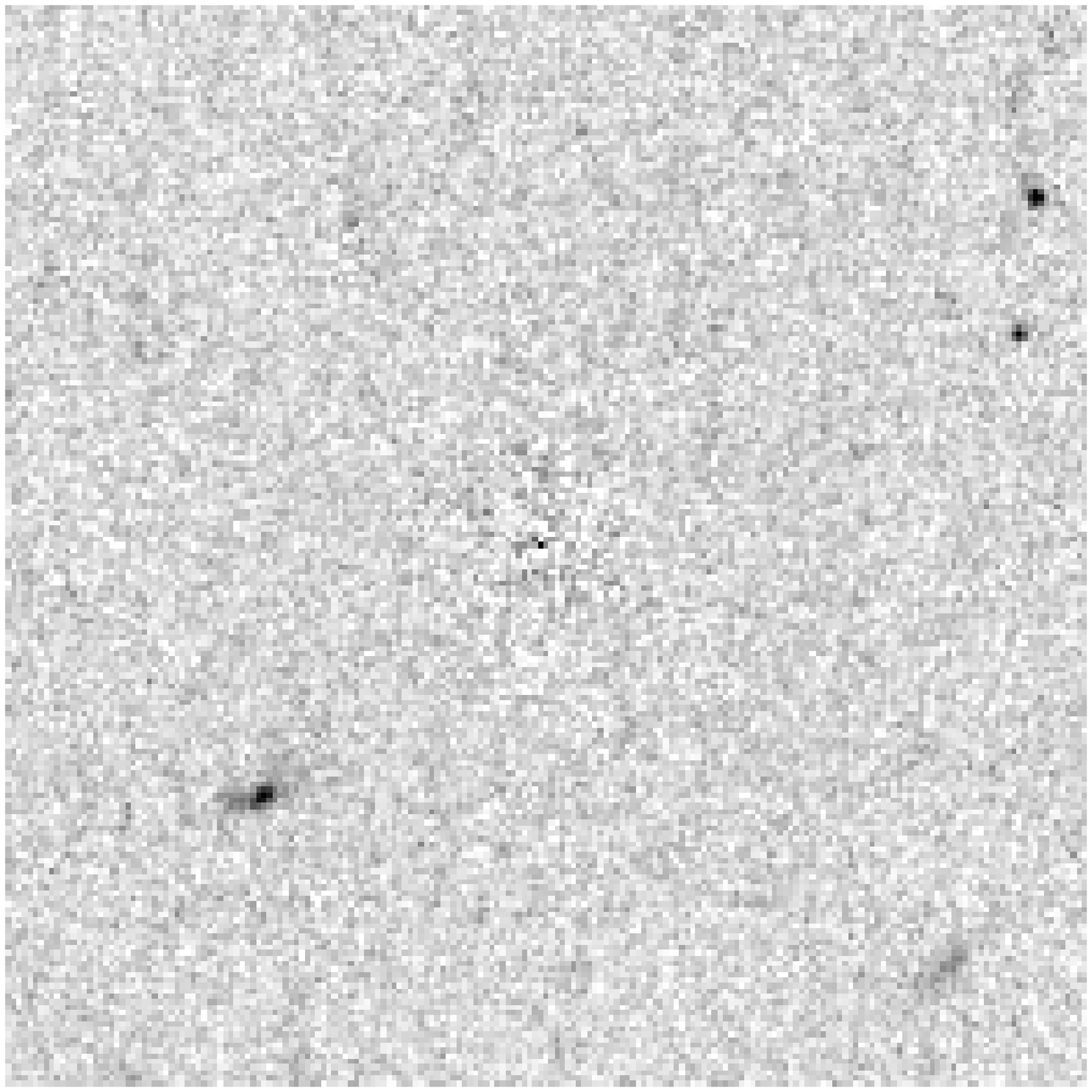}}
\\
\scalebox{0.19}[0.19]{\includegraphics{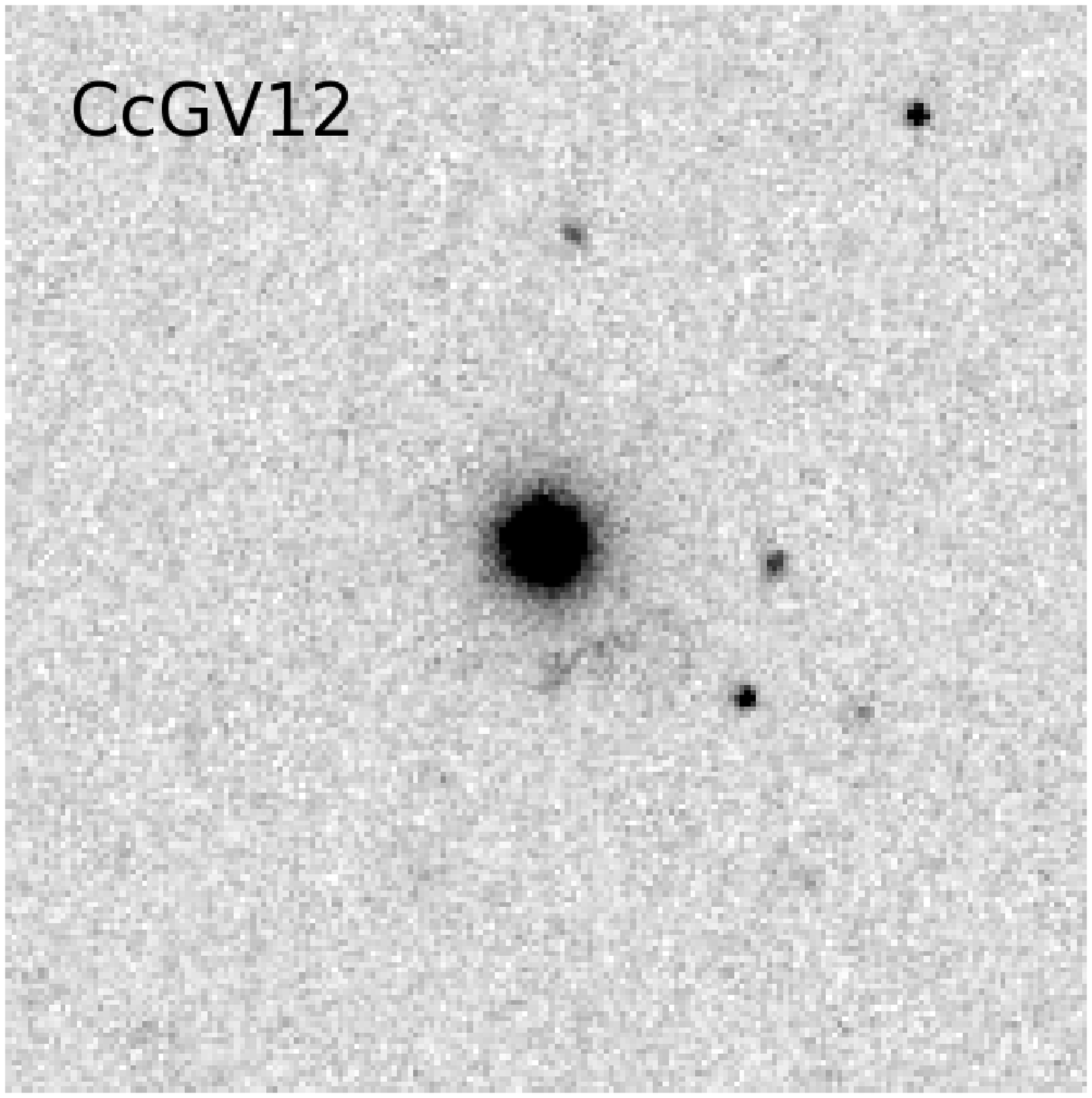}}
\scalebox{0.19}[0.19]{\includegraphics{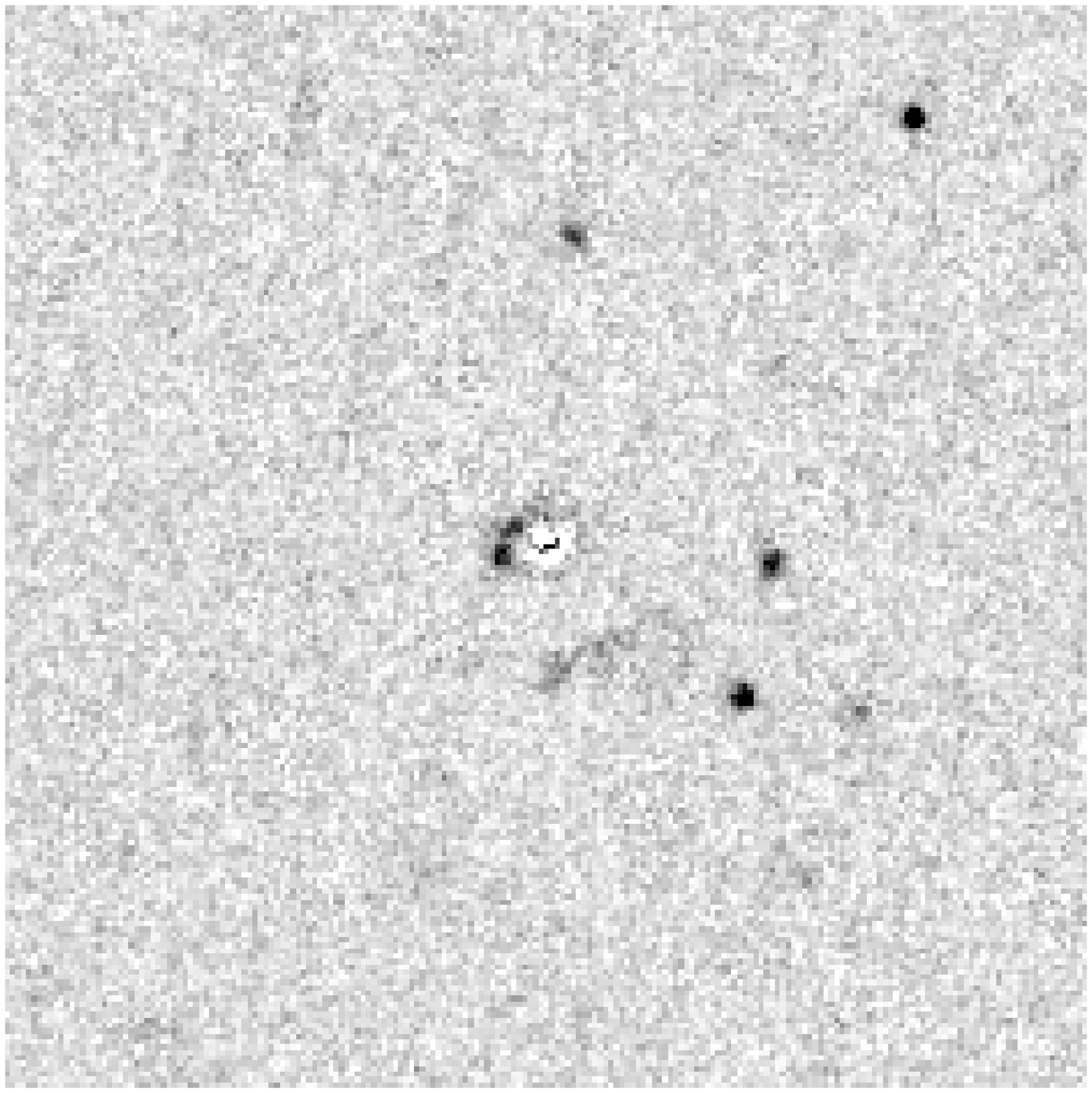}} 
\scalebox{0.19}[0.19]{\includegraphics{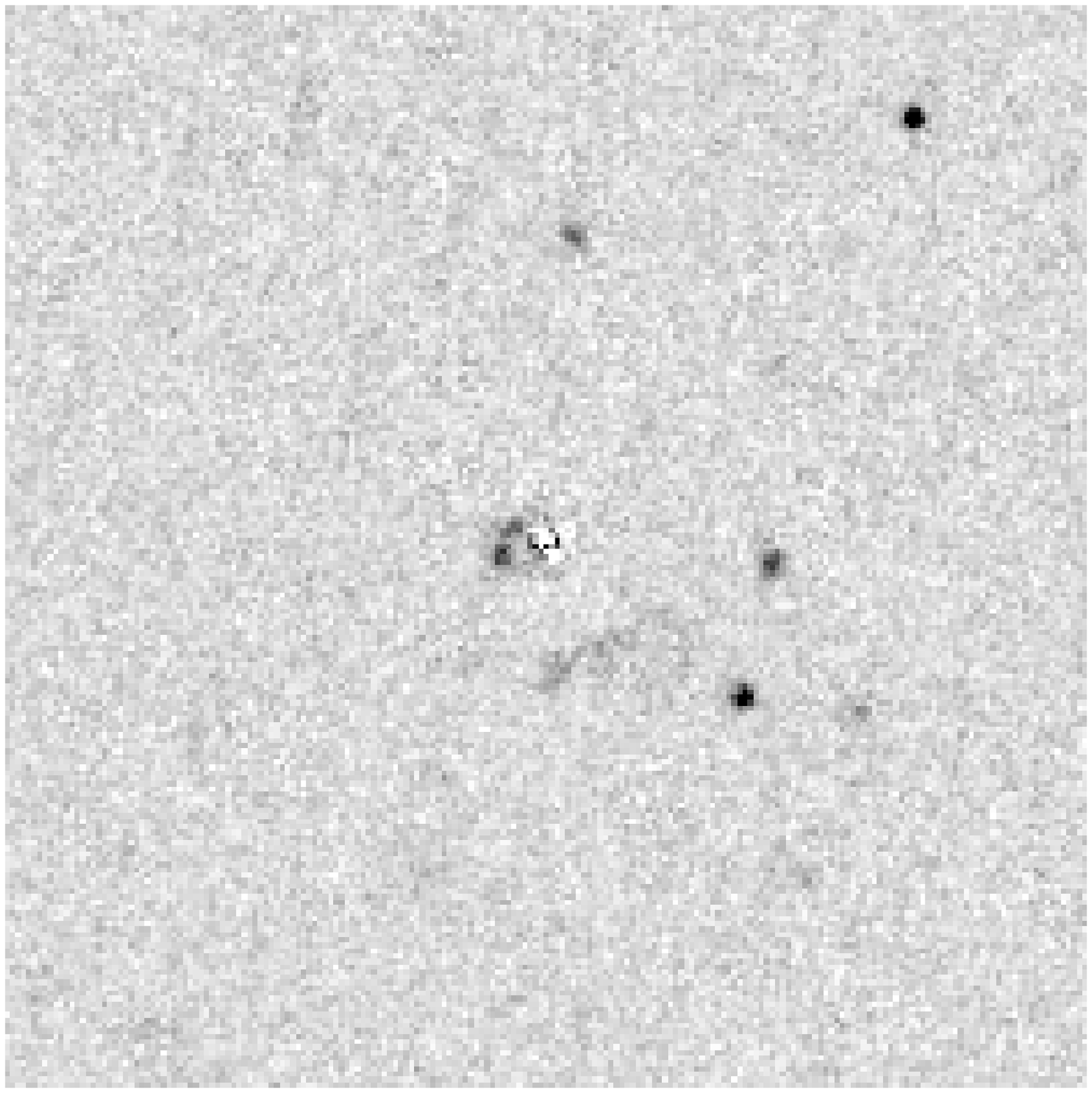}}
\\
\scalebox{0.19}[0.19]{\includegraphics{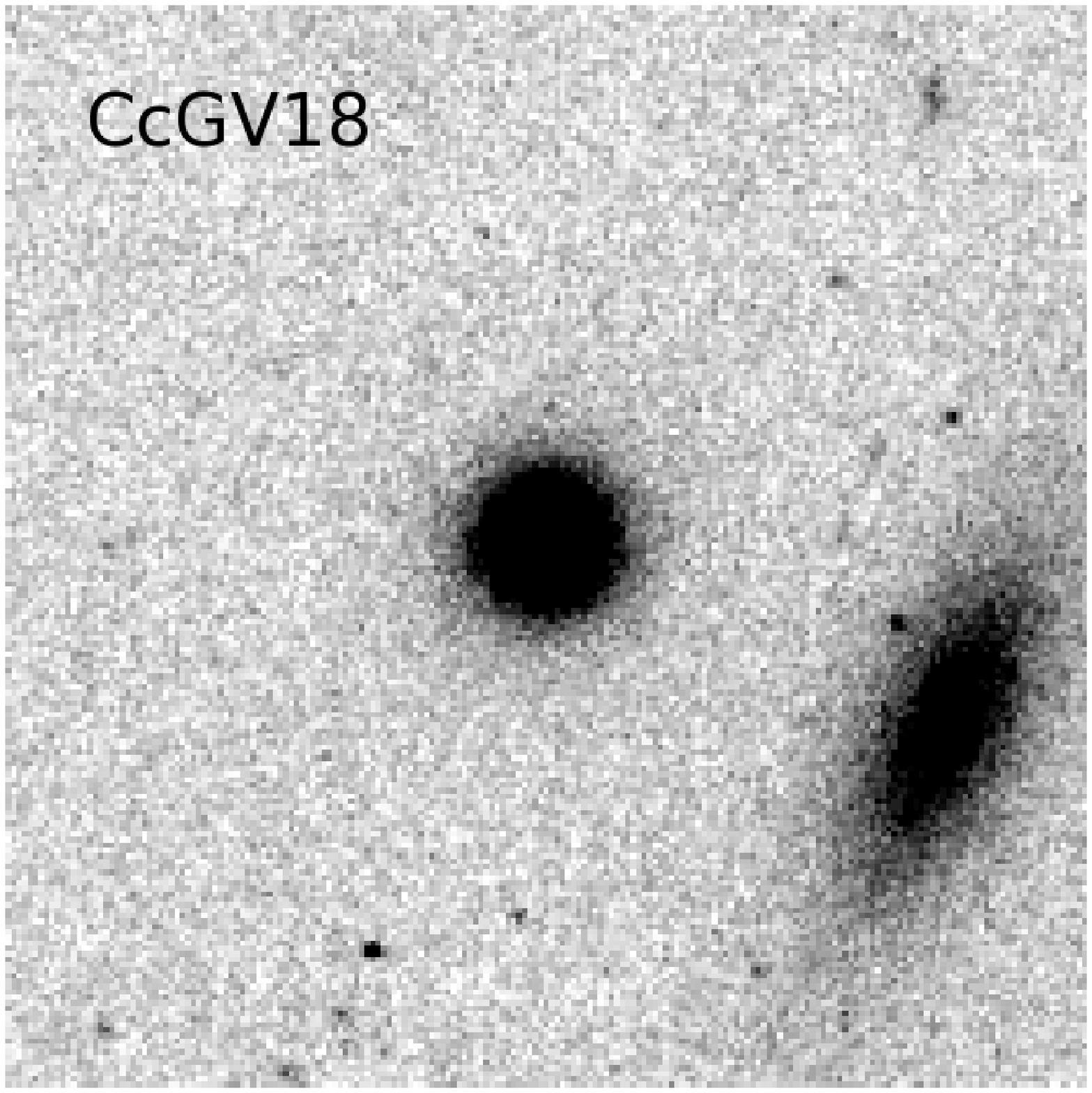}}
\scalebox{0.19}[0.19]{\includegraphics{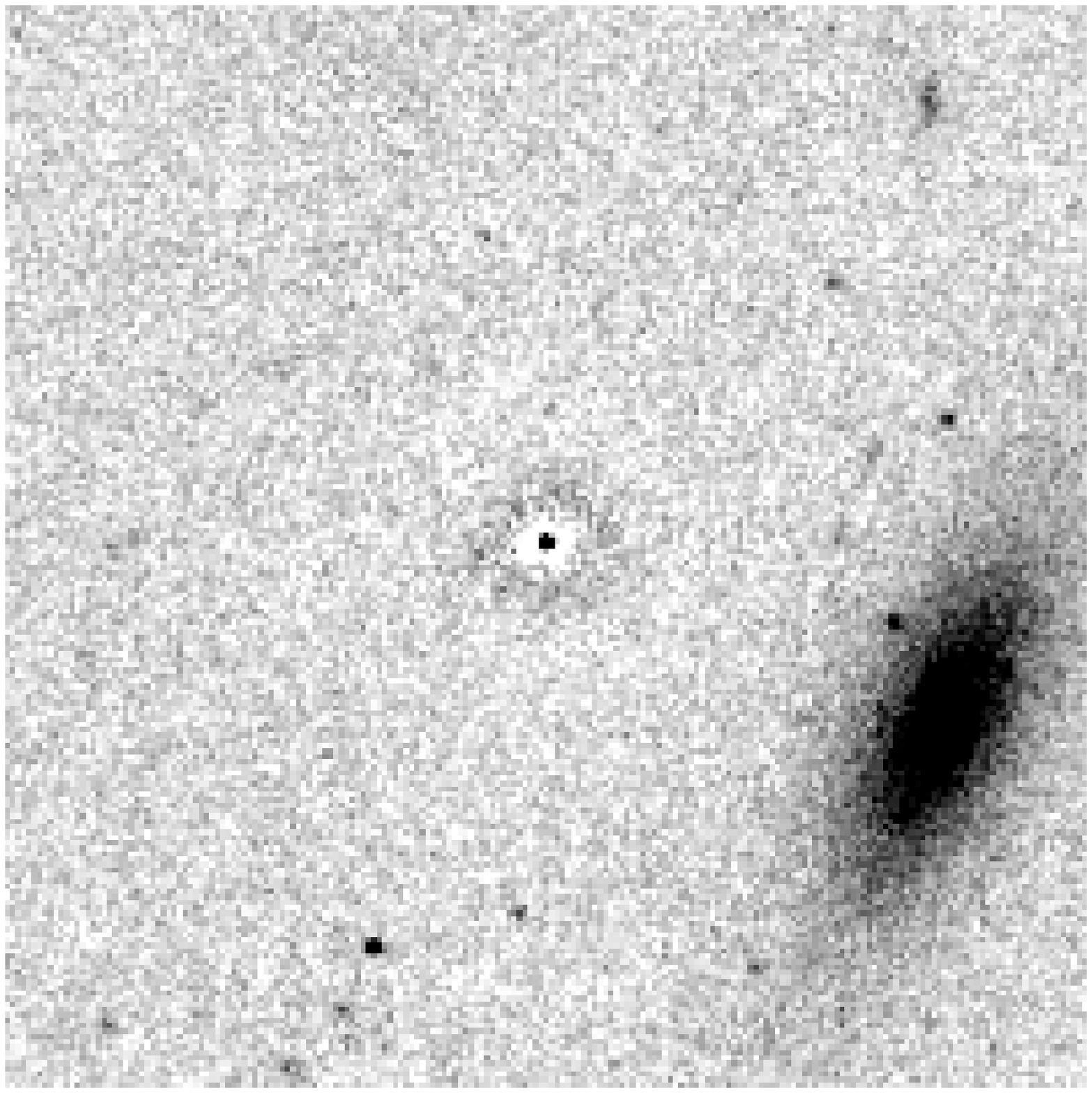}} 
\scalebox{0.19}[0.19]{\includegraphics{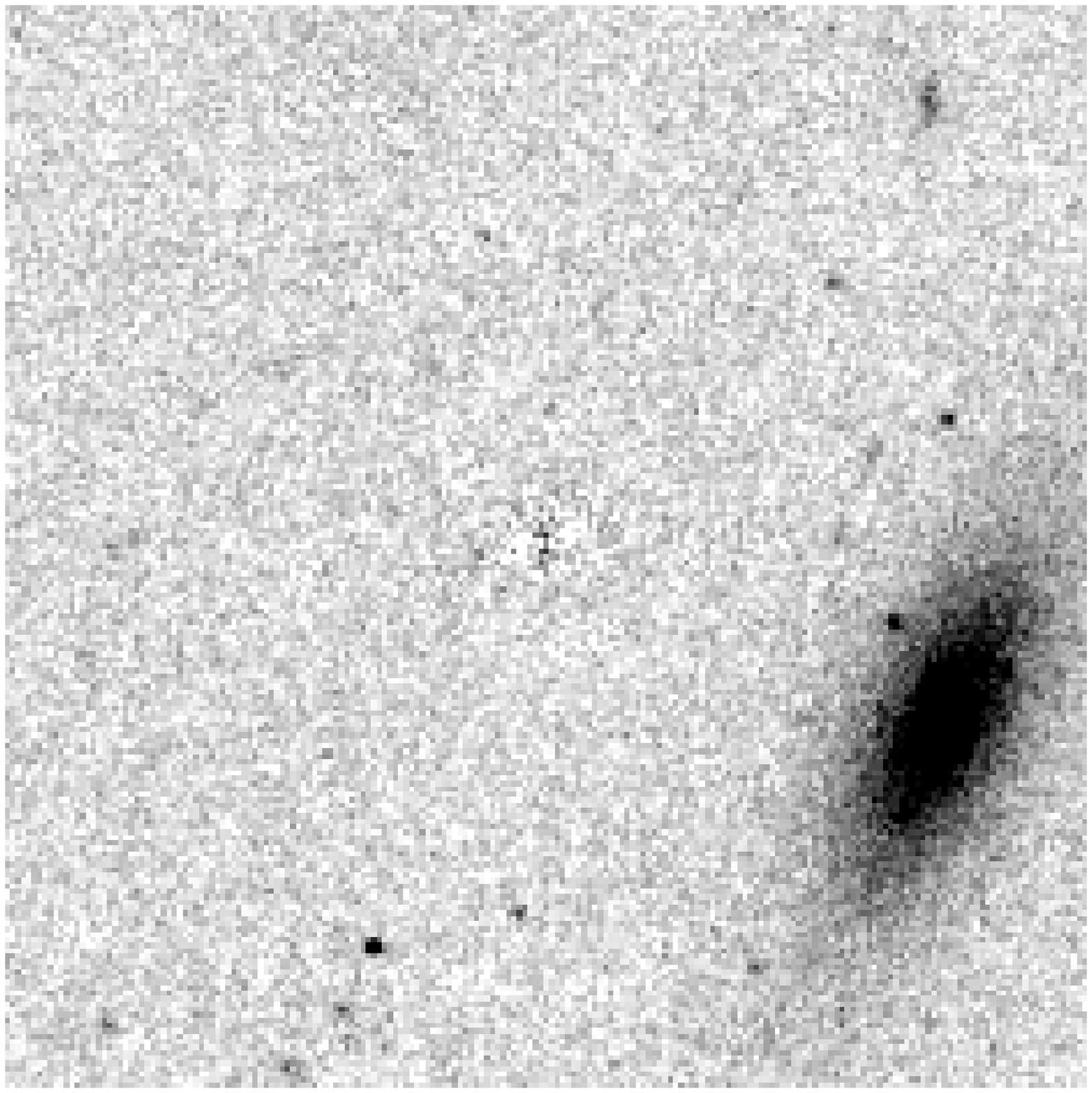}}
\\
\scalebox{0.19}[0.19]{\includegraphics{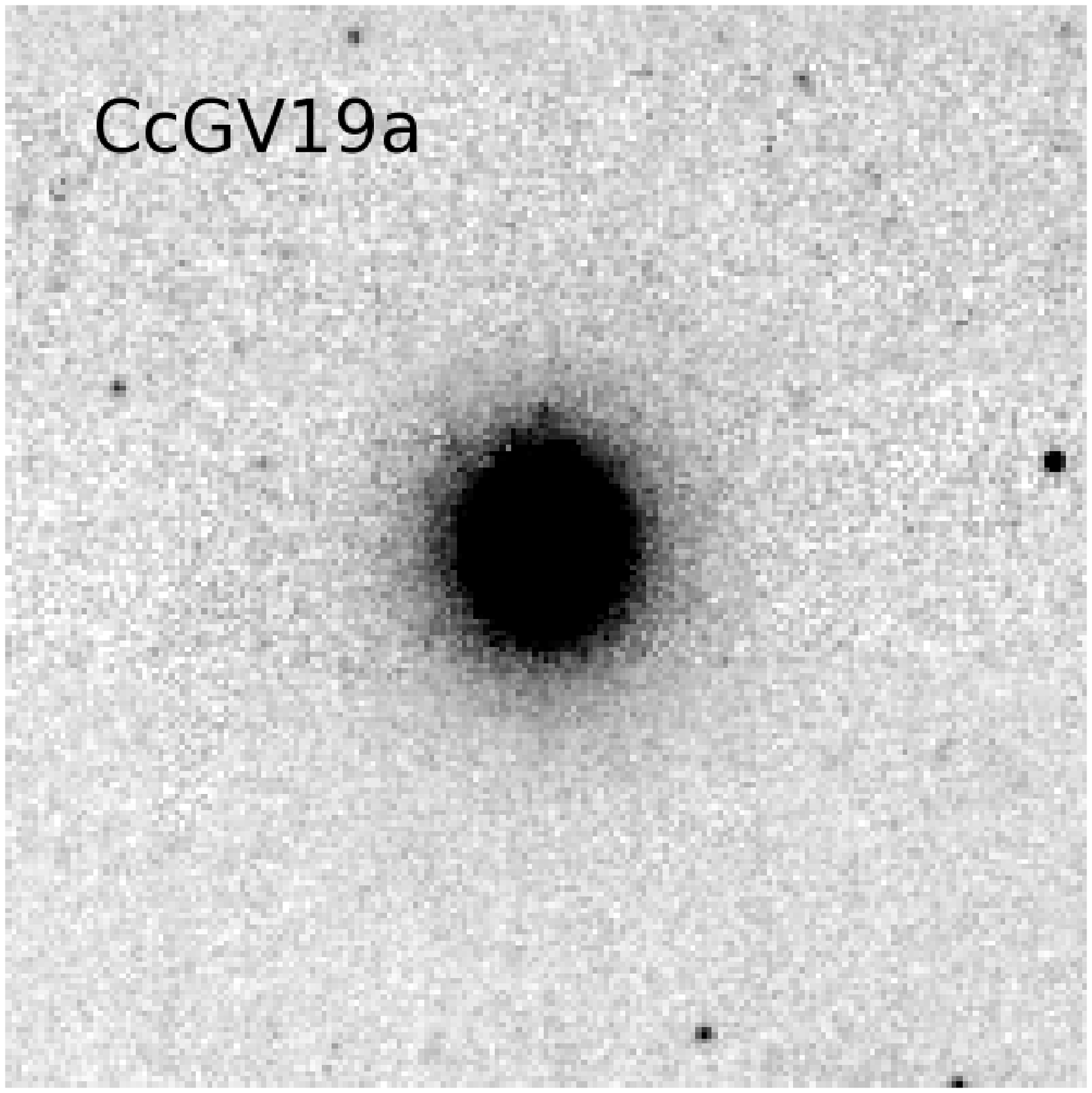}}
\scalebox{0.19}[0.19]{\includegraphics{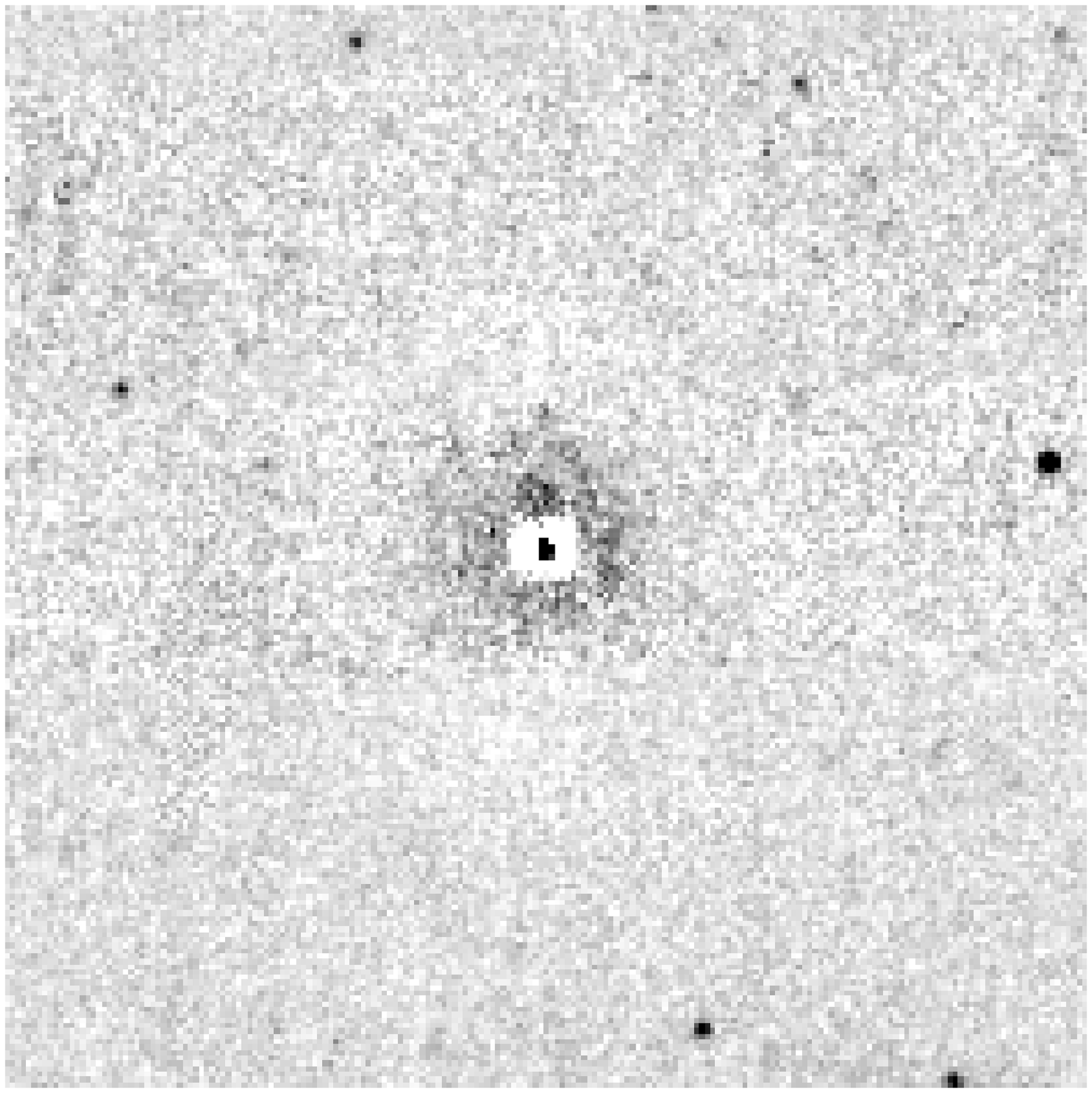}} 
\scalebox{0.19}[0.19]{\includegraphics{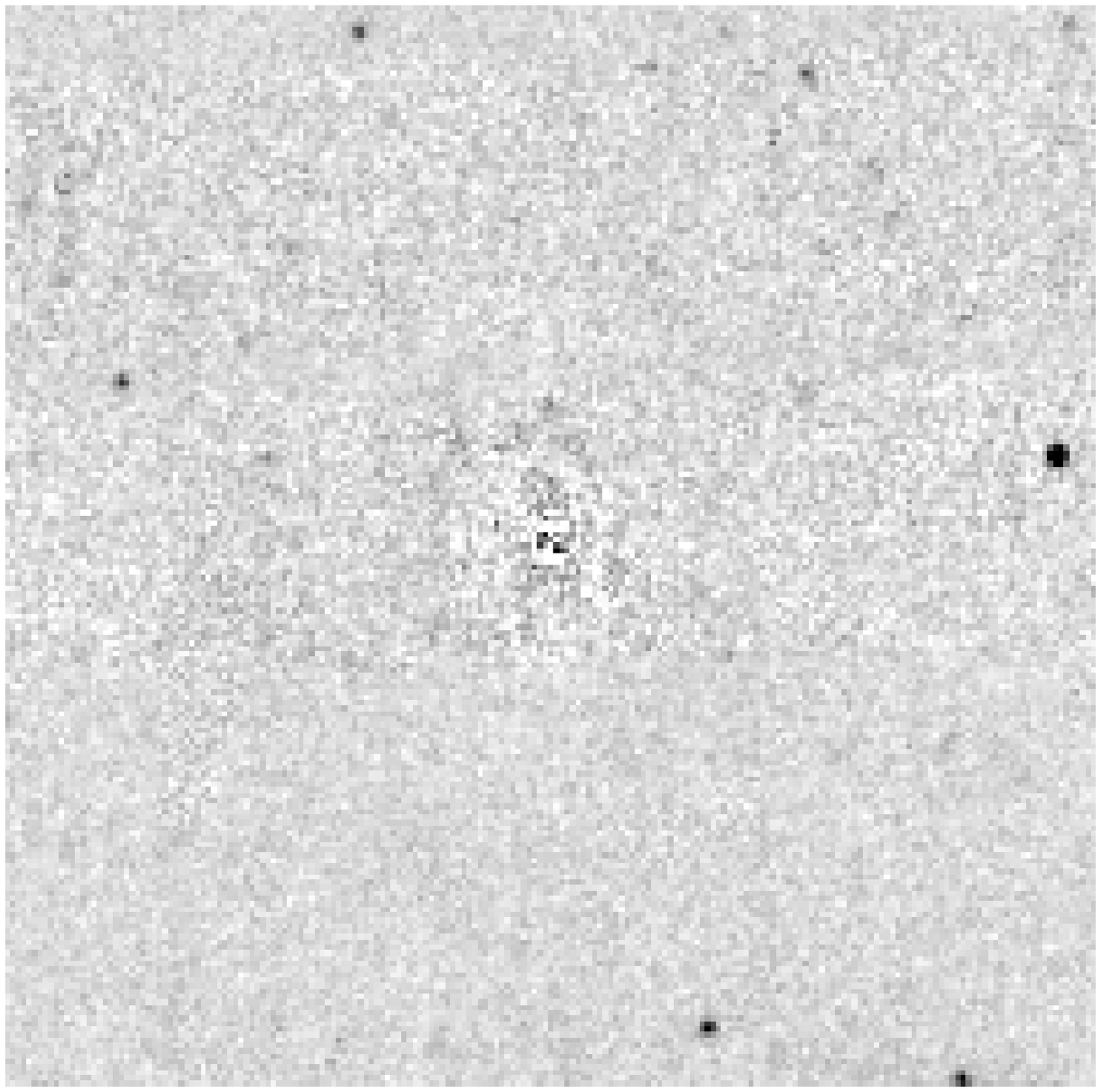}}
\\
\scalebox{0.19}[0.19]{\includegraphics{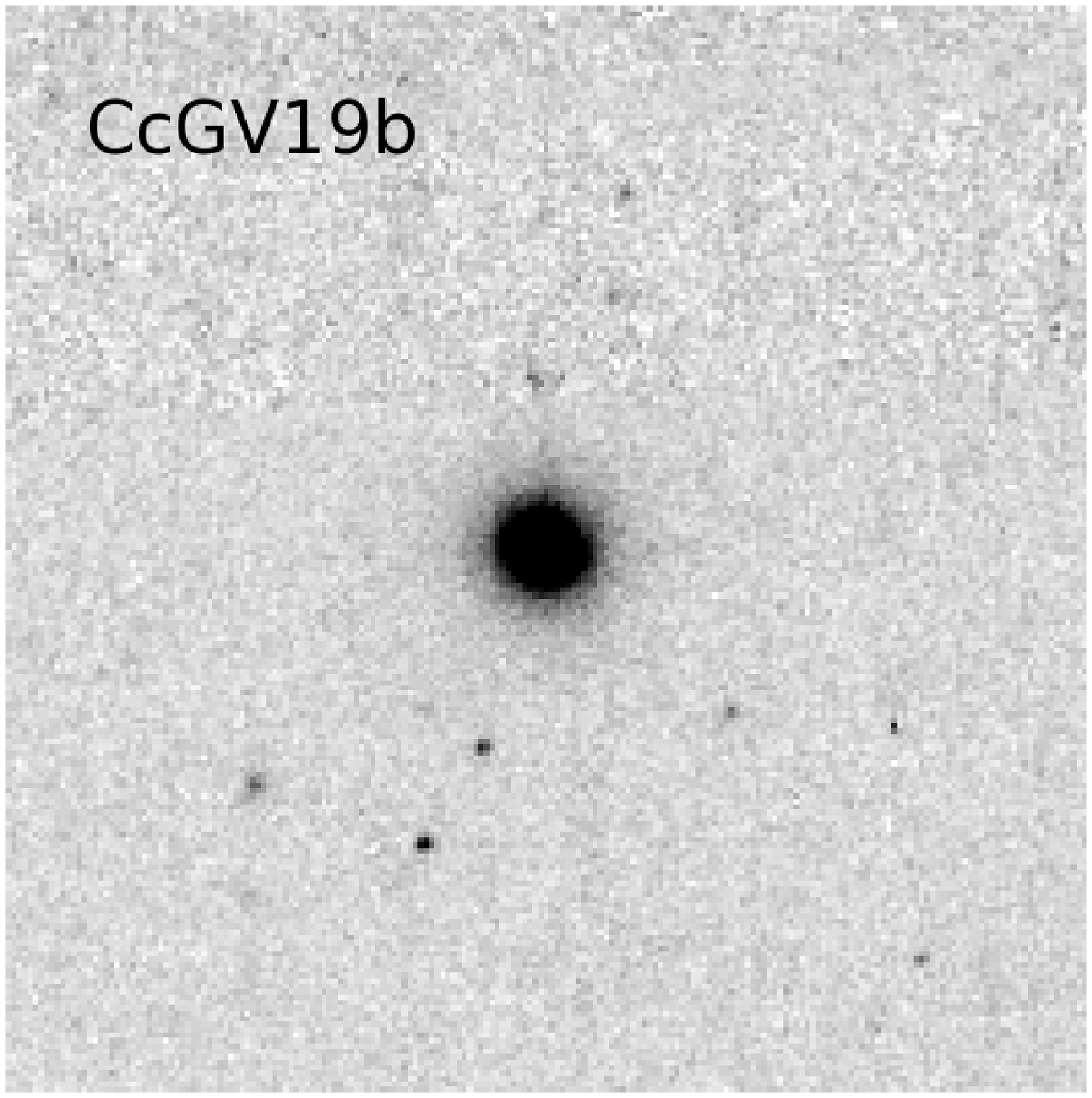}}
\scalebox{0.19}[0.19]{\includegraphics{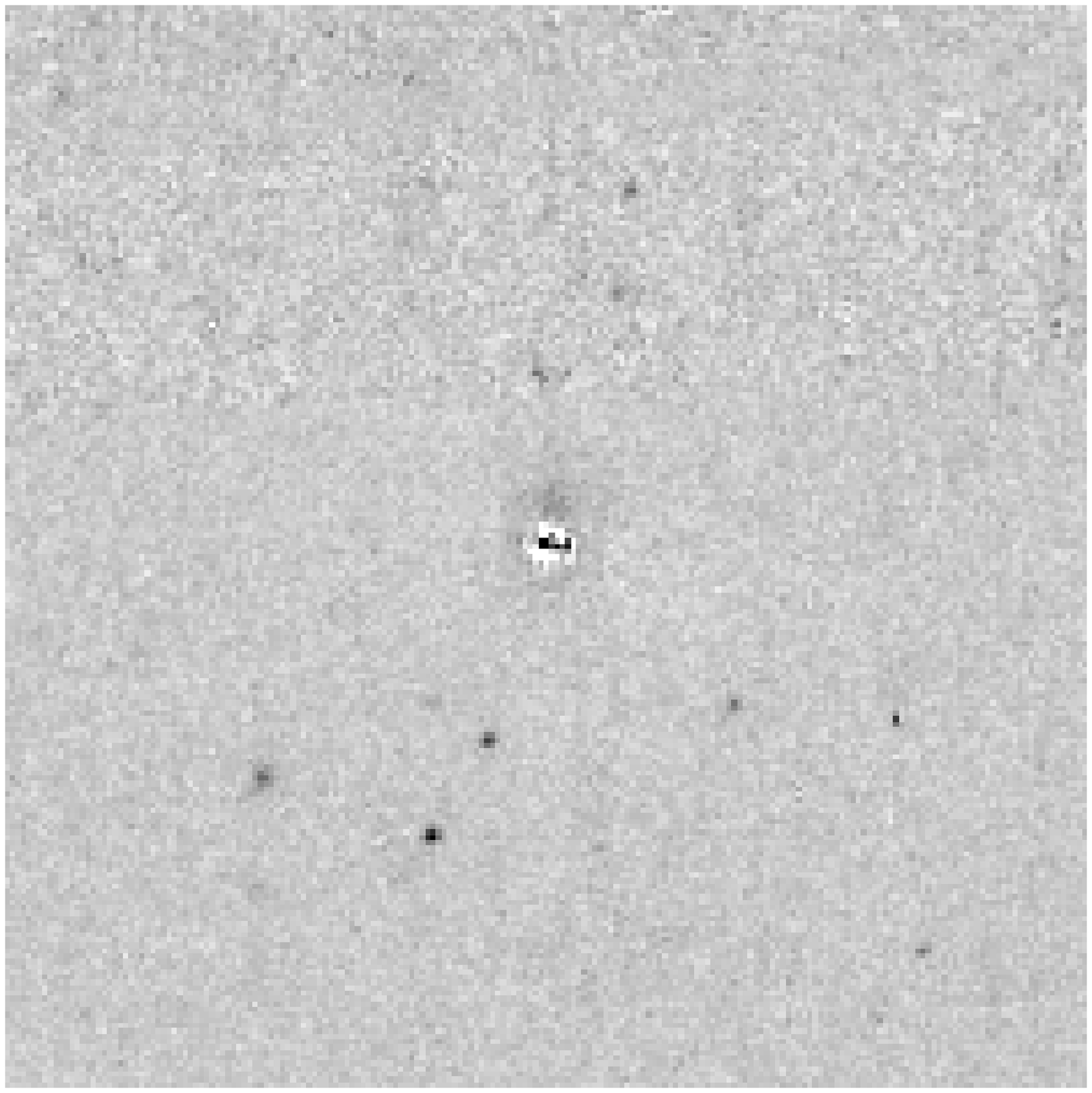}} 
\scalebox{0.19}[0.19]{\includegraphics{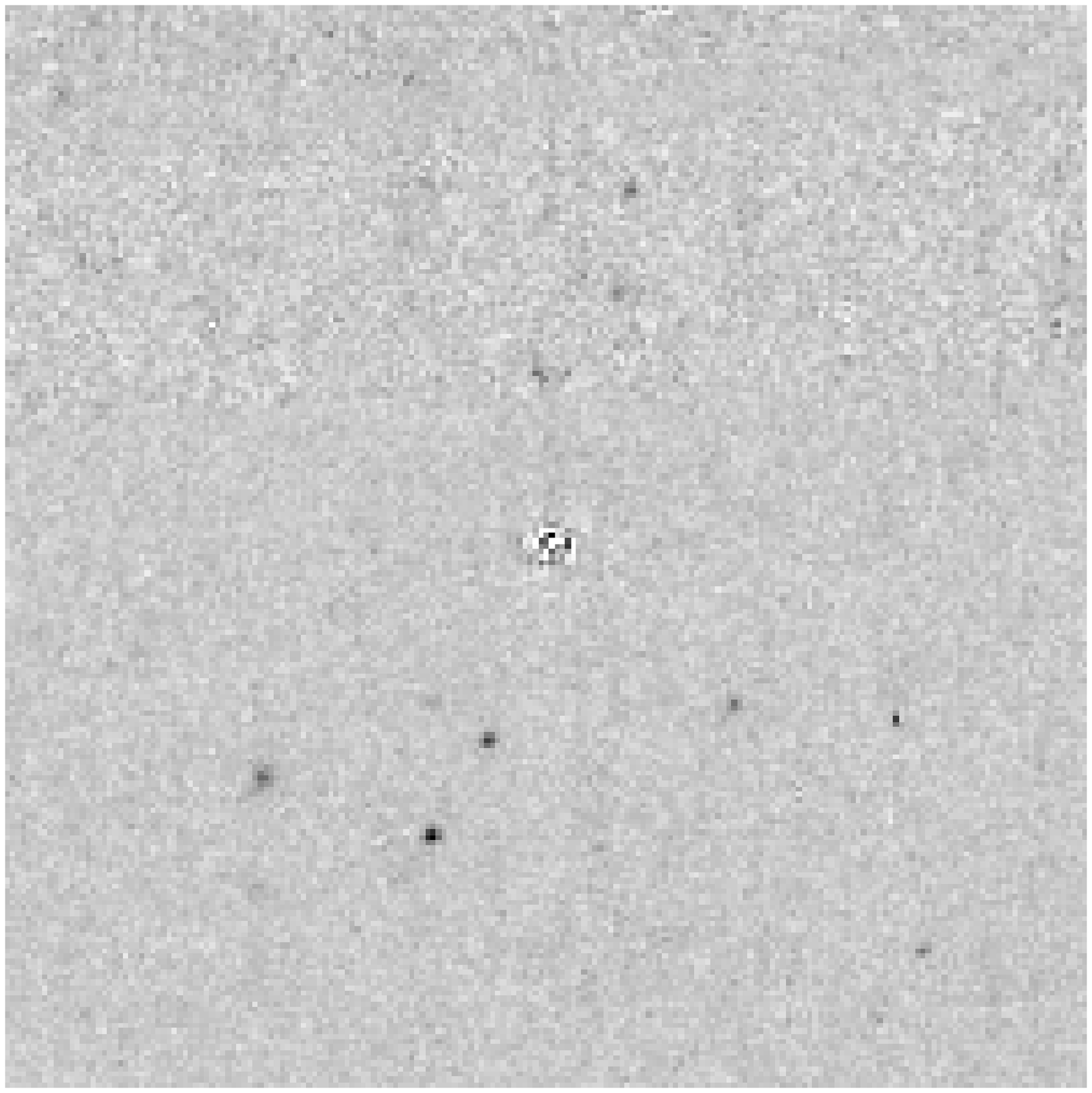}}
\\
\caption{Compact galaxy images in the F475W filter (left column), residuals from subtracting the best fitting seeing convolved single S{\'e}rsic model (middle column) and overall best fitting model (right column). All images are 10$^{\prime\prime}$x 10$^{\prime\prime}$ (4.6 kpc x 4.6 kpc).}.
\label{cutouts}
\end{minipage}
\end{figure*}

\begin{figure*}
\begin{minipage}{135mm}
\centering
\scalebox{0.32}[0.32]{\includegraphics{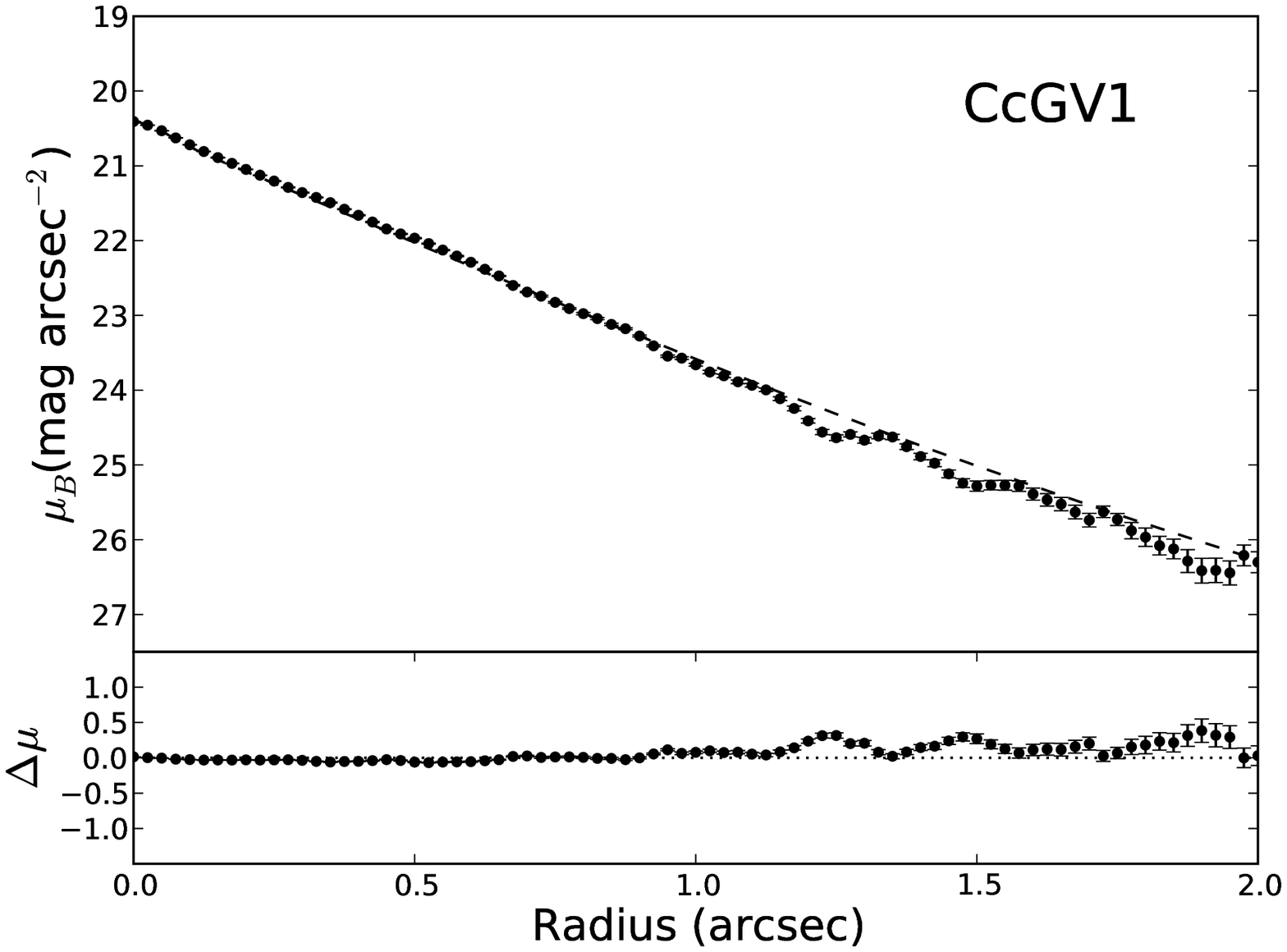}} \scalebox{0.32}[0.32]{\includegraphics{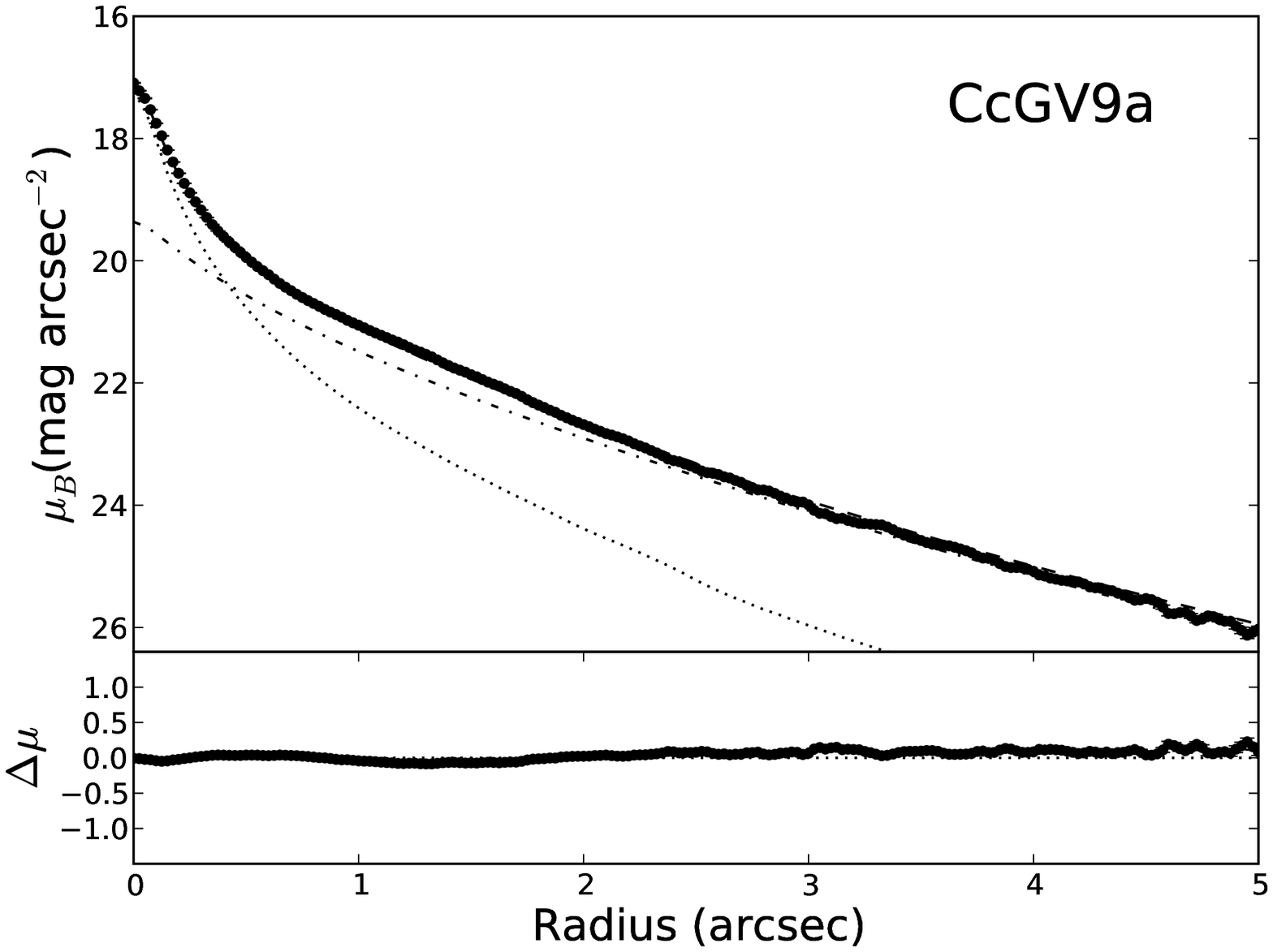}}\\
\scalebox{0.32}[0.32]{\includegraphics{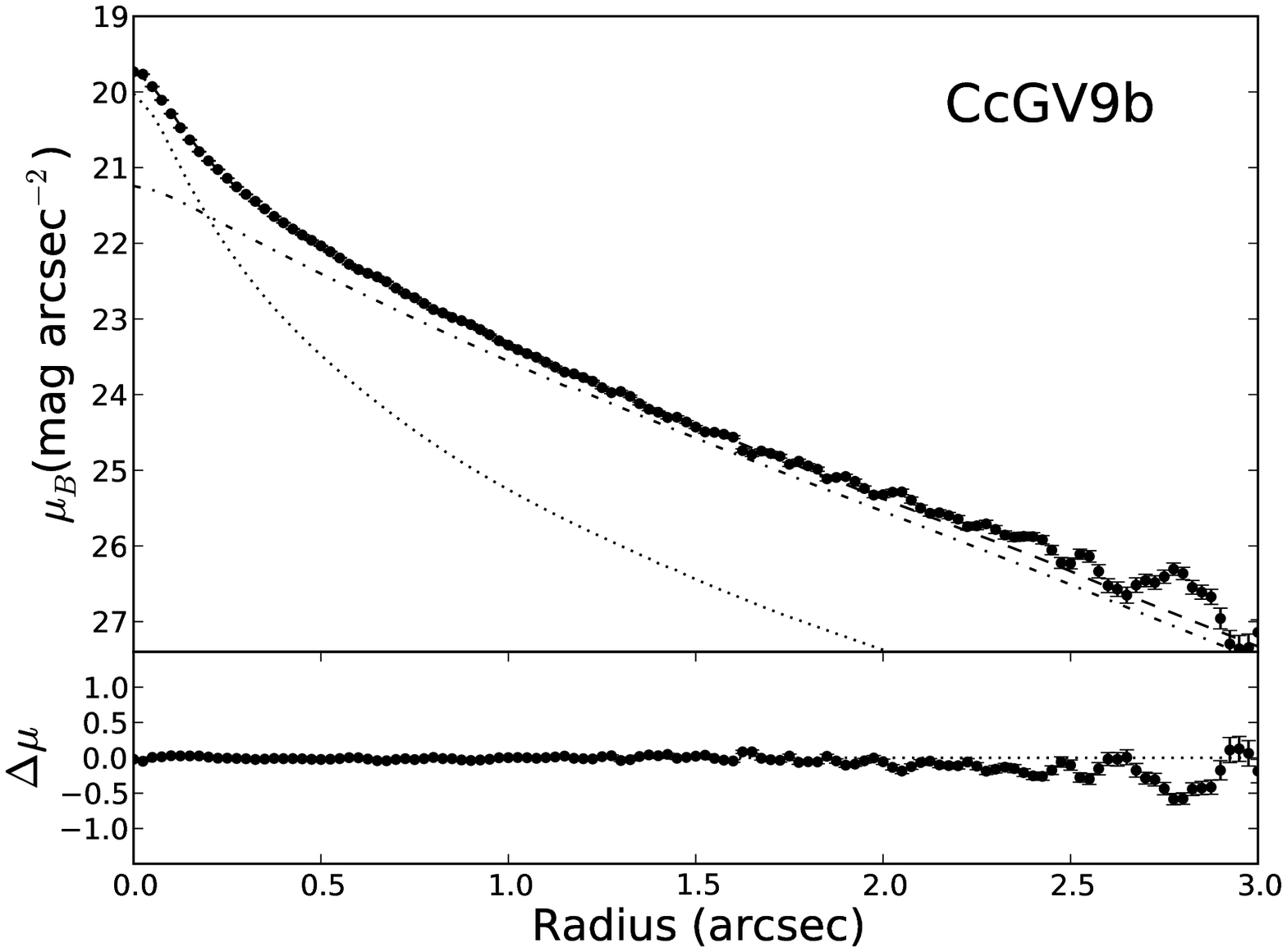}} \scalebox{0.32}[0.32]{\includegraphics{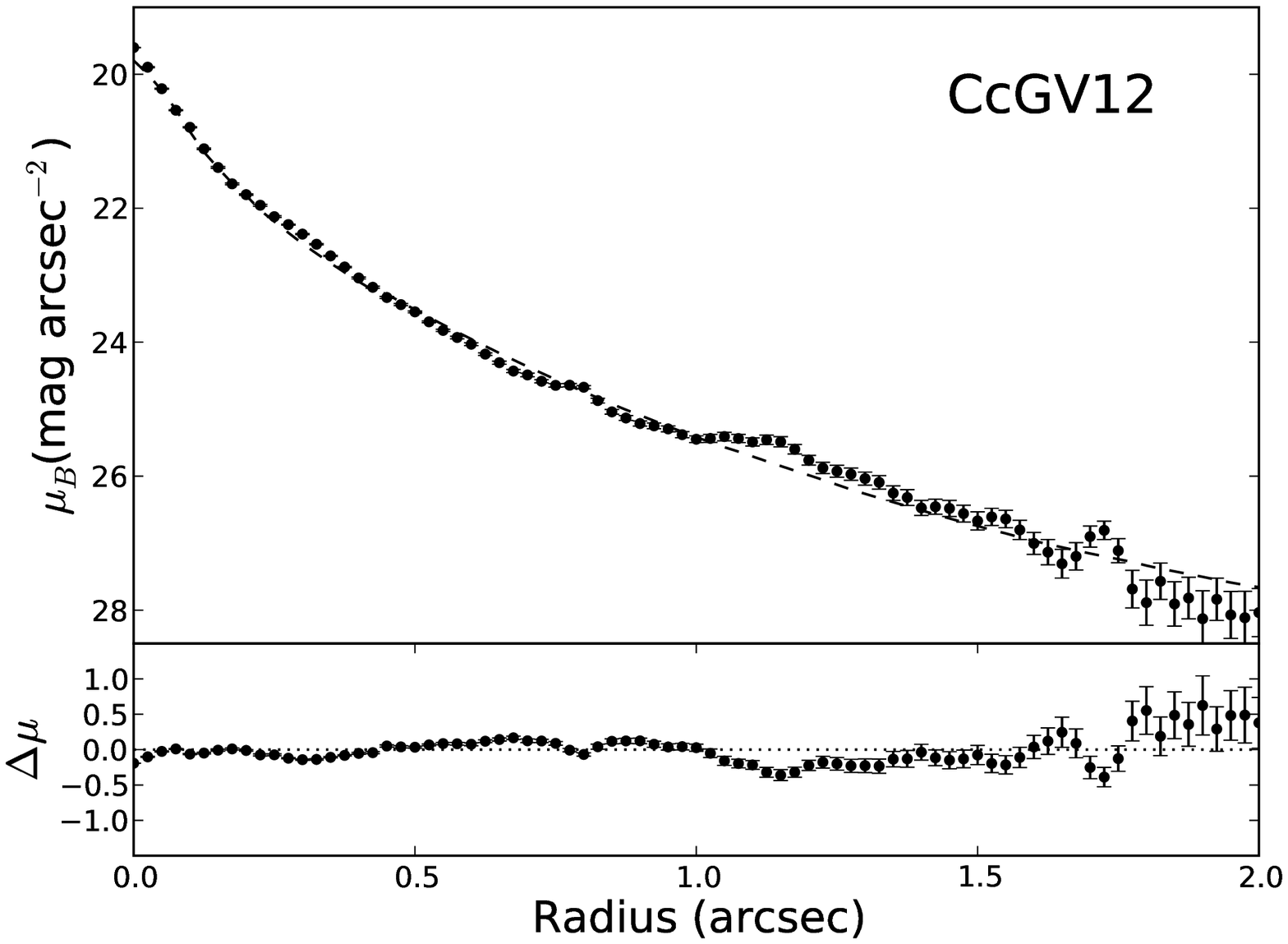}}\\
\scalebox{0.32}[0.32]{\includegraphics{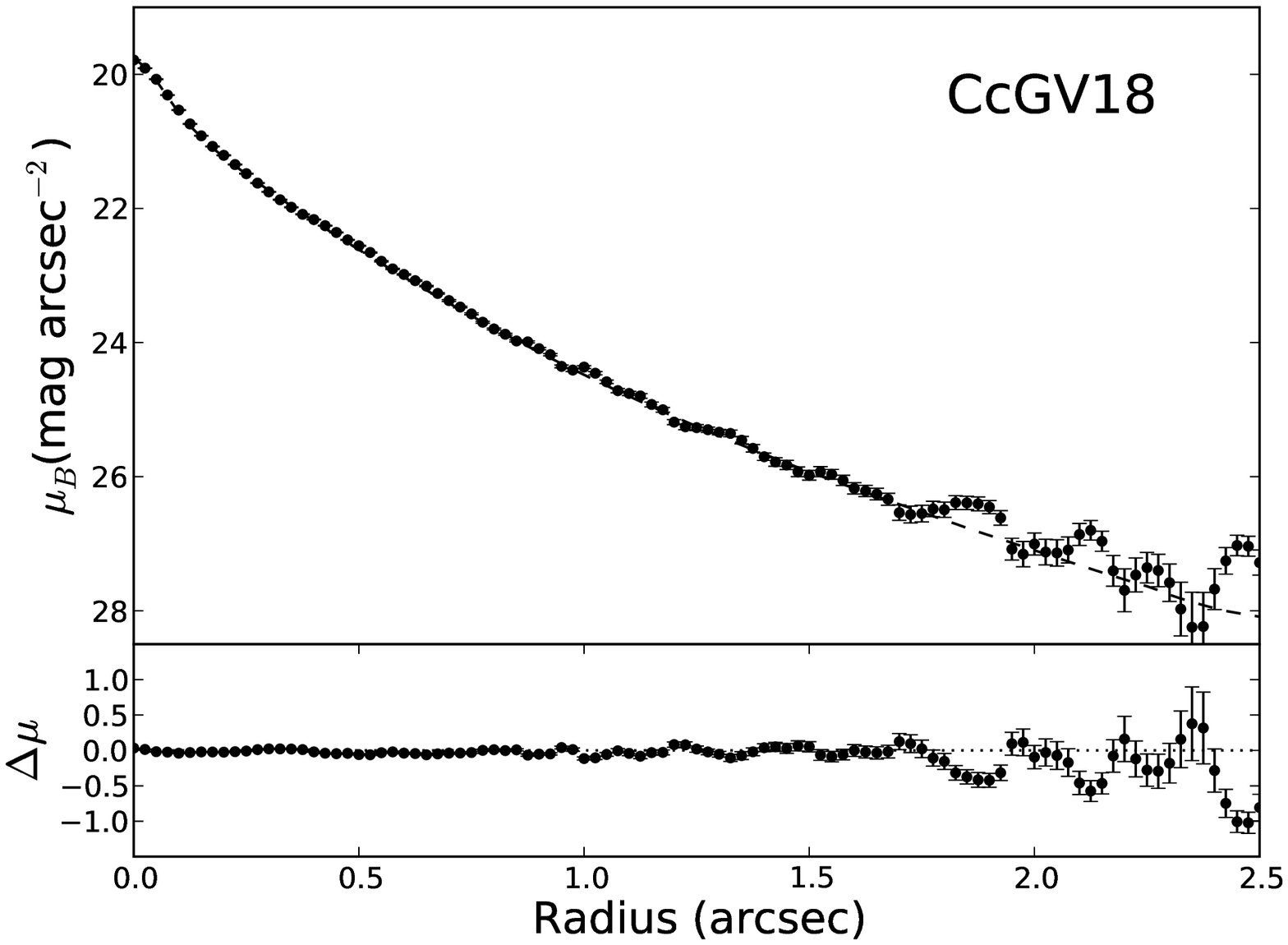}} \scalebox{0.32}[0.32]{\includegraphics{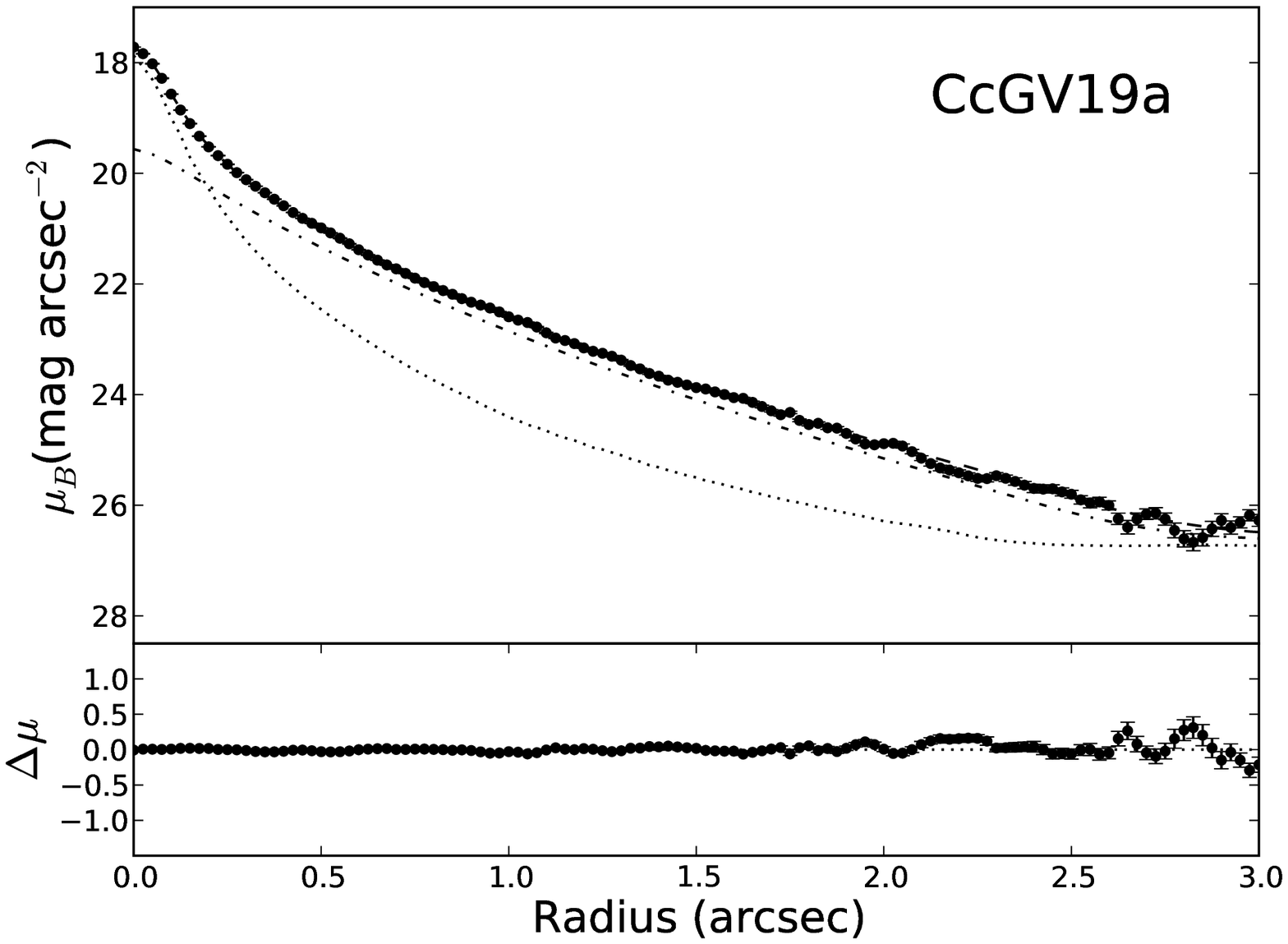}}\\
\scalebox{0.32}[0.32]{\includegraphics{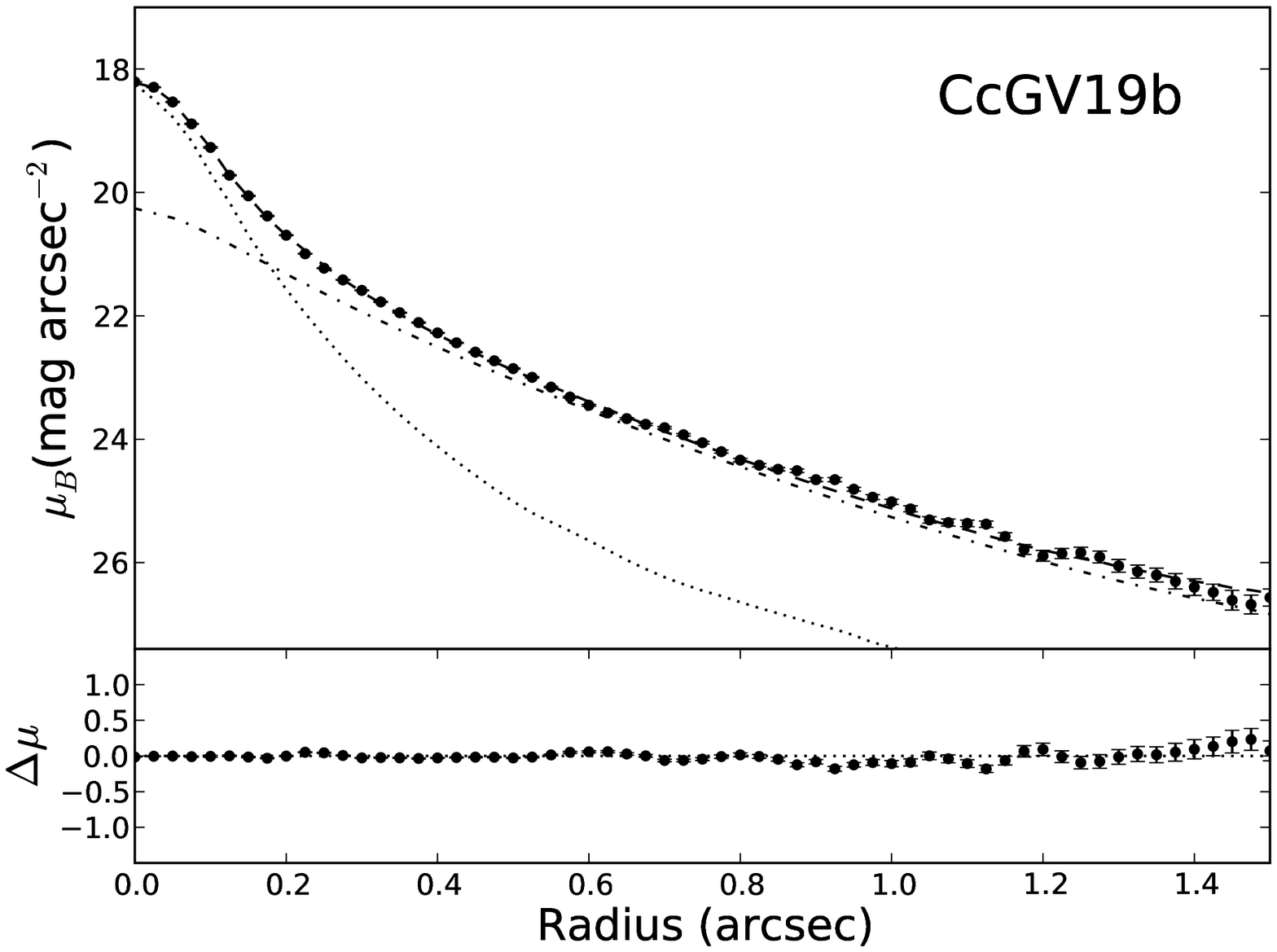}}\\
\caption{Surface brightness profiles for the sample galaxies. The observed profile and best fitting model are plotted as filled circles and dashed lines respectively. For galaxies where the best fitting model consists of two S{\'e}rsic profiles, the inner and outer components are denoted by dotted and dash-dotted lines respectively.}
\label{profs}
\end{minipage}
\end{figure*}

A well constructed PSF is of primary importance when fitting models to the inner regions of galaxy surface brightness profiles. Galfit convolves the input PSF with the model currently being tested before comparing it to the real image and determining a goodness of fit for that model. If the object being fit is approaching the resolution limit of the instrument, as is often the case when attempting to decompose our compact galaxies into two components, the dependence of the derived structural parameters on the PSF increases.

It is well known that the PSF varies with position across the two ACS/WFC chips and so a separate PSF must be created for every desired location on the detector. The standard recourse is to either use artificial PSFs generated by the Tiny Tim software package\footnote{See http://www.stsci.edu/software/tinytim/} or else some sort of empirical approximation. Here we explore both methods to obtain suitable input PSFs.

To create our artificial PSFs at the required position for each galaxy we make use of DrizzlyTim\footnote{Written by Luc Simard.}. Briefly, the task transforms the $x$--$y$ coordinates of each galaxy's centroid on the drizzled science frame back to the system of the individual geometrically distorted FLT images. It then executes TinyTim with the specified filter and a PSF diameter of 5$^{\prime\prime}$, and injects the distorted output PSFs into blank FLT files. Next these files are passed through Multidrizzle using the same configuration as used for the science frames to produce an undistorted PSF at the observed location of each object. Finally the PSF is cutout and centred according to the requirements of Galfit\footnote{Input PSFs for Galfit must be centred correctly, $N/2+1$ for an even number of pixels per side and $N/2+0.5$ for an odd number.}.

In the derivation of our empirical PSFs it would be ideal to use stars observed at the same time as our galaxies. Unfortunately owing to the high galactic latitude of Coma there are an insufficient number to correctly sample the detector's PSF variations. As a work around we select recent observations of Galactic globular cluster 47 Tucanae taken during program 10737 (P.I. J. Mack). The images are reduced using Multidrizzle with the same setup as our science data. The IRAF implementation of Daophot II \citep{stetson87,stetson90} is then used to construct a semi-empirical PSF that varies quadratically across the CCDs from $\sim$500 stars. The PSF has a Moffat function as its analytic base with six empirical look-up tables used to correct the function to the observed data. To sample the extended wings of the ACS/WFC PSF we include non-saturated pixels from centrally saturated stars. After the initial run of the task we use the current version of the Daophot PSF to subtract all stars used to derive the PSF from the 47 Tuc science image. Stars which have a high reduced $\chi^{2}$ as determined by Daophot during the subtraction, and so are likely having a detrimental effect during the initial phase of PSF creation, are removed from the input star list and the PSF derivation repeated. Following the completion of this procedure we generate a PSF at the precise pixel position of each galaxy, again centred correctly for input to Galfit.

With both PSFs in hand we test their consistency when used to derive structural parameters. Results are comparable when fitting the sample with a one component model. However, when fitting a two component model we find significant disagreement in $n$ and $R_{e}$ (see equation (1) below) for the majority of our galaxies. To get a handle on which derived PSF best represents the actual instrumental PSF at the time of observation, we select high S/N stars within $\sim$40$^{\prime\prime}$ of each target galaxy as new input PSFs. Running Galfit once more indicates a broad consistency between parameters determined using the empirical and stellar PSFs. Based on this result we opt to use the Daophot derived PSFs when obtaining structural parameters for our sample.

\subsubsection{Image Masks}

When fitting galaxies which have a near neighbour whose light distribution does not visibly interfere with the target galaxy, but may still have some influence on its structural parameters due to its presence in the fitting box, we choose to mask the secondary. The masks used are based on the results returned by our SExtractor analysis of the Coma imaging. For each neighbour galaxy we take SExtractor's total aperture and scale it linearly by a factor of 1.5 (2.25 in area) in a similar approach to that used by \cite{haussler07}. In certain situations where there is a significant blending of a target and large nearby galaxy, such as CcGV19a which sits in the halo of NGC 4874, we subtract off the neighbour using the IRAF/ELLIPSE task.

\subsubsection{Light Profile Parametrisation}

Recent work by \cite{graham02}, \cite{chilingarian07} and \cite{chilandmamon08} has found cEs to require a two component fit to their surface brightness profiles. To search for similar structure in our galaxies we fit single and double \cite{sersic68} models which are of the form,
\begin{equation}
	I(R) = I_{e}\exp\left[-\kappa\left(\left(\frac{r}{R_e}\right)^{1/n} - 1\right)\right],
\end{equation}

\noindent where $I_{e}$ is the intensity at the half-light or effective radius $R_{e}$. The parameter $n$ defines the shape of the profile with $n$=1 being an exponential profile and $n$=4 the standard $R^{1/4}$ law \citep{devaucouleurs48}. The shape parameter is closely connected to $\kappa$ by the expression $\kappa\approx 2n - \frac{1}{3}$.

The decision to fit a double S{\'e}rsic model was made to avoid unnecessary constraints being placed during two-component fitting. Indeed as it is currently unclear as to the progenitor of cEs, there would appear to be no reason to enforce a classic bulge-disk decomposition.

To test the resolvability of the inner components which Galfit detects, we also make use of the task's ability to fit a point source + S{\'e}rsic model. The advantage here is that Galfit uses the input PSF, making no further analytic approximations, to model the central component. This model describes the case of an unresolved nucleus within a S{\'e}rsic halo. The quality of the resultant fit is assessed based on the Galfit computed reduced $\chi^{2}$ and visual inspection of the residual map obtained by subtracting the PSF convolved model from the data.

\begin{table*}
\centering
\begin{minipage}{175mm}
\caption{Structural Parameters for CcGs. Col. 1: Galaxy identifier. Col. 2: Best fitting model S{\'e}rsic + S{\'e}rsic (S+S) or point source + S{\'e}rsic (PSF+S). Col. 3-10: Absolute magnitude and mean surface brightness within the effective radius $R_{e}$ in the B band (Vega), S{\'e}rsic n and effective radius $R_{e}$ for inner and outer components respectively. .}
\label{structab}
\begin{tabular}{lcr@{}c@{}lr@{}c@{}lr@{}c@{}lr@{}c@{}lr@{}c@{}lr@{}c@{}lr@{}c@{}lr@{}c@{}l}
\hline
\multicolumn{1}{c}{CCG} 
& \multicolumn{1}{c}{Model}
& \multicolumn{3}{c}{M$_{B,in}$}
& \multicolumn{3}{c}{\small{$<\!\mu_{B}\!>_{e,in}$}}
& \multicolumn{3}{c}{n$_{in}$}
& \multicolumn{3}{c}{R$_{e,in}$} 
& \multicolumn{3}{c}{M$_{B,out}$}
& \multicolumn{3}{c}{\small{$<\!\mu_{B}\!>_{e,out}$}}
& \multicolumn{3}{c}{n$_{out}$}
& \multicolumn{3}{c}{R$_{e,out}$} \\
\multicolumn{1}{c}{(ID)}
& \multicolumn{1}{c}{-}
& \multicolumn{3}{c}{(mag)} 
& \multicolumn{3}{c}{(mag arcsec$^{-2}$)}
& \multicolumn{3}{c}{-}
& \multicolumn{3}{c}{(pc)}
& \multicolumn{3}{c}{(mag)}
& \multicolumn{3}{c}{(mag arcsec$^{-2}$)}
& \multicolumn{3}{c}{-}
& \multicolumn{3}{c}{(pc)} \\
\hline
V1 & PSF+S & -8.51&$\pm$&0.13 & -& &- & -& &- & -& & -&  -14.07&$\pm$&0.01 & 21.42&$\pm$&0.01 & 1.04&$\pm$&0.01 & 252&$\pm$&2 \\
V9a & S+S & -15.72&$\pm$&0.21 & 18.84&$\pm$&0.22 & 3.49&$\pm$&0.23 & 172&$\pm$&36 & -16.22&$\pm$&0.13 & 20.95&$\pm$&0.13 & 1.64&$\pm$&0.07 & 575&$\pm$&10 \\
V9b & S+S & -12.72&$\pm$&0.24 & 21.26&$\pm$&0.25 & 2.95&$\pm$&0.26 & 145&$\pm$&33 & -14.05&$\pm$&0.07 & 22.27&$\pm$&0.07 & 1.17&$\pm$&0.03 & 359&$\pm$&3 \\
V12 & PSF+S & -10.42&$\pm$&0.04 & -& &- & -& &- & -& & -& -13.03&$\pm$&0.01 & 21.39&$\pm$&0.01 & 1.88&$\pm$&0.02 & 158&$\pm$&2 \\
V18 & PSF+S & -10.04&$\pm$&0.03 & -& &- & -& &- & -& & -& -13.59&$\pm$&0.01 & 21.48&$\pm$&0.01 & 1.34&$\pm$&0.01 & 201&$\pm$&1 \\
V19a & S+S & -14.07&$\pm$&0.09 & 18.46&$\pm$&0.1 & 4.38&$\pm$&0.19 & 76&$\pm$&9 & -14.89&$\pm$&0.04 & 20.63&$\pm$&0.04 & 1.37&$\pm$&0.03 & 254&$\pm$&2 \\
V19b & S+S & -12.90&$\pm$&0.20 & 16.82&$\pm$&0.21 & 2.60&$\pm$&0.52 & 23&$\pm$&5 & -13.19&$\pm$&0.10 & 21.09&$\pm$&0.13& 1.40&$\pm$&0.04 & 142&$\pm$&14 \\
\hline
\end{tabular}
\end{minipage}
\end{table*}

\subsubsection{Sky Background Determination}

Making a robust sky background estimation is of key importance when fitting observed galaxy surface brightness profiles with parametric models \citep{dejong96}. Using the S{\'e}rsic profile as an example, if the sky level is under estimated then the galaxy being fit appears to have more flux in its outer regions. This results in the model's wings becoming larger and either an artificially inflated $n$ value, a larger effective radius or both. If the sky is over estimated the reverse occurs. An additional complication comes from sky gradients generated by diffuse intracluster light. It is therefore clear that a reliable means to determine each galaxy's local sky background is critical when obtaining their structural parameters.

Galfit allows the sky to be fit as an extra parameter but this relies on the assumption that the model profile being used well describes the real light profile. Tests indicated that this assumption was adequate for one component fits, however, when employing a two component model it was found that one of the profiles was able to suppress the sky and consequently grow too large in both $n$ and $R_{e}$. To solve this problem we determine and fix the sky prior to profile fitting using a method similar to \cite{haussler07}.

Each galaxy's light distribution is measured by the IRAF/ELLIPSE task with a maximum semi-major axis for iterative fitting set to the SExtractor total aperture radius. In this way the annuli beyond this point are of fixed ellipticity and orientation. The profile gradient between these annuli is computed and the sky value determined as the mean of the annuli outward of the point at which the gradient reaches a prescribed small value. The radius at which this transition occurs is scaled by a factor 1.5 and results in a region of width 25-65 pixels (area $\sim$9500-65000 pixels$^{2}$) being used to estimate the sky, depending on galaxy size.

The above procedure makes no attempt to measure sky gradients. Galfit, when permitted to fit the sky freely, models it as a plane which may tilt. We take advantage of this functionality and allow the task to measure the plane's tilt during fitting, while keeping the mean sky level fixed.

\subsubsection{Results}

Table \ref{structab} presents the best fitting structural parameters for all seven CcGs. Native F475W magnitudes are corrected for Galactic extinction and K-correction before being converted to the B band using the same methods as before. The mean surface brightness parameters are additionally corrected for cosmological dimming. All magnitudes and surface brightness measurements are converted to a Vega zeropoint\footnote{http://www.stsci.edu/hst/acs/analysis/zeropoints} for comparison with previous work.

In Fig. \ref{cutouts} we present three cutout images of each galaxy. The left hand column is the observed data, the middle column is the residuals produced by subtracting the best fitting seeing convolved single S{\'e}rsic model from the data and the right column contains the equivalent but for the model which best describes each galaxy (see Table \ref{structab} column 2). To further display the quality of the best fits in Fig. \ref{profs} we use IRAF/ELLIPSE to produce surface brightness profiles of both real and model data and hence obtain one-dimensional residuals.

It is evident that the majority of our sample are well fit by their respective models. All galaxies were found to statistically prefer two components, either two S{\'e}rsic profiles (S+S) or a point source plus S{\'e}rsic profile (PSF+S), over a single S{\'e}rsic model. Only CcGV9a and CcGV12 have residual images showing remaining substructure. Based on our colour maps constructed in the next section, it is unlikely the residuals associated with CcGV12 are part of the galaxy given their bluer colour with respect to the rest of the galaxy at such radii. CcGV9a residuals are interestingly symmetric and are probably indicative of a weak bar.

Both CcGV1 and CcGV18 have comparable reduced $\chi^{2}$ when fit by PSF+S and S+S models. This likely indicates the inner component is almost resolved but is either too faint, too small or both to be modelled clearly. As such we opt for the simpler PSF+S model to parameterise these galaxies. Also of note is the inner component of CcGV19b which has an $R_{e}$ = 23 pc or $\sim$ 1 ACS/WFC pixel. Given its exceptionally small angular size both resolvability and repeatability checks were conducted using Galfit. 

Fitting a PSF+S model and comparing this with a S+S fit we find $\Delta\chi^{2}_{\nu}$ = 0.11 in favour of the latter. We also performed some simulations using Galfit to first create a set of artificial S+S models with 0.2 $\leq R_{e,in} \leq$ 1.0 pixel in steps of 0.05 pixel, convolved with the PSF used for CcGV19b. Next we fitted the models with both PSF+S and S+S profiles and estimate our limit for detection of a resolved inner component for this particular galaxy to be $R_{e,in} \sim$ 0.4 pixel or 9 pc. From this we conclude that the core of CcGV19b {\em is} sufficiently well resolved, likely due to its exceptional luminosity. We also note that the inner component of this galaxy is equally well fit by a King profile, often used when studying GCs. Fitting King + S{\'e}rsic we find that the derived structural parameters for both inner and outer components are consistent to well within the quoted errors in Table \ref{structab}.

Typically Galfit estimates errors based on the shape of the $\chi^{2}$ surface surrounding the $\chi^{2}_{min}$ (see \citealt{peng02} for further details) and in general provides reasonable uncertainty estimates. However in certain regimes there are likely to be further systematics which may become more significant such as in the case of CcGV19b where the observed profile is poorly sampled and CcGV1 where the point source component is extremely faint. By varying the starting values of $R_{e}$ and $n$ given to Galfit we check the stability of our derived parameters for the inner component of CcGV19b, which are reported by Galfit to have uncertainties of around a few percent. Such tests indicate a more realistic uncertainty of $\sim$ 20\% for both $R_{e}$ and $n$ while the component's magnitude varies by roughly $\pm$ 0.2 mag. For CcGV1, we use the visit's uncertainty map to produce new realisations of the frame and then pass them through Galfit. The standard deviation of the distribution of magnitudes for the point source component is then used as its quoted error in Table \ref{structab}.

\begin{figure*}
\begin{minipage}{178mm}
\centering
\scalebox{0.42}[0.52]{\includegraphics{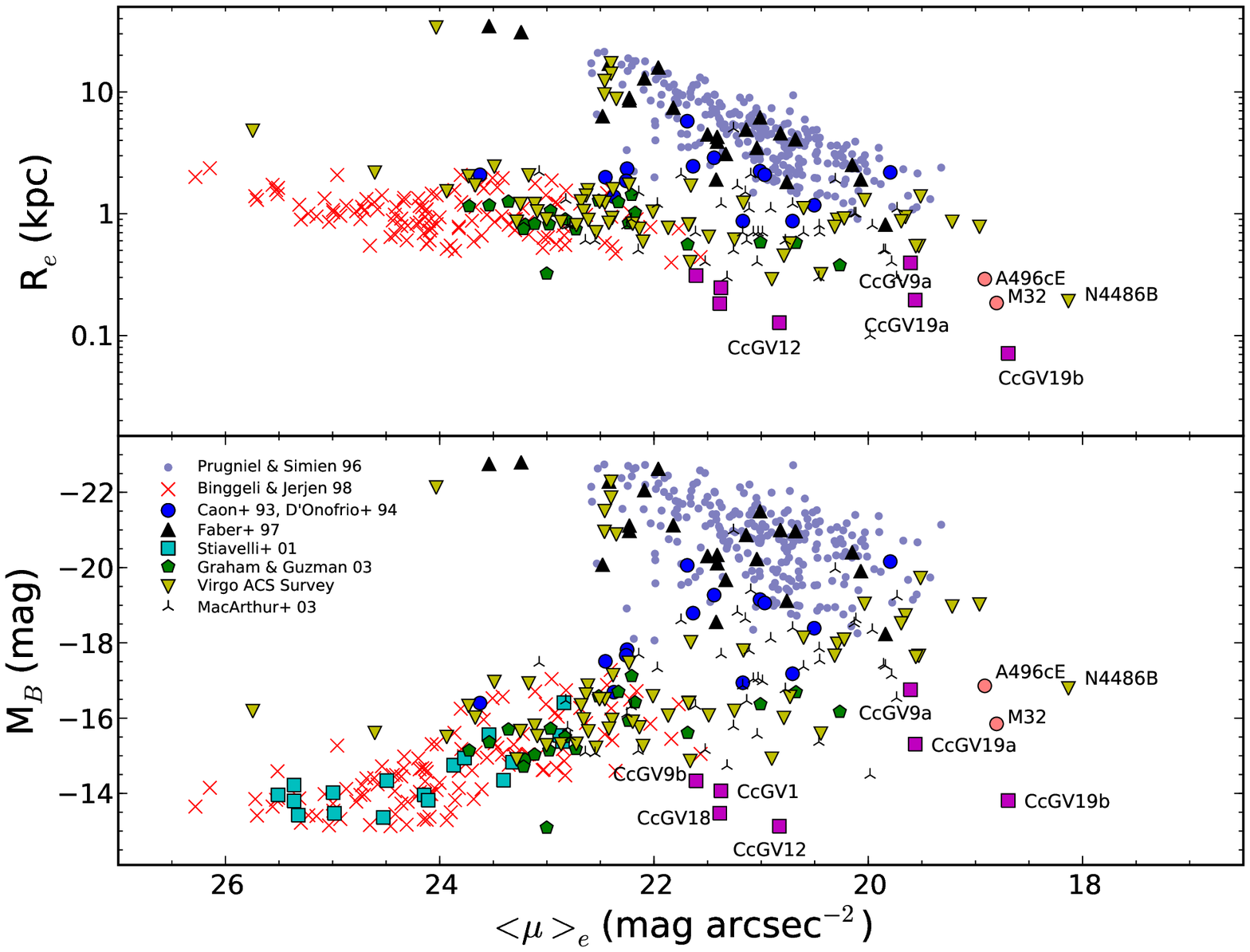}}
\scalebox{0.42}[0.52]{\includegraphics{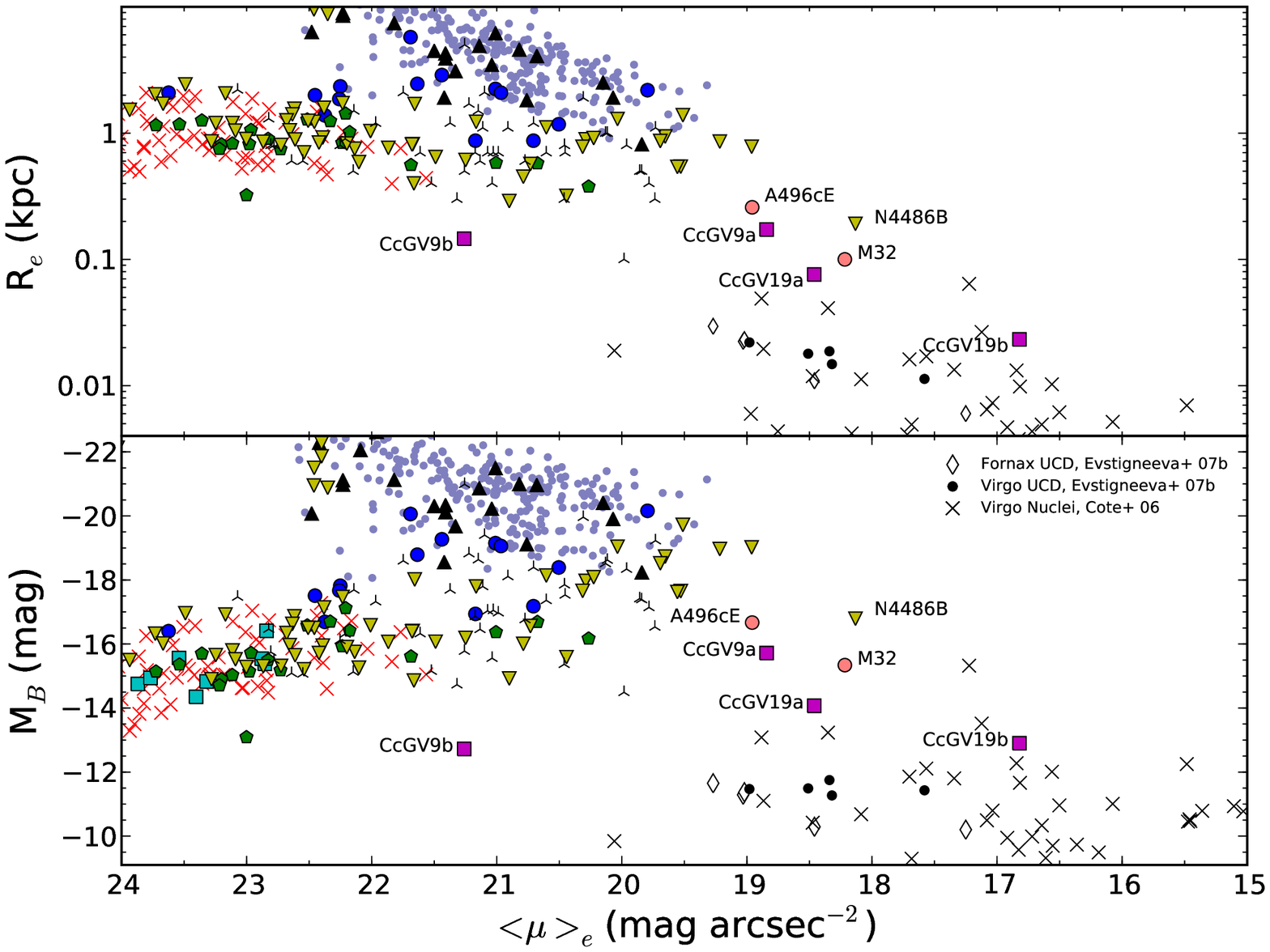}}
\caption{Structural parameters for dwarf to giant elliptical galaxies, bulges, nuclei and compact galaxies from the literature and this work (purple filled squares). The left panel displays the parameters for the combined or whole profile of all CcGs, M32 and A496cE while the right panel presents the inner components of our sample where they are resolved, M32 and A496cE. Coma compact galaxy points are not labelled in some panels for clarity's sake.}
\label{structplots}
\end{minipage}
\end{figure*}

To place our sample in context with other recent work in this area we present in Fig. \ref{structplots} updated versions of Figs. 9a \& g from \cite{gra&guz03} \citep[see also][]{chilingarian07}. These diagrams compare mean surface brightness within the effective radius $R_{e}$ with absolute magnitude in the B band and effective radius $R_{e}$ for dynamically hot stellar systems. They represent different versions of the well known $M_{B}$ - $\mu_{eff}$ \cite{kormendy77} relation which itself is a projection of the Fundamental Plane \citep{djorg&davis87}. For the full range of elliptical galaxies, dwarf to giant, data is taken from \cite{bingjer98}, \cite{caon93}, \cite{donofrio94}, \cite{faber97}, \cite{gra&guz03}, \cite{stiavelli01} and homogenised as per \cite{gra&guz03}. We additionally include data for Es and dEs from the Virgo Cluster ACS Survey \citep{ferrarese06}, with the g$_{F475W}$ photometric parameters transformed to the B band using similar methods to those described for our Coma data, and Es brighter than $M_{B}$ = -18 mag from \cite{prugniel96}. A sample of bulges is taken from \cite{macarthur03}. In terms of compact galaxies from the literature we include M32 \citep{graham02}, NGC 4486B \citep{ferrarese06} and A496cE \citep{chilingarian07}.

It is current practice to assume cEs that are best fit by two component models such as S{\'e}rsic + Exponential disk or S+S with outer n $\sim$ 1 are indeed bulge/disk decompositions. As such only their bulges would be plotted on diagrams like Fig. \ref{structplots}. That said, quite whether these outer components are in fact rotationally supported disks or just the remains of a dwarf/intermediate-luminosity elliptical progenitor galaxy are unclear. With this in mind we present two versions of both plots. The left panel of Fig. \ref{structplots} shows the position in the parameter space of the combined two component models for all CcGs, M32 and A496cE. In this case the ``whole" profile $R_{e}$ is obtained via numerical integration. The right panel of Fig. \ref{structplots} then presents the structural parameters for the S{\'e}rsic inner components of all Coma compacts where these are resolved, M32 and A496cE. The position of NGC 4486B does not change in these plots as it is modelled by a single S{\'e}rsic function in the literature. To aid comparison we also plot data obtained for Fornax and Virgo UCDs by \cite{evstigneeva07b} and galaxy nuclei in the Virgo Cluster from \cite{cote06}. For those UCDs which are best fit by an inner compact core and outer halo, we plot only the inner component. Here we note that both the UCDs and the galaxy nuclei have had their structural parameters derived from King profile fits rather than a S{\'e}rsic model. From the tests conducted for the inner component of CcGV19b and by inspection of the results presented in Tables 8 and 9 of \cite{evstigneeva07b}, we typically expect at most a $\sim$ 30\% difference in parameters derived from the two fitting functions (VUCD 3 excluded).

The combined profile diagram clearly shows that three of the CcGs, V9a, V19a and V19b, have structural parameters which are comparable, if marginally less extreme, to those of confirmed cEs. They are seen to extend the sequence defined by giant ellipticals toward fainter magnitudes/higher surface brightness in line with those compacts from the literature. The remaining four CcGs appear to follow a comparable sequence, albeit with lower surface brightness, originating from the dE region of the diagram.

Turning to the right panel, the inner components of CcGV9a and CcGV19a are seen to have similar structural properties to the inner components of A496cE and M32 respectively. The most extreme case is CcGV19b whose position on the diagram is well within the parameter space occupied by the more luminous/more massive nuclei of early-type galaxies in Virgo. CcGV9b on the other hand appears to have an inner component much less extreme in terms of surface brightness and relies instead on small physical size and faint magnitude to perturb it off the correlation seen for normal dEs. It is evident that the inner components of the Coma cEs, in combination with those from the literature, begin to fill up the region of the plot between Es/bulges and UCDs/nuclei. Additionally, given the observed sequence it is possible to speculate that, structurally speaking, the inner component of CcGV19b, and perhaps CcGV19a, are likely galaxy nuclei while that of CcGV9a appears more bulge like. 

\begin{figure*}
\begin{minipage}{135mm}
\centering
\scalebox{0.32}[0.32]{\includegraphics{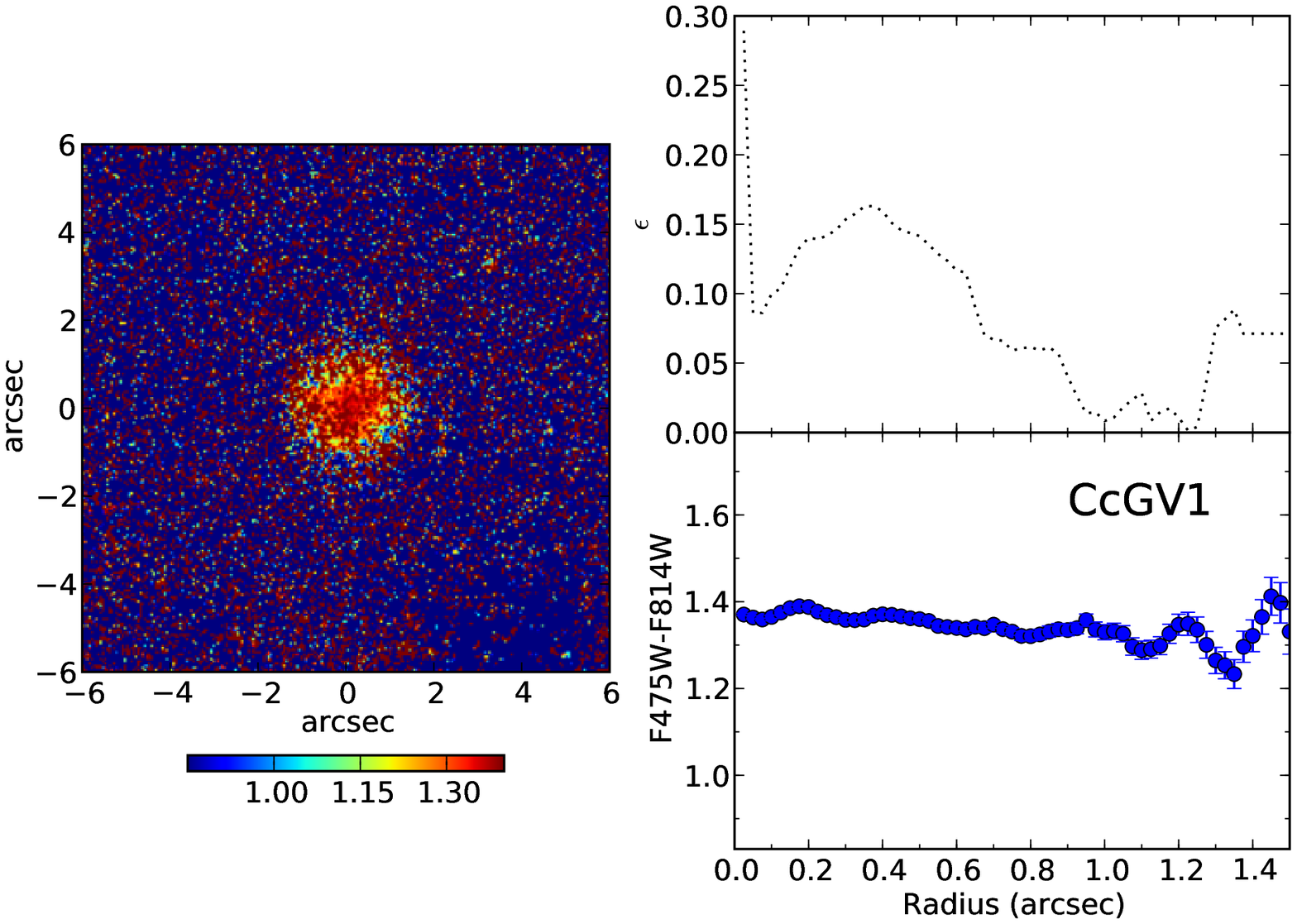}} \scalebox{0.32}[0.32]{\includegraphics{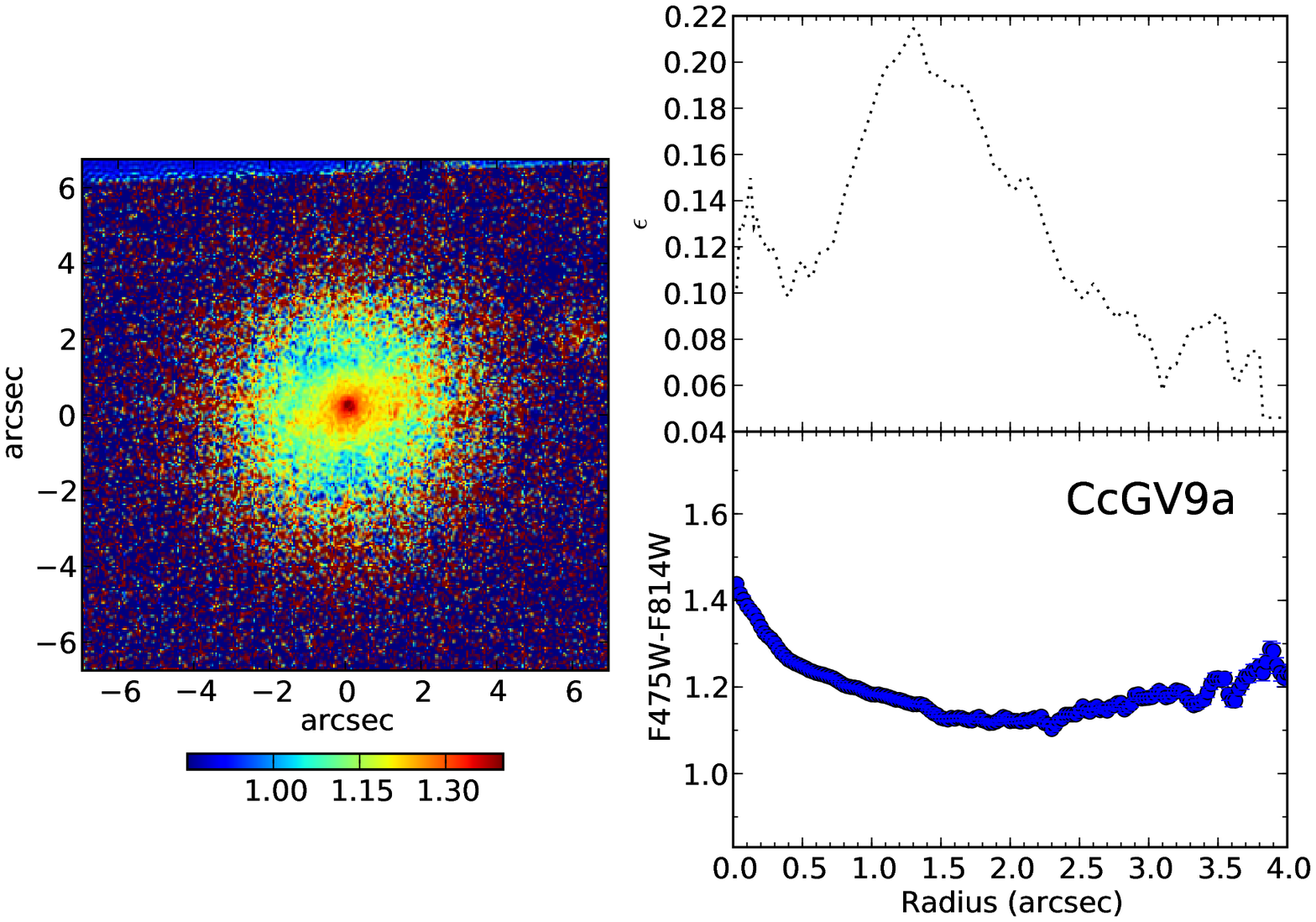}}\\
\scalebox{0.32}[0.32]{\includegraphics{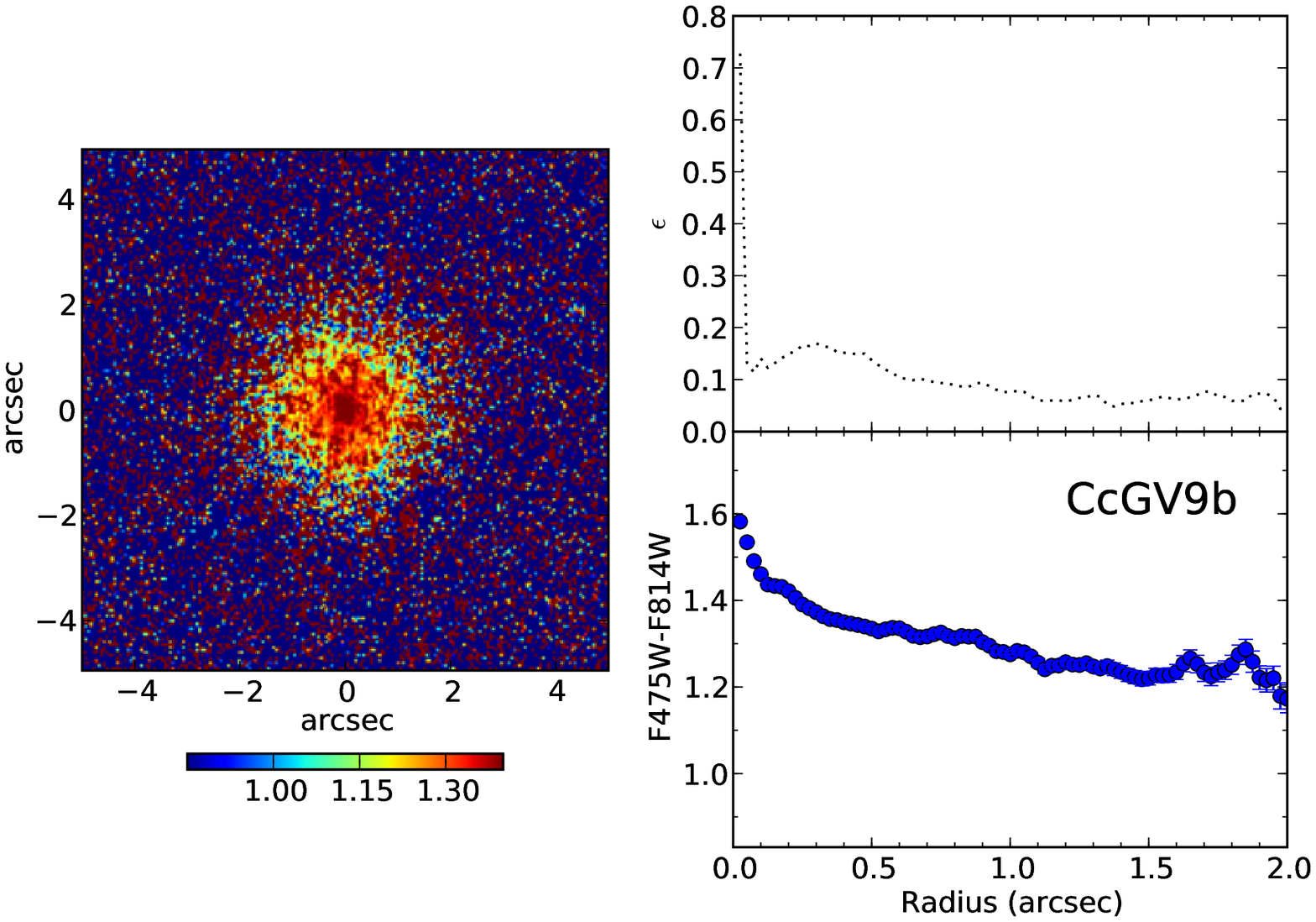}} \scalebox{0.32}[0.32]{\includegraphics{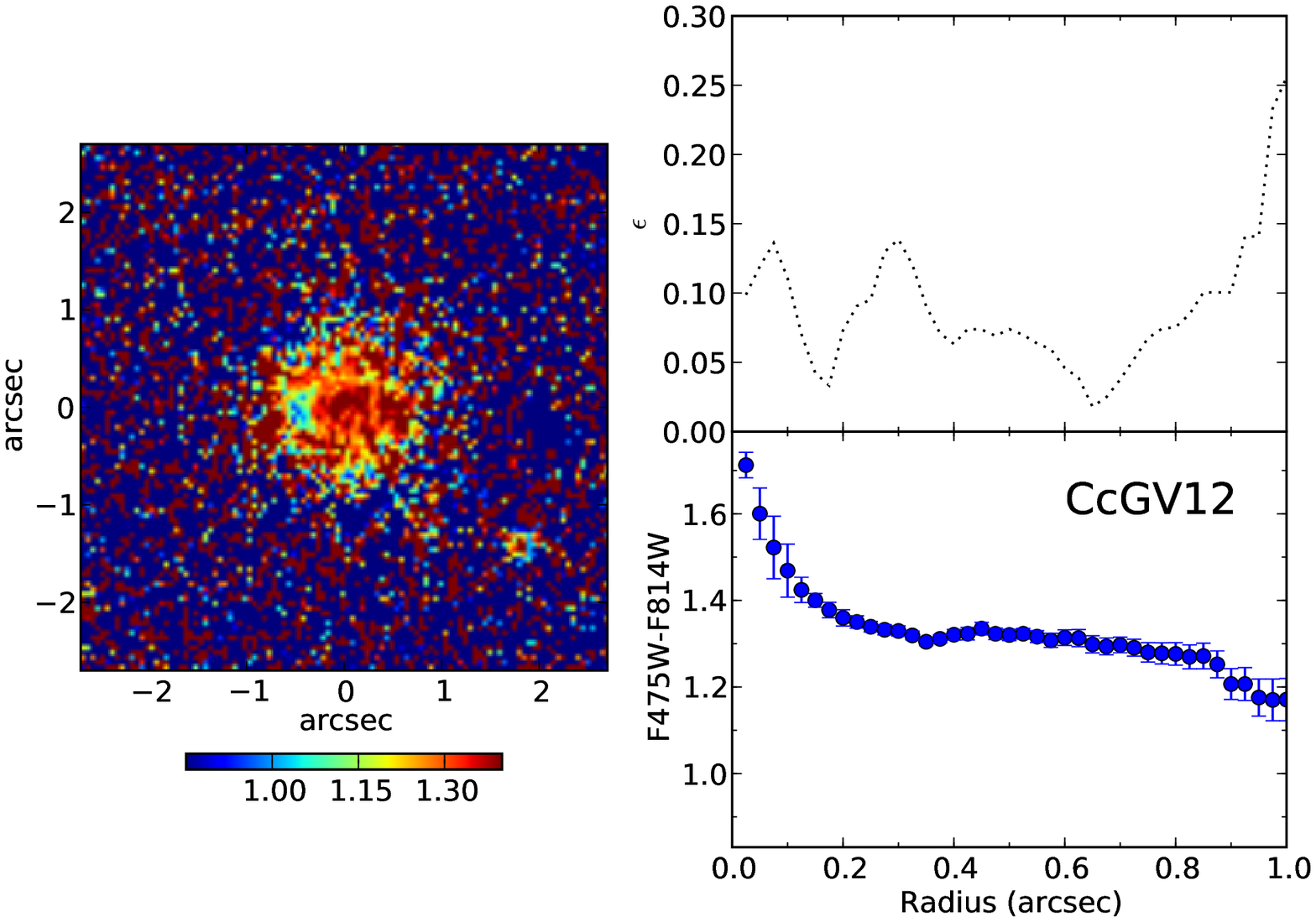}}\\
\scalebox{0.32}[0.32]{\includegraphics{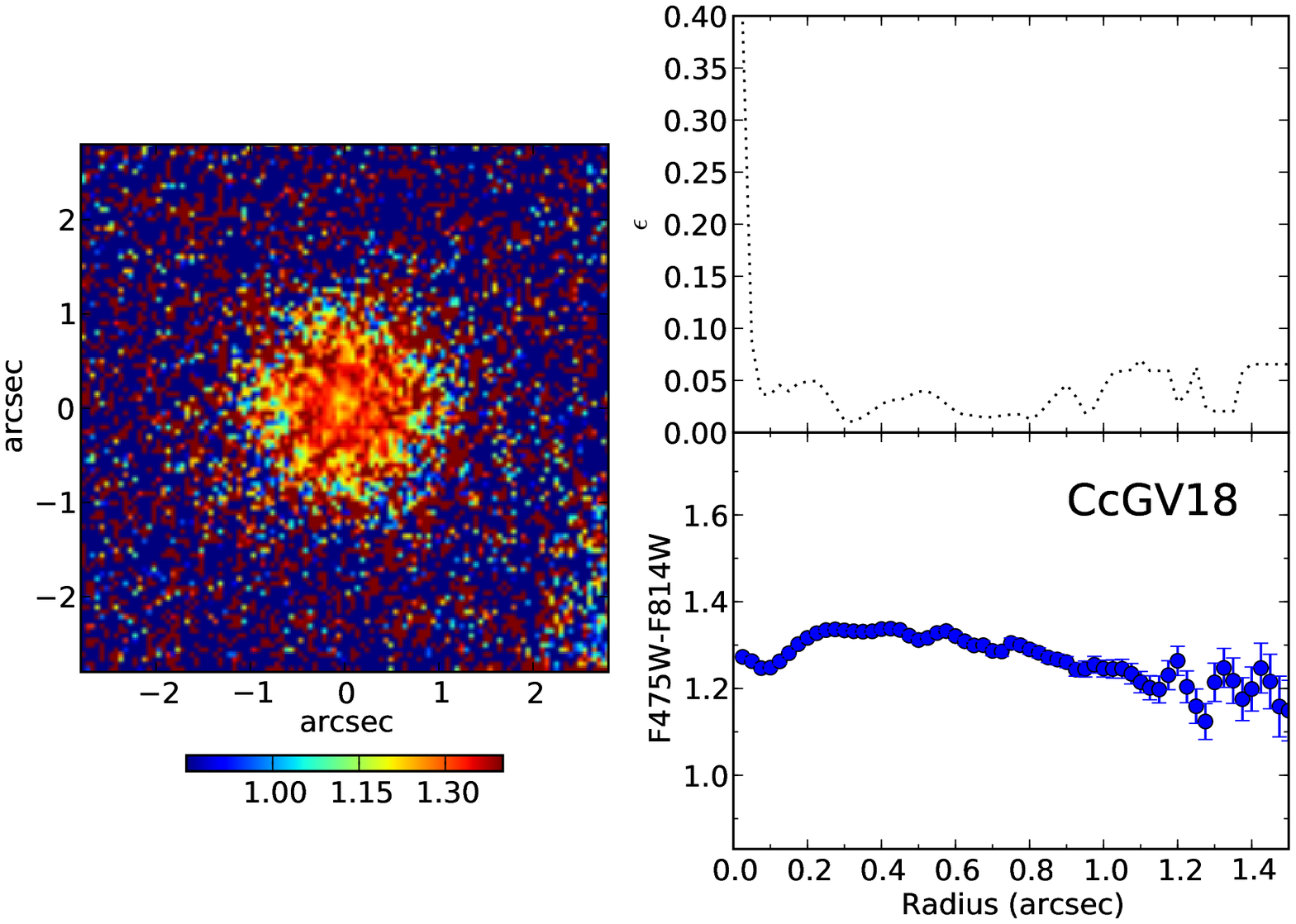}} \scalebox{0.32}[0.32]{\includegraphics{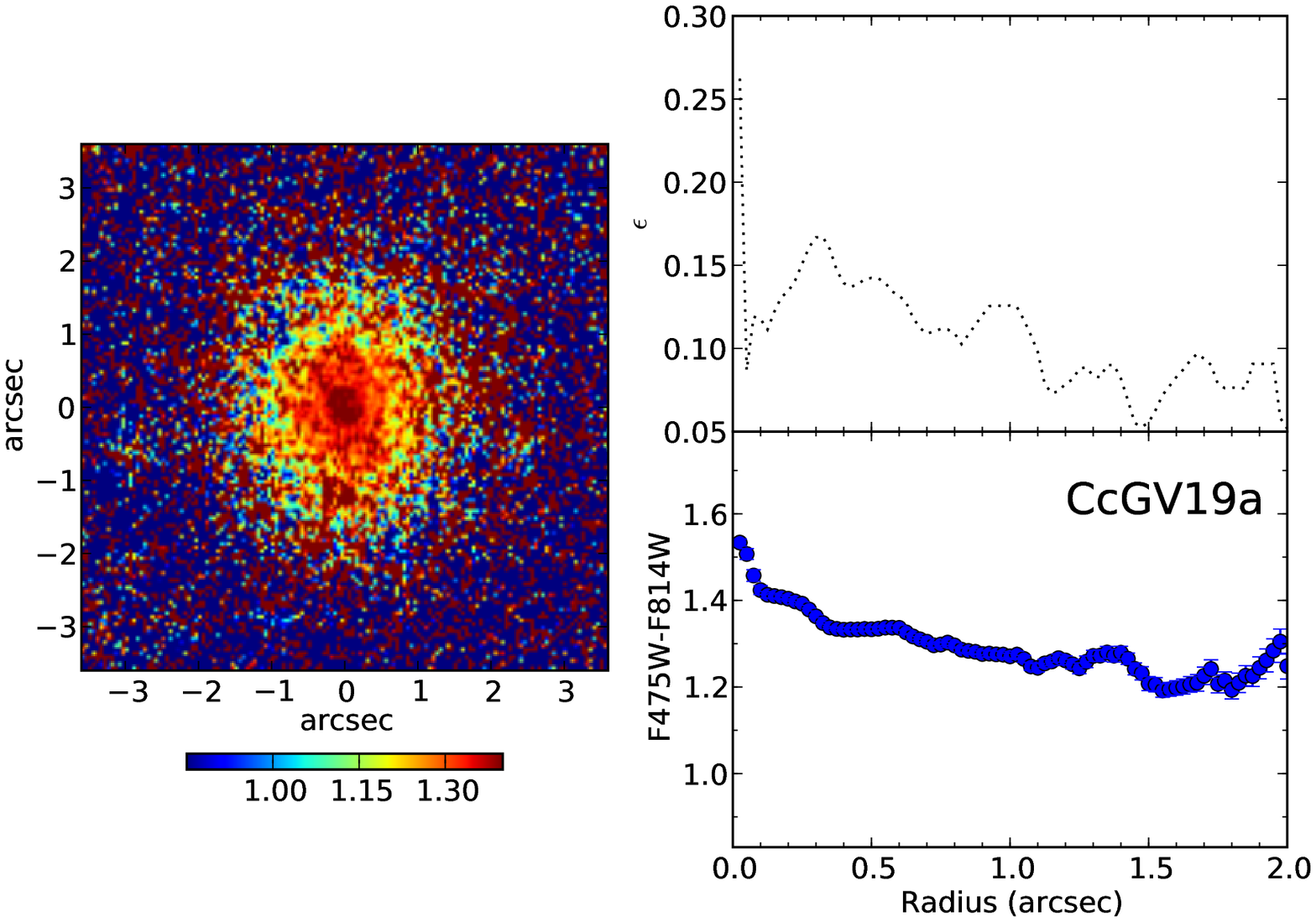}}\\
\scalebox{0.32}[0.32]{\includegraphics{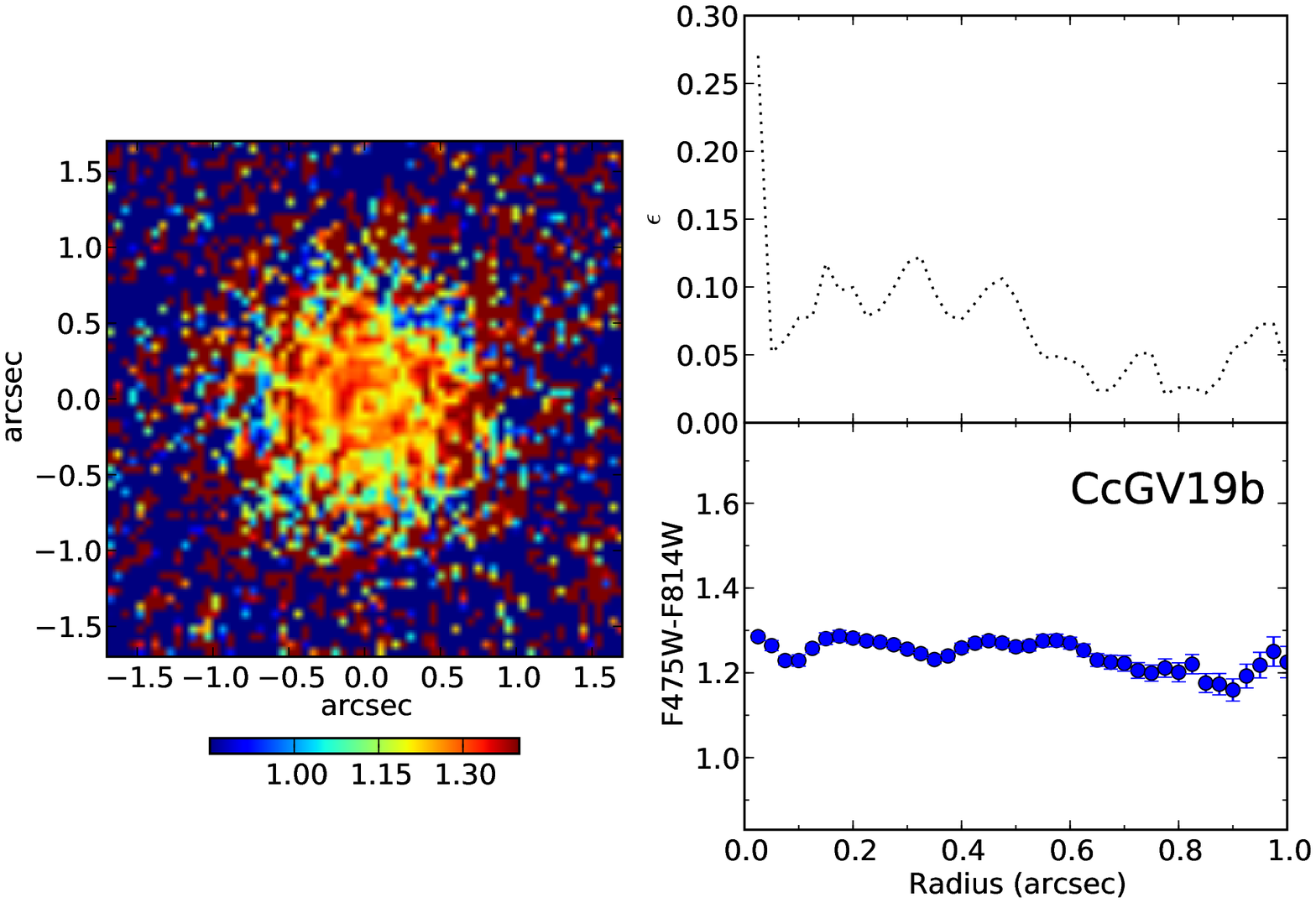}}\\
\caption{Colourmaps in F475W-F814W filters from the ACS imaging. We also plot each galaxies colour and ellipticity profile.}
\label{colourmaps}
\end{minipage}
\end{figure*}

\subsection{Colour Maps}

Colour gradients in early-type galaxies can be direct indicators as to their formation and evolution mechanisms. If one neglects the influence of any dust present in a given galaxy, such gradients are generated by variations in stellar population age and metallicity with radius. In terms of formation, monolithic collapse has been predicted to establish a steep negative metallicity gradient \citep{carlberg84} which is then progressively diminished to varying degrees by major and minor mergers during a galaxy's evolution \citep{kobayashi04}. Hence colour gradients in early-type galaxies are typically considered to be due to metallicity variations alone rather than age although the validity of this assumption is still somewhat in question \citep{silva94,baes07}.

Previously M32 has been found to have a flat colour profile in both the optical and near-infrared \citep{peletier93} which is, apparently, due to an exact compensation in metallicity and age gradients \citep{rose05}. Both A496cE and NGC 4486B have also been found to have no colour gradient, although given the scale used to display the colour axis of Fig. 81 in \cite{ferrarese06} it is difficult to assess the possibility for a moderate gradient as is often the case in early-type galaxies. On the other hand M59cO discovered by \cite{chilandmamon08} was observed to have a relatively strong gradient in the sense that the galaxy's core was $\sim$0.15 mag bluer than its halo.

To search for radial stellar population gradients in our sample we have constructed colour maps and profiles from our ACS imaging. In order to do this we must first degrade the resolution of the F475W image to that of the F814W. Following \cite{chilandmamon08} we smooth the F475W by the ratio of the fourier transforms of PSF$_{F814W}$ and PSF$_{F475}$ using the IRAF task PSFMATCH. This method acts to minimise resolution loss which would occur when convolving either science image with the PSF of the other. We then run IRAF/ELLIPSE on both images, taking the isophote table generated from the F475W run and using them as input isophotes for the F814W image. This process correctly ensures that our colour profiles are derived from matched annuli in both filters. In Fig. \ref{colourmaps} we present the colour profiles and maps constructed using this method as well as isophotal ellipticity profiles obtained from the F475W images.

Nearly flat profiles are observed for CcGV1 and CcGV19b which may point to similar stellar populations at all radii although this is not definitive given the age-metallicity degeneracy suffered by broadband colours (and the example of M32 noted above). CcGV18 has a shallow gradient, the colour becoming $\sim 0.2$ mag redder as one moves inward, though interestingly within the central 0.2$^{\prime\prime}$ there is a noticeable blueing in colour of order $\sim$ 0.1 mag, perhaps indicating a younger population present in its unresolved nucleus (as in dEs such as NGC205). The remaining four galaxies, which include the cEs CcGV9a and CcGV19a, all contain strong colour gradients with bluer colours at larger radii.

The colour map of CcGV9a clearly shows a red core region with higher ellipticity isophotes than those outward of 2$^{\prime\prime}$. The peak in the ellipticity profile at $\sim$ 1.5$^{\prime\prime}$ provides further evidence for the existence of a weak bar in this galaxy. There is a noticeable change in the colour profile slope at $\sim$ 0.3$^{\prime\prime}$ which corresponds well with the effective radius of the inner S{\'e}rsic component returned by Galfit. Beyond $\sim$ 2.2$^{\prime\prime}$ the profile turns over and gradually becomes redder by 0.15 mag. From the colour map alone this galaxy certainly appears to be at least a two component system with a significant colour difference between the inner and outer regions.

CcGV19a is observed to have a colour variation of $\sim$ 0.3 mag within just 2$^{\prime\prime}$. As with CcGV9a there is again a change, albeit poorly sampled, in colour profile slope near the effective radius of the inner component detected by Galfit, $\sim$ 0.1 to 0.15$^{\prime\prime}$. Beyond this transition the colour profile decreases in a remarkably linear fashion until $\sim$ 1.8$^{\prime\prime}$, while the isophotes maintain a fairly stable ellipticity $\epsilon$ = 0.1 $\pm$ 0.05.

The colour profile of CcGV9b has a slope change at a smaller radius than would be suggested by its inner component effective radius. This is perhaps due to the smaller difference between inner and outer component surface brightness, $\Delta\mu$ = 1 mag for CcGV9b but, $\Delta\mu \gtrsim$ 2 mag for CcGV9a and 19a, where $\Delta\mu$ is the difference between inner and outer component $<\!\mu_{B}\!>_{e}$. This aside, CcGV9b and 19a have very similar colour gradients, both becoming $\sim$ 0.3 mag bluer outwards, which may be evidence for comparable changes in stellar population with radius between the two. Whilst speculative, if this were true it would be an interesting result given the differences in their structural parameters.

Finally CcGV12 also contains a strong colour gradient which is mostly confined to within 0.2$^{\prime\prime}$. The structure identified in the Galfit residual is clearly visible in the colour map and is masked when the profile is constructed, using a simple binary mask of approximately comparable size. Once outside the nuclear regions the colour profile steadily moves to bluer colours with increasing radius. As with all the galaxies with strong gradients, it is likely this galaxy contains a central component which is older or more metal rich, or indeed both, relative to the outer halo.

\section{Spectroscopic Analysis}

\subsection{Velocity Dispersions and Masses}

To measure the velocity dispersions of our sample we employ the Penalized Pixel Fitting (PPXF) routine developed by \cite{cap&emsel04}. This technique works by fitting a set of template spectra convolved with a parametric line of sight velocity dispersion (LOSVD) to the observed galaxy spectrum in the pixel space. Both the target galaxy and template spectra are first rebinned to a linear scale in ln(wavelength) and then a model $M$ of the target's spectrum $G$ is constructed such that,

\begin{equation}
M = P_{m} \sum_{k=1}^{k}w_{k}(B(\sigma,V) * T_{k}) + \sum_{l=0}^{l}b_{l}P_{l}
\end{equation}

\noindent where $P_{m}$ and $P_{l}$ are multiplicative and additive Legendre polynomials respectively of order $m$ and $l$ which are used to correct the continuum shape of the templates during the fit, $T_{k}$ is a set of template spectra, $B(\sigma,V)$ is the broadening function used to parametrise the LOSVD and $w_{k}$ and $b_{l}$ the weights given to each template. Finding the best fitting parameters for $B$ is then a matter of minimising,

\begin{equation}
\chi^{2} = \sum_{n=1}^{n}\left(\frac{G - M}{\Delta G}\right)^{2}
\end{equation}

\noindent where $\Delta G$ is the per pixel uncertainty on $G$ obtained during reduction. While it is possible to fit for higher moments of the LOSVD \citep{vandermarel93,gerhard93}, given the intermediate velocity resolution of our data we choose to fit for $\sigma$ alone setting $B$ as a gaussian function. For $T_{k}$ we use the simple stellar population (SSP) models of Vazdekis et al. (in preparation\footnote{http://www.iac.es/galeria/vazdekis/}) which are based on the MILES empirical stellar spectral library \citep{sanchblaz06a}. These were chosen due to their large spectral coverage of 3540-7410\AA\ and intermediate resolution of 2.3\AA\ FWHM which is better (by a factor of 2) than that of our data, as required if they are to be used as templates in PPXF. To determine the order of $P_{m}$ and $P_{l}$ we fix $l$ = 2 and increment $m$ = 0-10 in steps of unity. We then select $m$ based on a trade off between the stability of $\sigma$ and the increasing computational time required as the complexity of $P_{m}$ grows. Typical values are $m$ = 3-6 and for $P_{l}$ we find that $l$ = 2 satisfies our criteria.

\begin{figure*}
\centering
\scalebox{0.42}[0.38]{\includegraphics{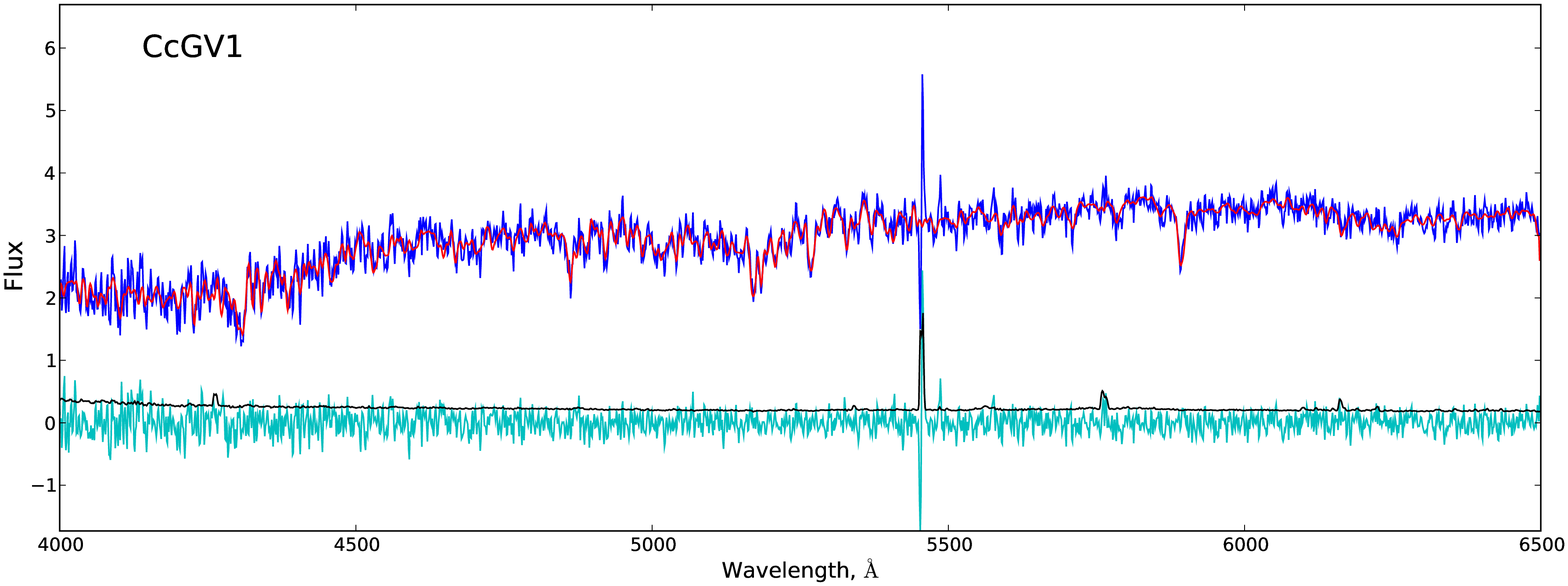}} \\
\scalebox{0.42}[0.38]{\includegraphics{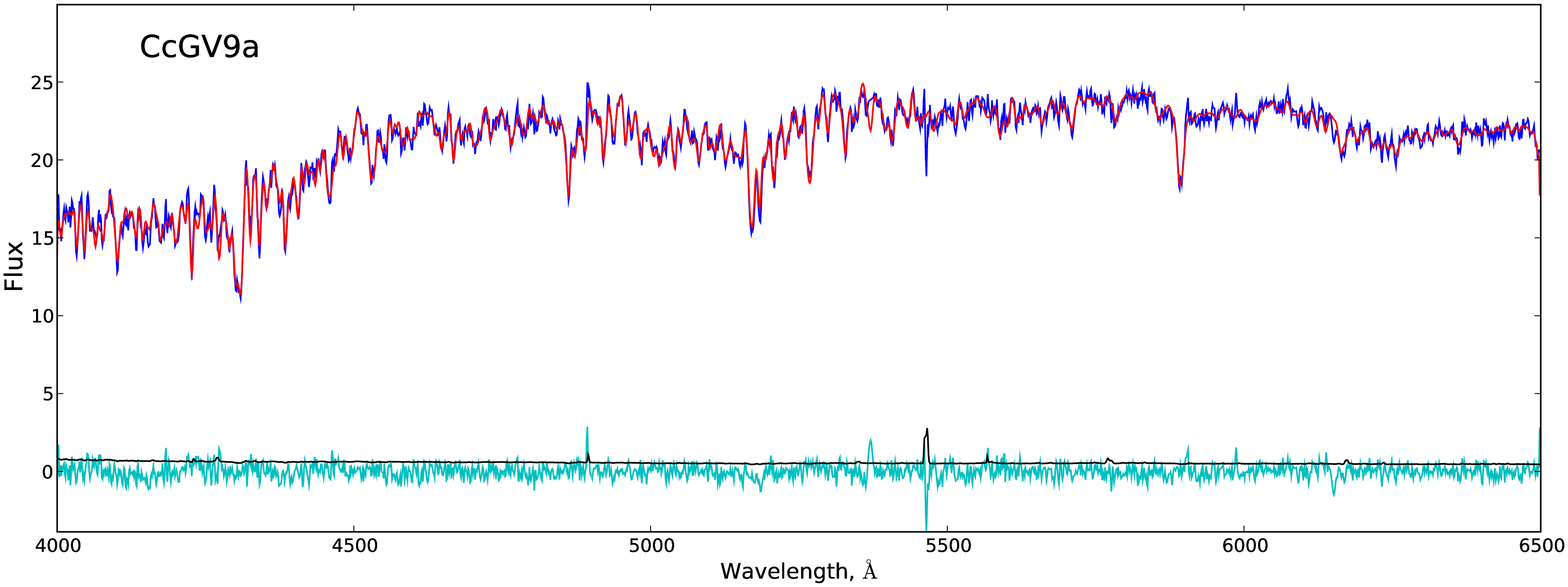}}\\
\scalebox{0.42}[0.38]{\includegraphics{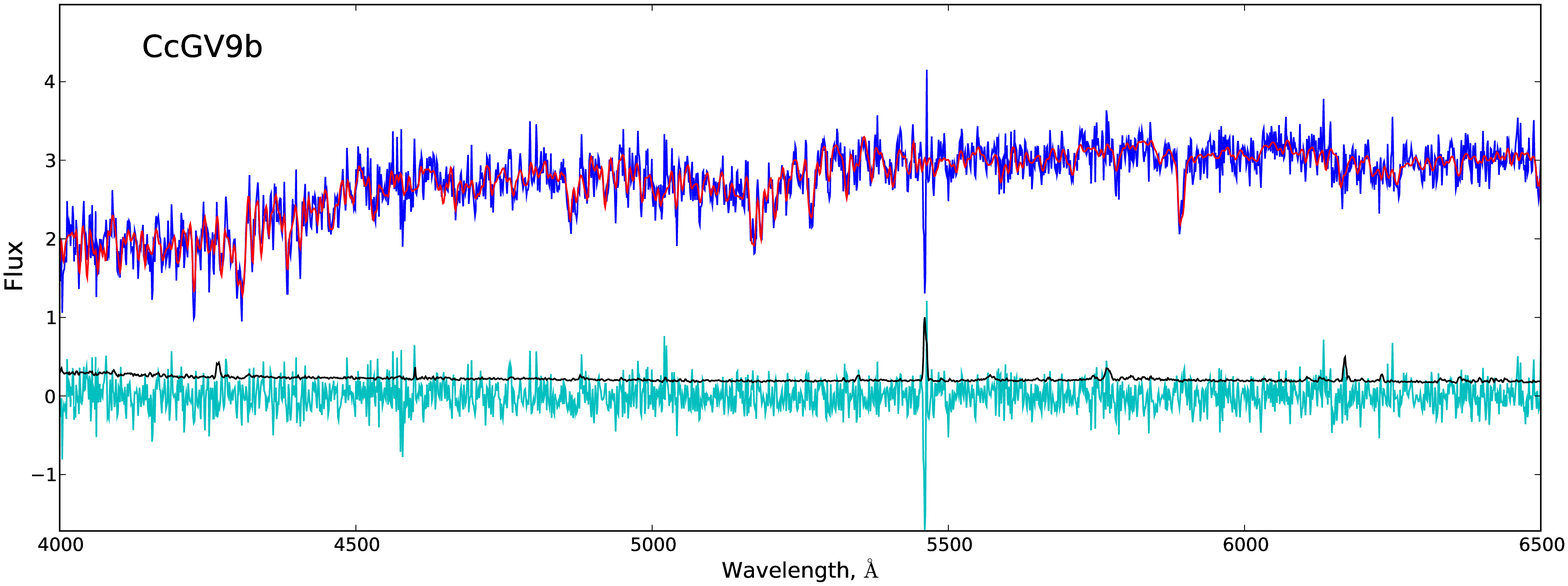}}\\ 
\scalebox{0.42}[0.38]{\includegraphics{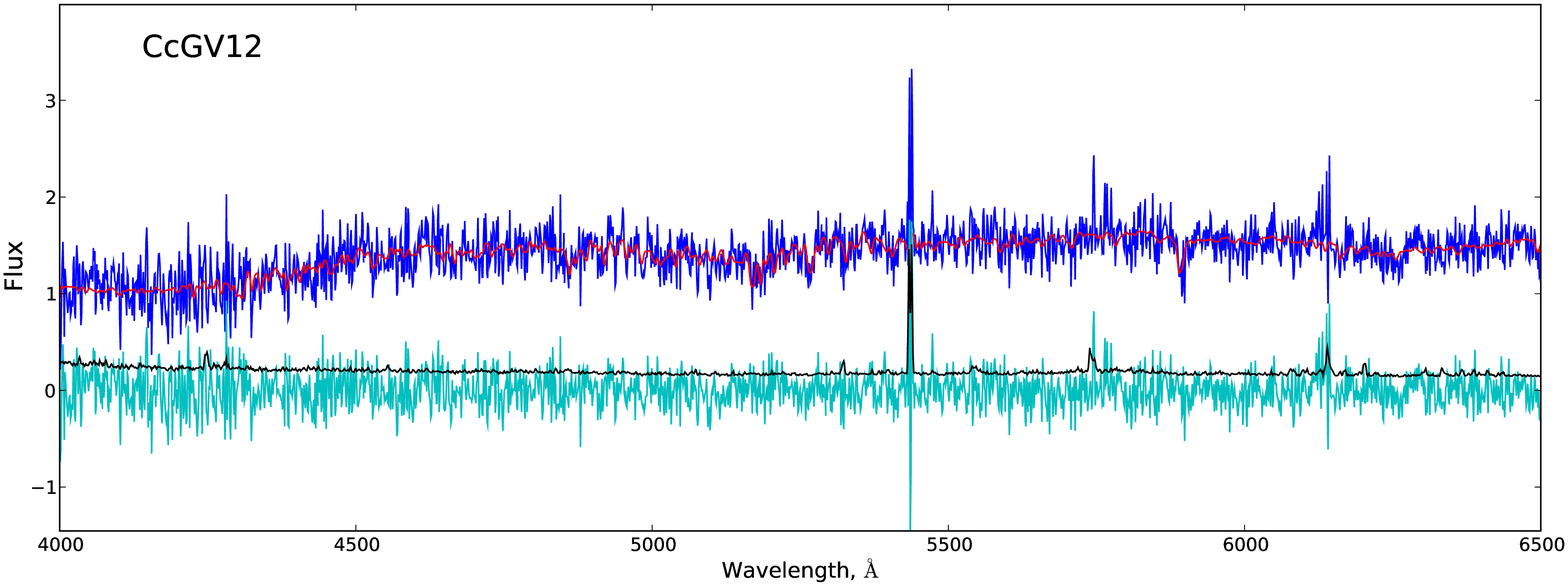}}\\
\caption{CcG spectra as observed by Hectospec. Observed data, PPXF best fits, residuals and 1$\sigma$ error spectra are denoted by blue, red, cyan and black lines respectively.}
\label{spectra}
\end{figure*}

\addtocounter{figure}{-1}
\begin{figure*}
\centering
\scalebox{0.43}[0.39]{\includegraphics{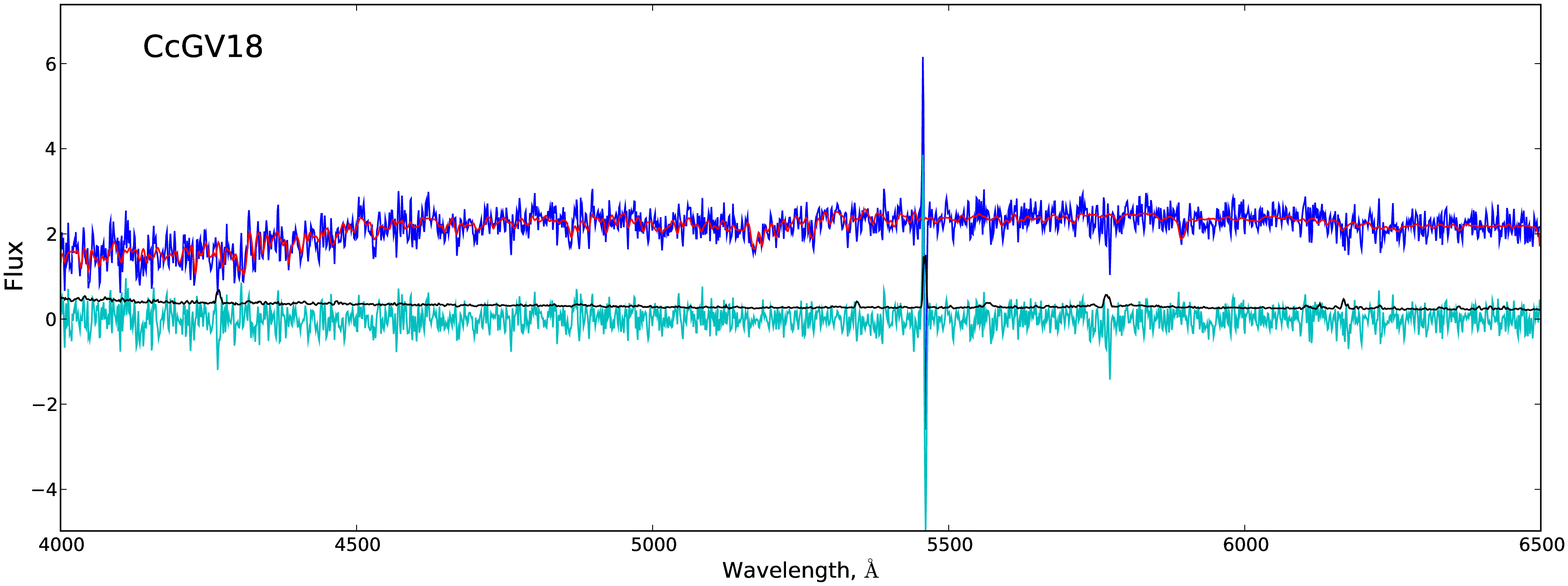}} \\ 
\scalebox{0.43}[0.39]{\includegraphics{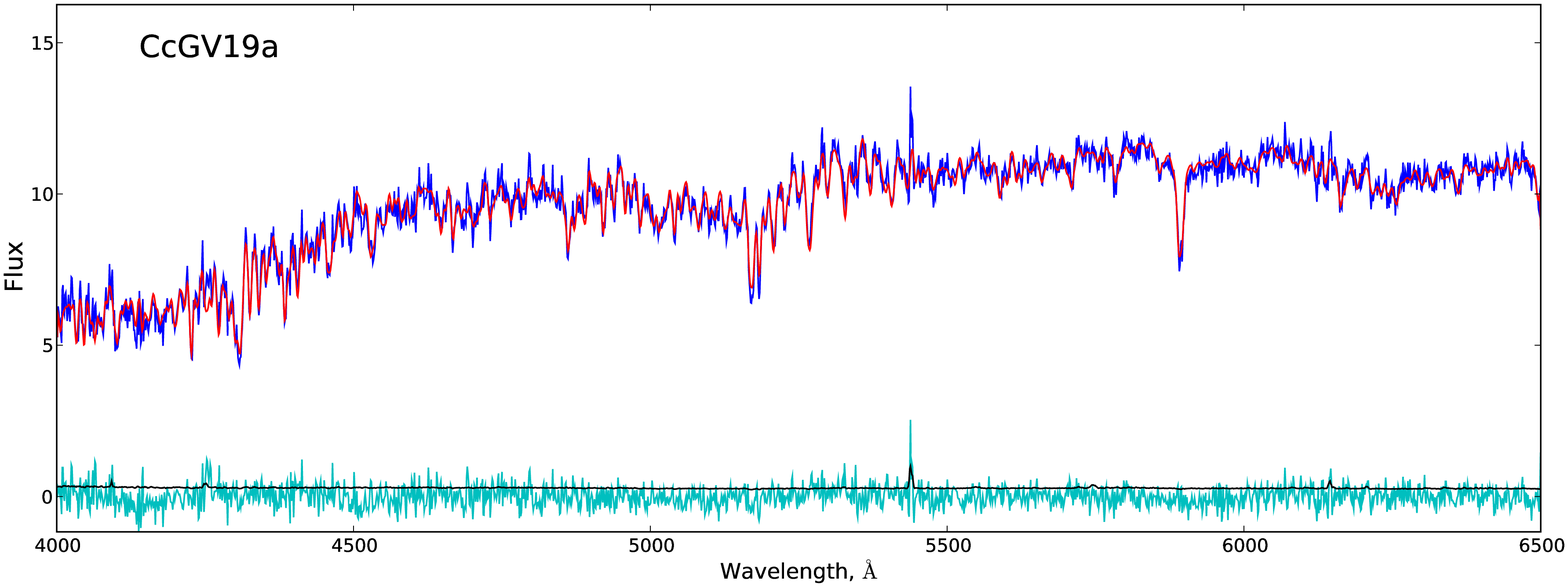}}\\
\scalebox{0.43}[0.39]{\includegraphics{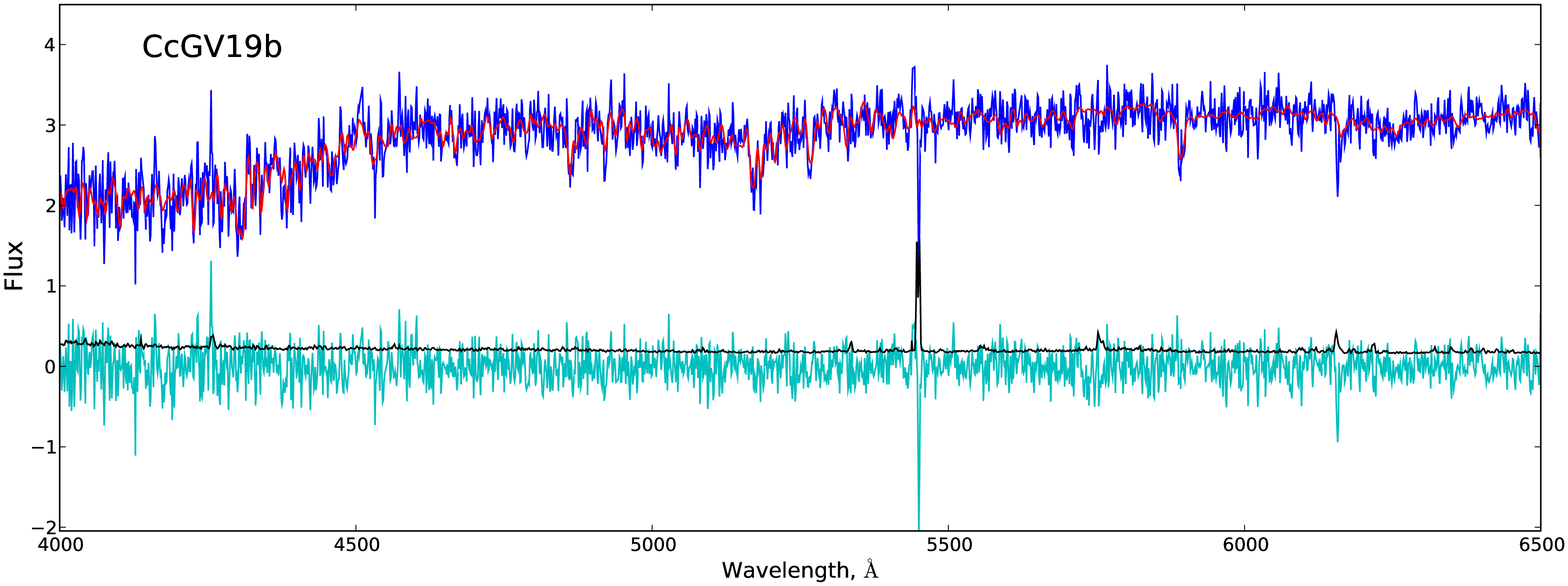}}\\
\caption{continued.}
\end{figure*}

The actual fitting proceeds by smoothing the template spectra to match the instrumental resolution and rebinning data, templates and uncertainty spectra onto a logarithmic scale. Next the fitting range is set to 4000-6500\AA\ (in the rest frame), a region chosen so as to avoid the saturated H and K lines in the blue and the increasing sky subtraction uncertainties in the red. Taking advantage of fitting in pixel as opposed to Fourier space, we mask poorly subtracted sky lines and the NaD resonance line at 5892\AA. As our spectra were taken only with measuring redshifts in mind, their signal-to-noise (S/N) is often less than ideal. However, we make an attempt to fit the entire sample using PPXF and to suitably quantify the uncertainty on each LOSVD for completeness. Fig. \ref{spectra} and Table \ref{veldisp} present the spectra of our sample together with best fitting model, residuals of data minus model and 1$\sigma$ uncertainty spectra and the velocity dispersion resulting from the fits.

\begin{table*}
\begin{minipage}{135mm}
\centering
\caption{Median S/N per \AA\ and measured velocity dispersions for our sample based on Hectospec 2007 and 2008 observations and a fitting range of 4000-6500\AA. The fourth and fifth columns present the estimated central and global velocity dispersions for each galaxy while the final two columns display the virial mass estimate and corresponding M/L within 5$R_{e}$.}
\label{veldisp}
\begin{tabular}{lr@{.}lr@{.}l@{$\pm$}r@{.}lr@{.}l@{$\pm$}r@{.}lr@{.}l@{$\pm$}r@{.}lr@{.}l@{$\pm$}r@{.}lr@{.}l@{$\pm$}r@{.}l}
\hline
\multicolumn{1}{c}{CcG}
& \multicolumn{2}{c}{S/N}
& \multicolumn{4}{c}{$\sigma$}
& \multicolumn{4}{c}{$\sigma_{0}$}
& \multicolumn{4}{c}{$\sigma_{global}$} 
& \multicolumn{4}{c}{M$_{dyn}$} 
& \multicolumn{4}{c}{(M/L$_{B}$)$_{dyn}$}
\\
\multicolumn{1}{c}{(ID)}
& \multicolumn{2}{c}{-}
& \multicolumn{4}{c}{(km s$^{-1}$)}
& \multicolumn{4}{c}{(km s$^{-1}$)}
& \multicolumn{4}{c}{(km s$^{-1}$)} 
& \multicolumn{4}{c}{(10$^8$ M$_\odot$)} 
& \multicolumn{4}{c}{(M$_\odot$/L$_\odot$)}
\\
\hline
V1 & 13 & 2 & 85&2&13&9 & 91 & 2 & 14 & 9 &         80 & 9 & 13 & 2 &37&9& 12&4& 57&1&18&6\\
V9a & 38& 4 & 97&3&5&0  & 102 & 3 & 5 & 4 &        90 & 7 & 4  & 8 &75&6&    7&9 &9&6   & 1&4\\
V9b & 12 & 7 & 25&1&15&8 & 26 & 6 & 16 & 8 &       23 & 6 & 14 & 9 &4&0&    5&1 &4&8  &  6&1\\
V12 & 7 & 3 & 42&2&30&0 & 46 & 4 & 33 & 0 &          41 &  2 & 29 & 3 &5&0&    7&1 &18&1  &  25&7\\
V18 & 7& 1 & 66&3&25&3  & 71 & 9 & 27 & 4 &          63 &  8 & 24 & 3 &17&3&   13&2 &45&3  &  34&5\\
V19a & 33&9 & 76&1&5&1 	& 82 & 3 & 5 & 6 &         73 &  0 & 4 & 9 &24&2&    3&3 &11&7 &   1&6\\
V19b & 13&9 & 65&8&17&7 & 74 & 1 & 19 & 9 &         65 &  7 & 17 & 6 & 7&2&    3&8 &13&8  &  7&5\\
\hline
\end{tabular}
\end{minipage}
\end{table*}

We test the robustness of our measurements in a variety of ways. To estimate the random error present on our velocity dispersions we perturb the best fitting model spectrum for each galaxy by the relevant uncertainty spectrum and run PPXF again on the new pseudo-observation. Following 100 executions of this process we take the standard deviation of the distribution of velocity dispersions as our 1$\sigma$ errors quoted in Table \ref{veldisp}. As we use the best fitting model for this process, template mismatch errors are not propagated into this uncertainty. 

Systematics in terms of the routine's ability to resolve dispersions significantly below our velocity resolution, particularly from data of low S/N, are again tested for via simulations. We take a high S/N model spectrum of comparable metallicity and age as our galaxies (see next section) from the Vazdekis library and smooth it to our instrumental resolution. Next we convolve the spectrum with a gaussian of width equal to the measured velocity dispersion of a particular galaxy in our sample and add the relevant noise. Starting with CcGV12, CcGV18 and CcGV9b we repeatedly generate such fake observations with known velocity dispersion and run them through PPXF. These galaxies possess the lowest S/N and smallest measured velocity dispersions of our sample. We find CcGV12 and CcGV18 to have systematic uncertainties in their velocity dispersions of 10\% while CcGV9b has 25\% uncertainty on its measurement. To place this in context, CcGV1 has a $\sim$1\% systematic error. 

As our wavelength fitting range is larger than typically used when measuring velocity dispersions, we also test PPXF over a range of 4400-5400\AA. In all cases we find that the differences are consistent within the random error presented in Table \ref{veldisp}. Therefore, along with the resolvability issues discussed above, we make no attempt to correct our velocity dispersions for systematics as it is clearly the random error that dictates the uncertainty in our measurements. Finally we note that CcGV9a is the only galaxy in our sample that has had its velocity dispersion measured elsewhere. SDSS DR6 observed the galaxy as having a velocity dispersion of 87$\pm$9 km s$^{-1}$ through a 3$^{\prime\prime}$ diameter fibre which is consistent with our result within the joint uncertainties.

\begin{figure}
\scalebox{0.42}[0.42]{\includegraphics{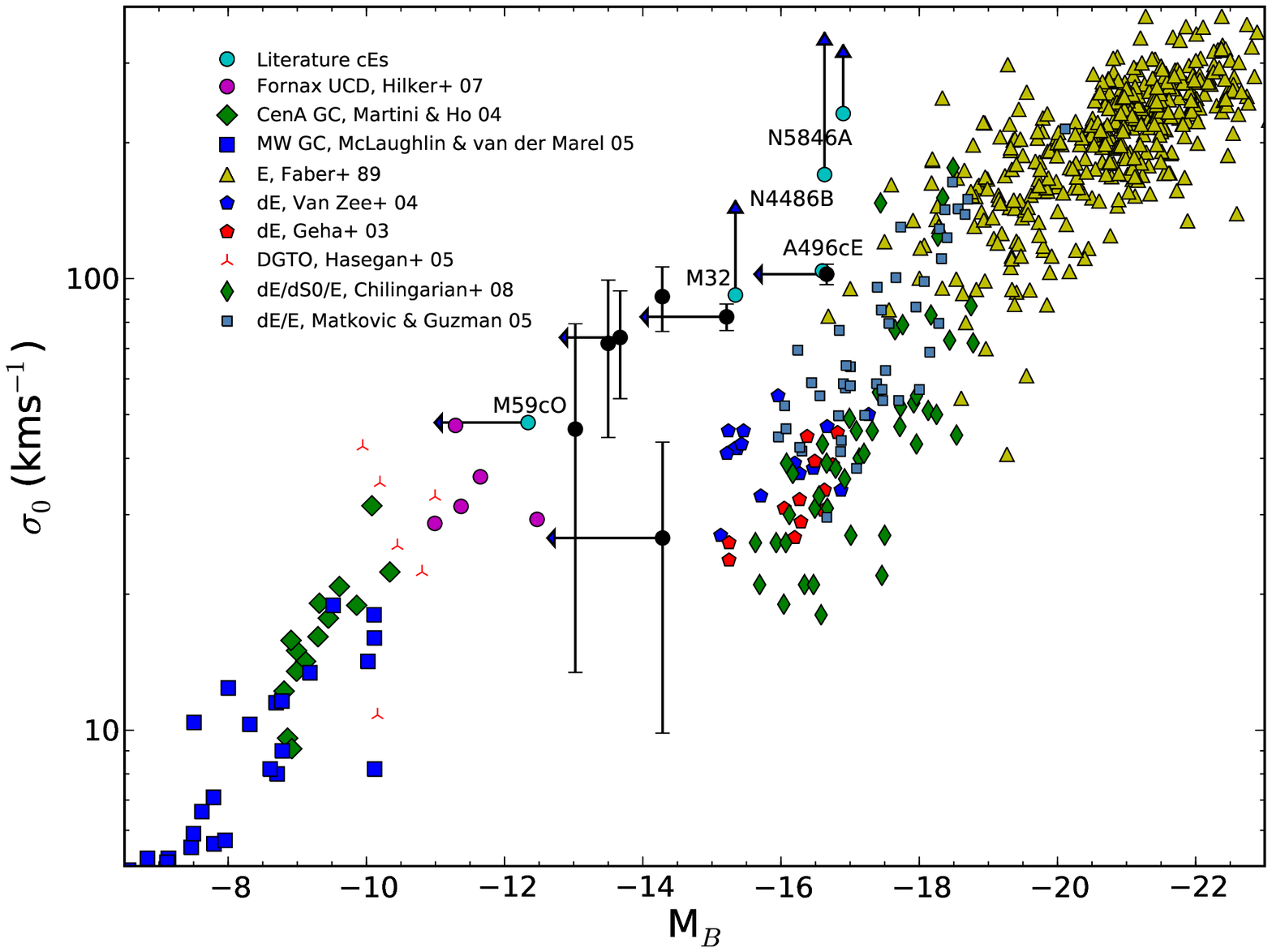}}
\caption{Faber-Jackson relation for elliptical galaxies, UCDs, DGTOs, GCs and literature compact galaxies. Coma compacts are plotted as black filled circles. The horizontal arrow ends show the absolute luminosity of the central components of our sample where resolved. The vertical arrow ends correspond to the central velocity dispersion measurements obtained for M32, NGC 4486B and NGC 5846A by \citet{davidge08}.}
\label{fj}
\end{figure}

In Fig. \ref{fj} we present the \cite{faberjackson76} relation for a compilation of hot stellar systems. For elliptical galaxies we take data from \cite{faber89}, \cite{geha03}, \cite{vanzee04}, \cite{matkovic05} and \cite{chil08}. We plot data for GCs in NGC 5128 from \cite{martini&ho04} and the Milky Way from \cite{mcl&vdm05} as well as UCDs in Fornax from \cite{drinkwater03} and \cite{hilker07} and DGTOs in Virgo from \cite{hasegan05}. Finally we include several cEs from the literature. For M32 we take data from \cite{bender96}, for NGC 4486B from \cite{smith00}, for NGC 5846A from \cite{muller99} and the SDSS, for A496cE from \cite{chilingarian07} and M59cO from \cite{chilandmamon08}. B band absolute magnitudes for M32 and NGC 4486B are those reported in Table 1 of \cite{chilingarian07} and references therein. It should be noted that the three archetypal compact galaxies, M32, NGC 4486B and NGC 5846A, have had their central velocity dispersions measured a number of times in the literature with a variety of different instruments. Using HyperLeda we have selected up to date representative measurements from ground based studies which provide a suitable comparison to the data presented in the rest of the diagram. Recent high spatial resolution adaptive optics observations by \cite{davidge08} of these three galaxies are also plotted for completeness and correspond to the upper ends of the arrows attached to these galaxies. Using the prescription of \cite{jorgensen95}, we correct our velocity dispersion measurements to a suitably small aperture (total $R_{e}/4$) to estimate their central values. In all cases the correction is small, $<$ 13\%, and comparable to the measurement errors. The estimated central velocity dispersions for our sample are tabulated in Table \ref{veldisp} and plotted on Fig. \ref{fj}. The head of each arrow denotes the position on the diagram of the central component of each cE where resolved. For M32 and A496cE only the bulge absolute magnitude is plotted. We note that conventional galaxies are largely absent from the parameter space of Fig. \ref{fj} at M$_{B}$ $<$ -15 mag, likely owing to their low surface brightness (see Fig. \ref{structplots}).

Fig. \ref{fj} shows that, along with the cEs from the literature, the majority of compact galaxies presented here conform to a tight sequence beginning at DGTOs and UCDs and extending to well above normal elliptical galaxies. Both CcGV9a and CcGV19a have velocity dispersions that are a factor of two larger than dEs at their respective magnitudes. It is noted that intermediate luminosity ellipticals do almost reach CcGV9a/A496cE, though they are clearly extreme outliers. The Coma compacts, CcGV9b aside, essentially begin to fill in the parameter space between M59cO and M32. From the diagram only CcGV9b appears to differ from the trend. Unfortunately due to the large uncertainty on this galaxy's velocity dispersion it is hard to comment in a definitive way. However it would seem possible that this galaxy is some sort of intermediate between dEs and the compact sequence.

With the velocity dispersions in hand we are able to use the virial mass estimator \citep{spitzer87},

\begin{equation}
M_{vir} \approx \beta\frac{R_{e}\sigma^{2}}{G}
\end{equation}

\noindent where $\sigma$ is the global projected velocity dispersion, $R_{e}$ the projected half light (mass) radius and $\beta$ the virial coefficient. We estimate $\sigma$ by correcting our data via the expression of \cite{jorgensen95} to an aperture of radius 5$R_{e}$ and tabulate our estimated global dispersions in Table \ref{veldisp}. For all profiles considered this contains $>$90\% of the flux and by inference mass of the system. Total galaxy $R_{e}$ is again computed by numerical integration. The choice of scaling factor, here set as $\beta$ = 10, warrants some further comment. This factor incorporates a number of corrections including those for projection effects and the specific density distribution of the system in question. Recent work by \cite{cappellari06}, which compared complex dynamical models with equation (4), found that for a sample of 25 E/S0 galaxies $\beta$ $\sim$ 5 with $\sigma$ measured within an aperture of radius $R_{e}$. This result is specifically based on one component R$^{1/4}$ profiles and is demonstrated in their Fig. 13, which shows the relation between their computed masses and those from equation (4) assuming $\beta$ = 5. It is interesting to note that in this figure M32, the lowest mass galaxy in their sample, is a substantial outlier above the relation, well beyond quoted errors. Disregarding other cEs, the next step down in the compact stellar system hierarchy are UCDs. Studies of these objects in both Fornax and Virgo have found that $\beta~\sim$ 10 reproduces well the results of more rigorous methods \citep{hilker07,evstigneeva07b}. Therefore, as it would seem $\beta$ could in fact be anywhere between these two extremes for our objects, we opt to follow previous compact galaxy work such as \cite{chilandmamon08} and set $\beta$ = 10. 

Dynamical masses and mass to light ratios (M/L) for our galaxies can be found in Table \ref{veldisp}. Uncertainties are those obtained by propagating the error on each galaxy's velocity dispersion and total B band magnitude. Aside from random errors, there are a number of caveats which apply when using the virial mass estimator. The expression assumes a constant M/L with radius, an isotropic velocity distribution and that the galaxies are spherical. Given these factors and the inherent uncertainty in scaling the aperture velocity dispersions to global values, the masses and corresponding M/L presented in Table \ref{veldisp} should be treated as best estimates given the available data.

\subsection{Stellar Population Parameters}

The standard approach when attempting to recover luminosity weighted age and metallicity for unresolved stellar populations is that of line index measurements. Pioneered by \cite{burstein84} and \cite{gorgas93} these indices target key information rich absorption features present in integrated galaxy spectra. Following the introduction of Lick Indices in particular \citep{faber85}, it was shown that when combined with predictions from SSP models they are able to break the age-metallicity degeneracy for such systems inherent in broad band photometric models \citep{worthey94}. More recent work has focused on updating and refining SSP models to correctly quantify the effects of non-solar abundance patterns \citep{thomas03,schiavon07} which have been observed in early-type galaxies for some time \citep{peletier89,worthey92,trager97}. Typically such $\alpha$-enhancement has been detected as an over abundance of Mg relative to Fe in luminous elliptical galaxies and has been seen as evidence for shorter star-formation episodes in more massive galaxies. This stems from the enrichment history of the interstellar medium out of which the current generation of stars formed. On short timescales this is dominated by Type II supernova, yielding more Mg than Fe relative to the solar neighbourhood, but on longer timescales by Type Ia supernova which redress the balance.

In this section we use the models of \cite{schiavon07} and the EZ-Ages code devised by \cite{graves08} to measure the age, metallicity and key abundance indicator [Mg/Fe] of our compact galaxies. Briefly, the Schiavon models are based on the flux-calibrated \cite{jones99} spectral library and allow for variable abundance ratios in terms of Mg, C, N and Ca to be present in the stellar atmospheres. Additionally the model spans a range of 1 to 17.7 Gyr and -1.3 to +0.2 in [Fe/H]. Note that the models are cast at given [Fe/H], rather than total metallicity $Z$. Here we use a Salpeter initial stellar mass function and scaled-solar isochrones.

EZ-Ages employs the model via a sequential grid inversion method to return consistent stellar population parameters. The process begins by making an initial measurement of age and metallicity from the Lick H$\beta$ and $<$Fe$>$ indices of the galaxy in question, where $<$Fe$>$ is the average of the Fe5270 and Fe5335 indices. These are selected as they provide good age (H$\beta$) and metallicity response (Fe5335 and Fe5270), while being essentially insensitive to abundance pattern variations \citep{schiavon07}. From the H$\beta$ -- $<$Fe$>$ index-index grid a two dimensional linear interpolation converts the position in index space to the fiducial age and metallicity values. 

Next the code moves onto determining the [Mg/Fe] ratio and to do this creates a new grid of H$\beta$--Mgb5177 (or Mg2). If the age and metallicity obtained from the inversion of this grid do not match the initial estimates it is assumed a non-solar [Mg/Fe] is present and the model is recomputed with a new [Mg/Fe] ratio. Further iterations occur until the age and metallicity derived from both grids match within some tolerance. EZ-Ages then goes on to repeat this process replacing Mg for C, N and Ca in turn with the relevant index used to compare back to the fiducial age and metallicity estimates. Finally the new abundance pattern is used to compute a new model and derive a new fiducial age and metallicity with the entire fitting sequence repeated until the fit does not improve.

Unfortunately the S/N of our spectra is once again an issue and we restrict our line index analysis to the five galaxies with the highest quality spectra to avoid uncertainties that would span entire index-index diagrams. Furthermore, while a full abundance pattern is fit, as the uncertainties associated with the initial parameters are propagated in turn through the fitting scheme, we limit our analysis to age, metallicity and [Mg/Fe].

\begin{table}
\centering
\caption{Absorption line index data used in our analysis of the CcGs stellar population parameters.}
\label{indices}
\begin{tabular}{lc@{$\pm$}cc@{$\pm$}cc@{$\pm$}cc@{$\pm$}c}
\hline
\multicolumn{1}{c}{CcG}	
 & \multicolumn{2}{c}{Hbeta} 
 & \multicolumn{2}{c}{Mgb} 
 & \multicolumn{2}{c}{Fe5270} 
 & \multicolumn{2}{c}{Fe5335} \\
\multicolumn{1}{c}{(ID)}	
 & \multicolumn{2}{c}{(\AA)} 
 & \multicolumn{2}{c}{(\AA)} 
 & \multicolumn{2}{c}{(\AA)} 
 & \multicolumn{2}{c}{(\AA)}
  \\
\hline
V1 & 1.94 & 0.45 & 3.64 & 0.44 & 3.30 & 0.46 & 3.23 & 0.52 \\
V9a & 1.74 & 0.14 & 4.09 & 0.16 & 2.16 & 0.25 & 1.86 & 0.19 \\
V9b & 2.62 & 0.45 & 3.72 & 0.45 & 2.67 & 0.48 & 2.23 & 0.56 \\
V19a & 1.62 & 0.17 & 4.50 & 0.17 & 3.00 & 0.18 & 2.41 & 0.21 \\
V19b & 1.71 & 0.42 & 3.88 & 0.40 & 1.87 & 0.45 & 2.09 & 0.53 \\
\hline
\end{tabular}
\end{table}

To measure the desired indices we use the IDL code provided with EZ-Ages. In addition to the galaxy and uncertainty spectra the script requires the velocity dispersion for each galaxy so that the spectra may be correctly smoothed to the Lick resolution which itself varies for each index (see \cite{schiavon07} Table 1). The required smoothing in each case is $\sigma_{smooth}=\sqrt{\sigma_{lick}^2-\sigma_{res}^2-\sigma_{gal}^2}$ where $\sigma_{res}$ is the Hectospec instrumental resolution and $\sigma_{gal}$ the galaxy velocity dispersion. Index errors are computed using the equations of \cite{cardiel98}. The relevant indices used in our analysis are tabulated in Table \ref{indices}.

\begin{figure}
\scalebox{0.42}[0.42]{\includegraphics{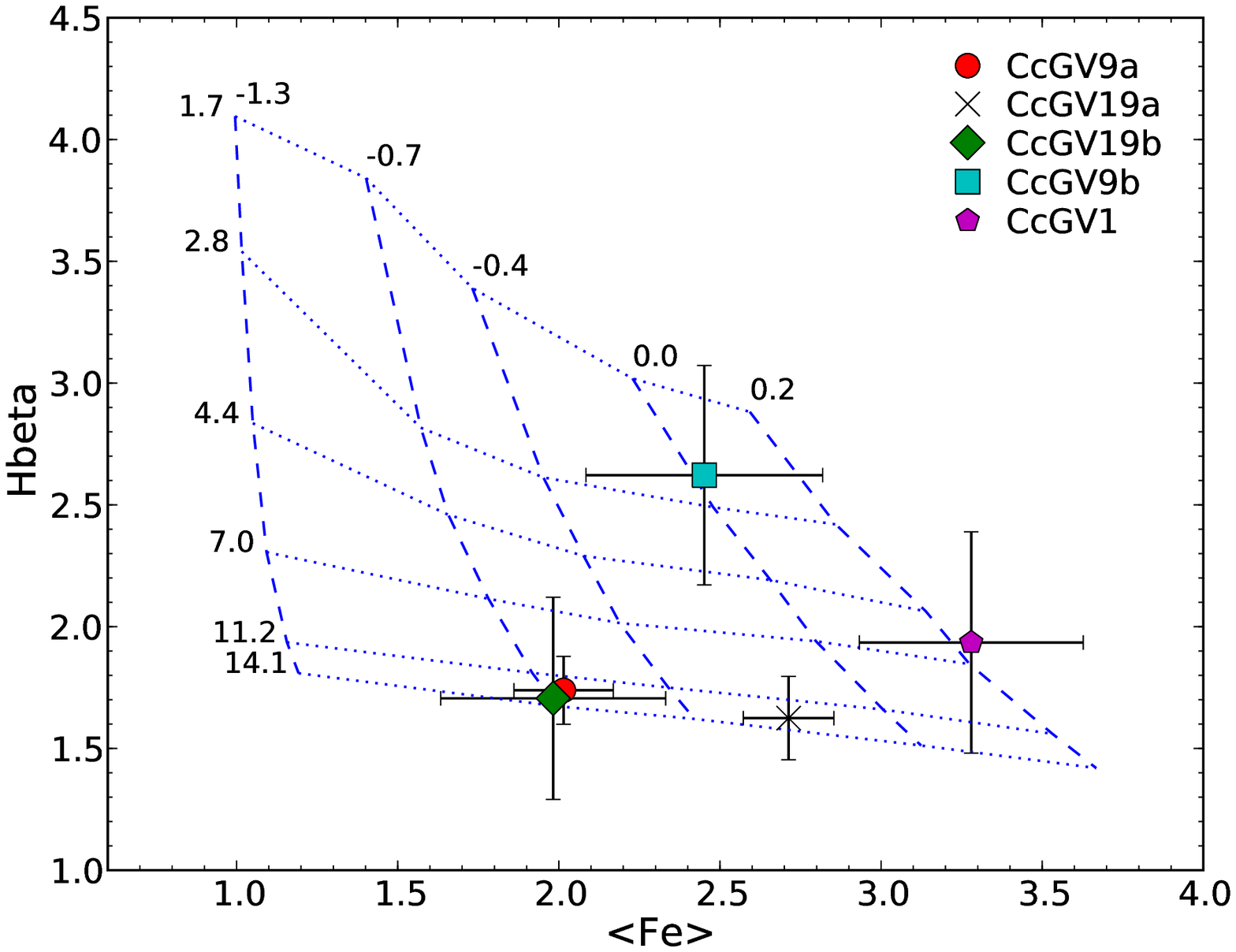}}
\caption{Example of the index-index diagram used by EZ-Ages to measure fiducial age and metallicity before going on to fit a full abundance pattern. The grid is for a solar abundance model and so is not tailored specifically for any galaxy as would be the case during normal fitting. Vertical dashed lines are iso-metallicity lines with [Fe/H] = -1.3, -0.7, -0.4, 0.0 an 0.2 and horizontal solid lines are constant age 1.7, 2.8, 4.4, 7.0, 11.2, 14.1 Gyr.}
\label{refgrid}
\end{figure}

Fig. \ref{refgrid} presents an example H$\beta$--$<$Fe$>$ diagram used by EZ-Ages to determine fiducial ages and metallicities. It is important to note that the index grid in this figure is generated from a model with a solar abundance pattern and so is slightly offset with respect to the grids that would be obtained from each galaxy's final best fitting abundance pattern. Hence to avoid overlaying five only marginally different grids we opt to display the data this way for reference only.

\begin{table}
\centering
\caption{Derived stellar population parameters and stellar M/L ratios for the Coma compact galaxies with sufficient S/N. Upper error bounds set to $max$ indicate the uncertainty estimate exceeds the models coverage in that direction.}
\label{agemetmg}
\begin{tabular}{lr@{}c@{}lr@{}c@{}lr@{}c@{}lr@{}c@{}l}
\hline
\multicolumn{1}{c}{CcG}
& \multicolumn{3}{c}{Age (Gyr)}
& \multicolumn{3}{c}{[Fe/H]} 
& \multicolumn{3}{c}{[Mg/Fe]}
& \multicolumn{3}{c}{(M/L$_{B}$)$_{\star}$}\\
\multicolumn{1}{c}{(ID)}
& \multicolumn{3}{c}{(Gyr)}
& \multicolumn{3}{c}{(dex)}
& \multicolumn{3}{c}{(dex)}
& \multicolumn{3}{c}{(M$_{\odot}$/L$_{\odot}$)}
\\
\hline
  V1 & 6.3 &$\pm$& 3.6 & 0.15&\multicolumn{2}{@{}l}{$^{+max}_{-0.21}$} &-0.04 &$\pm$&0.16 & 5.6 &\multicolumn{2}{@{}l}{$^{+max}_{-2.3}$}\\
  V9a & 12.0 &$\pm$& 2.1 & -0.51 &$\pm$& 0.15 & 0.52 &$\pm$ &0.18 & 4.2 &\multicolumn{2}{@{}l}{$^{+0.9}_{-0.8}$}\\
  V9b & 2.5 &$\pm$ &1.9 & 0.03&\multicolumn{2}{@{}l}{$^{+max}_{-0.25}$} & 0.3 &$\pm$& 0.3&1.95 &\multicolumn{2}{@{}l}{$^{+max}_{-1.5}$}\\
  V19a & 11.9 &$\pm$ &2.4 & -0.12 &$\pm$& 0.1 & 0.28 &$\pm$& 0.1 & 6.0&\multicolumn{2}{@{}l}{$^{+1.2}_{-1.1}$}\\
  V19b & 13.1 &$\pm$ &3.6 & -0.56& $\pm$ &0.29 & 0.48 &$\pm$& 0.36 & 4.2 &\multicolumn{2}{@{}l}{$^{+2.1}_{-1.2}$}\\
\hline
\end{tabular}
\end{table}

\begin{figure*}
\begin{minipage}{190mm}
\flushleft
\scalebox{0.28}[0.28]{\includegraphics{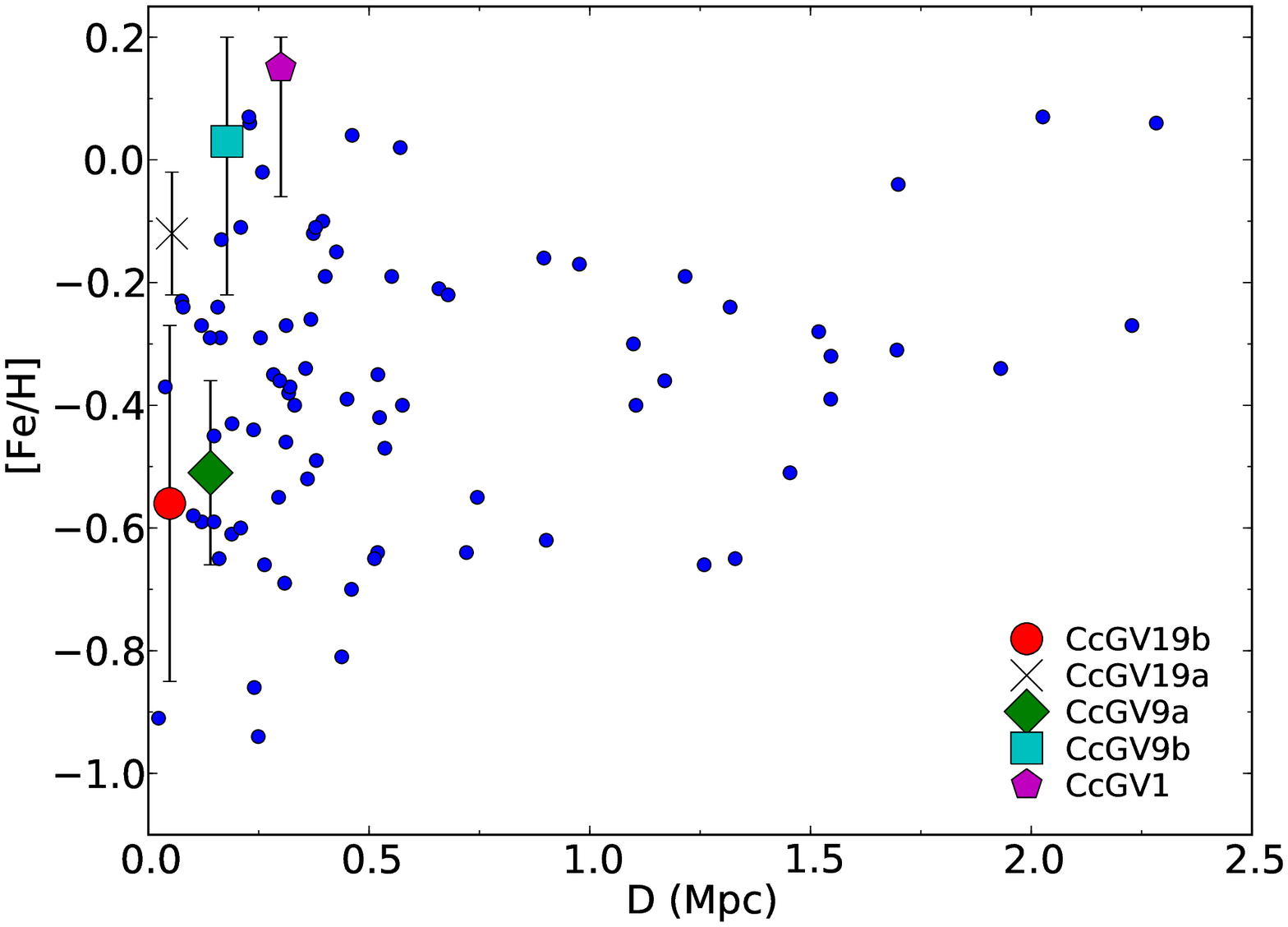}} 
\scalebox{0.28}[0.28]{\includegraphics{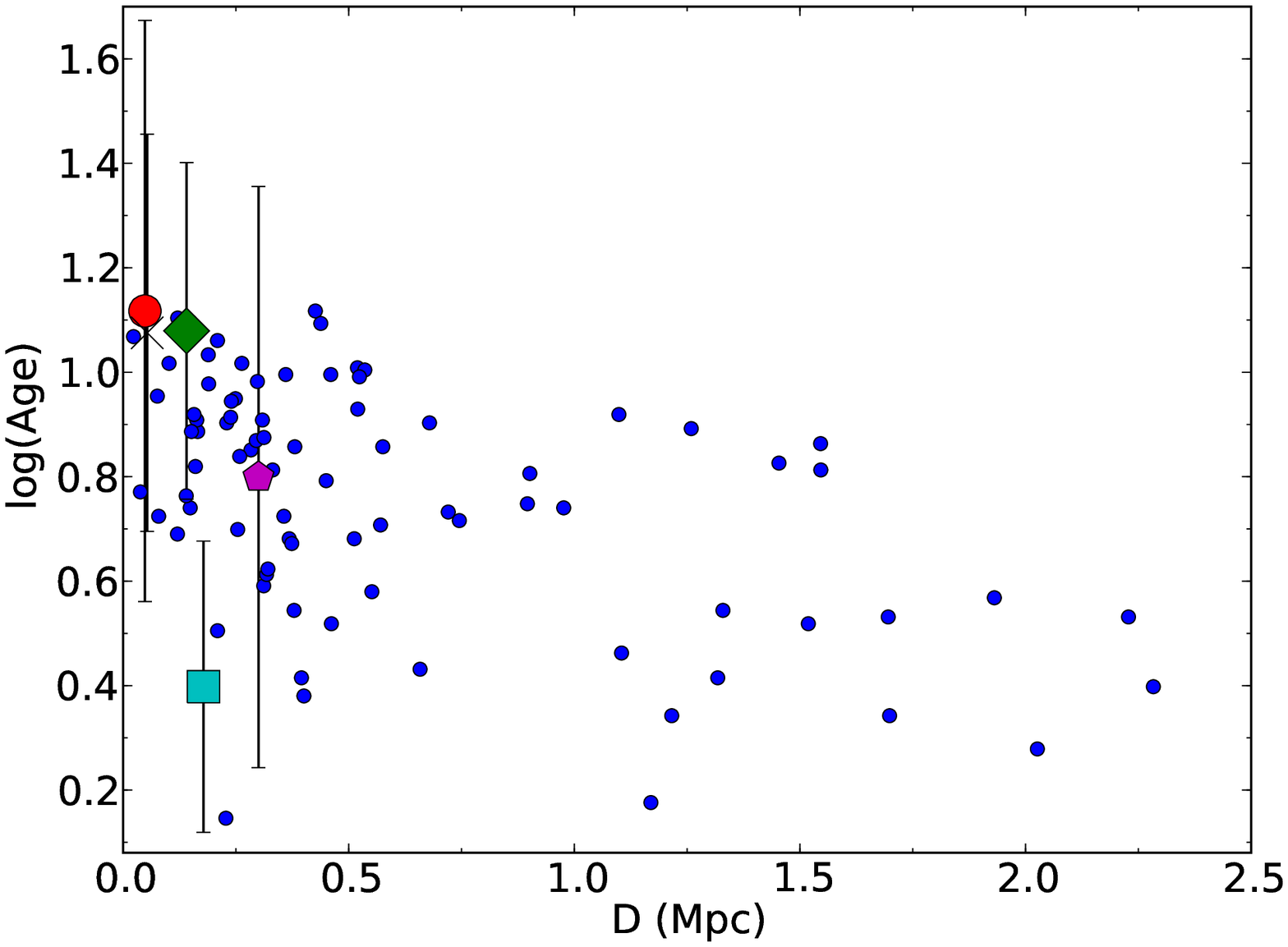}} 
\scalebox{0.28}[0.28]{\includegraphics{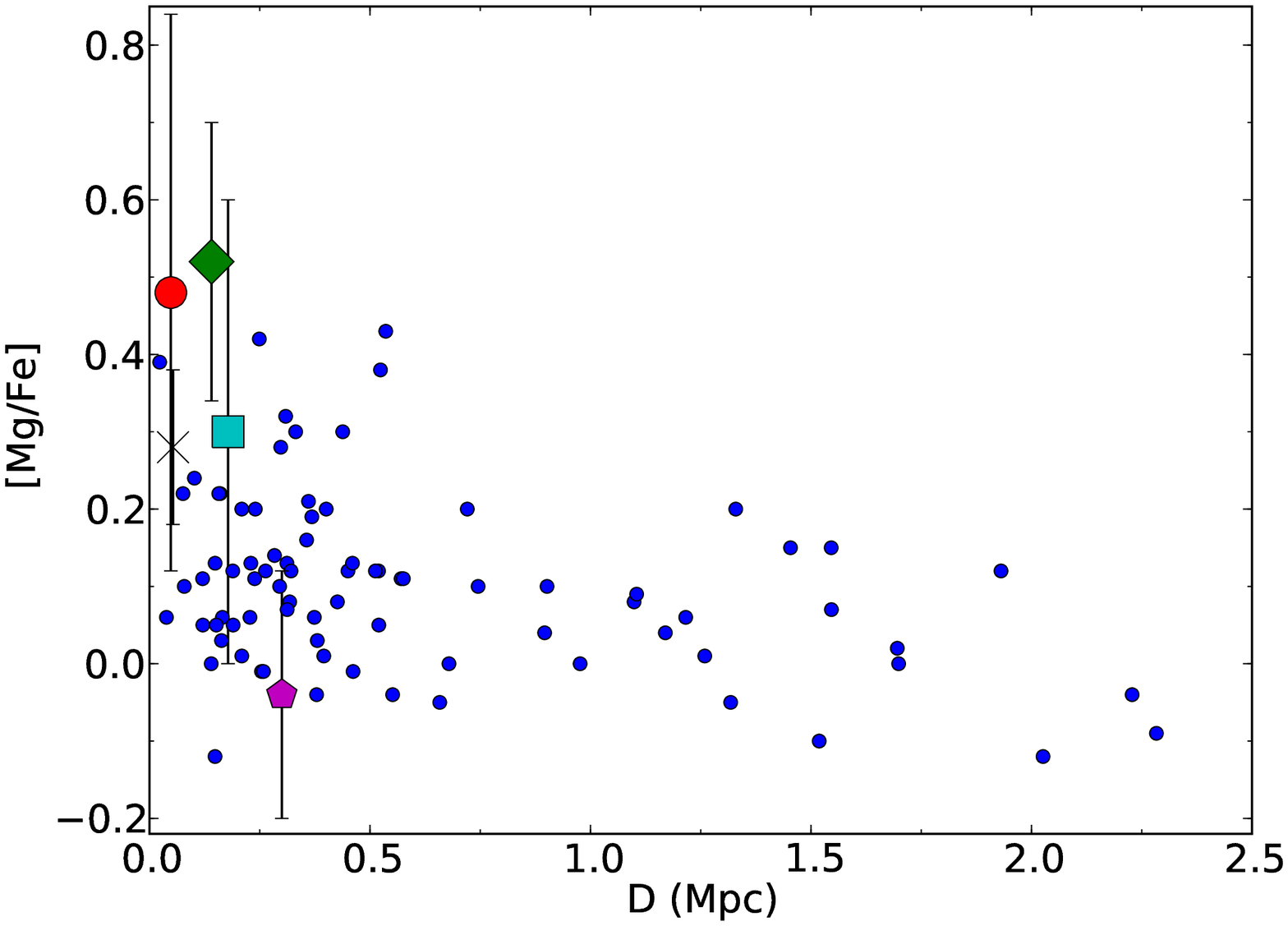}}\\ 
\scalebox{0.28}[0.28]{\includegraphics{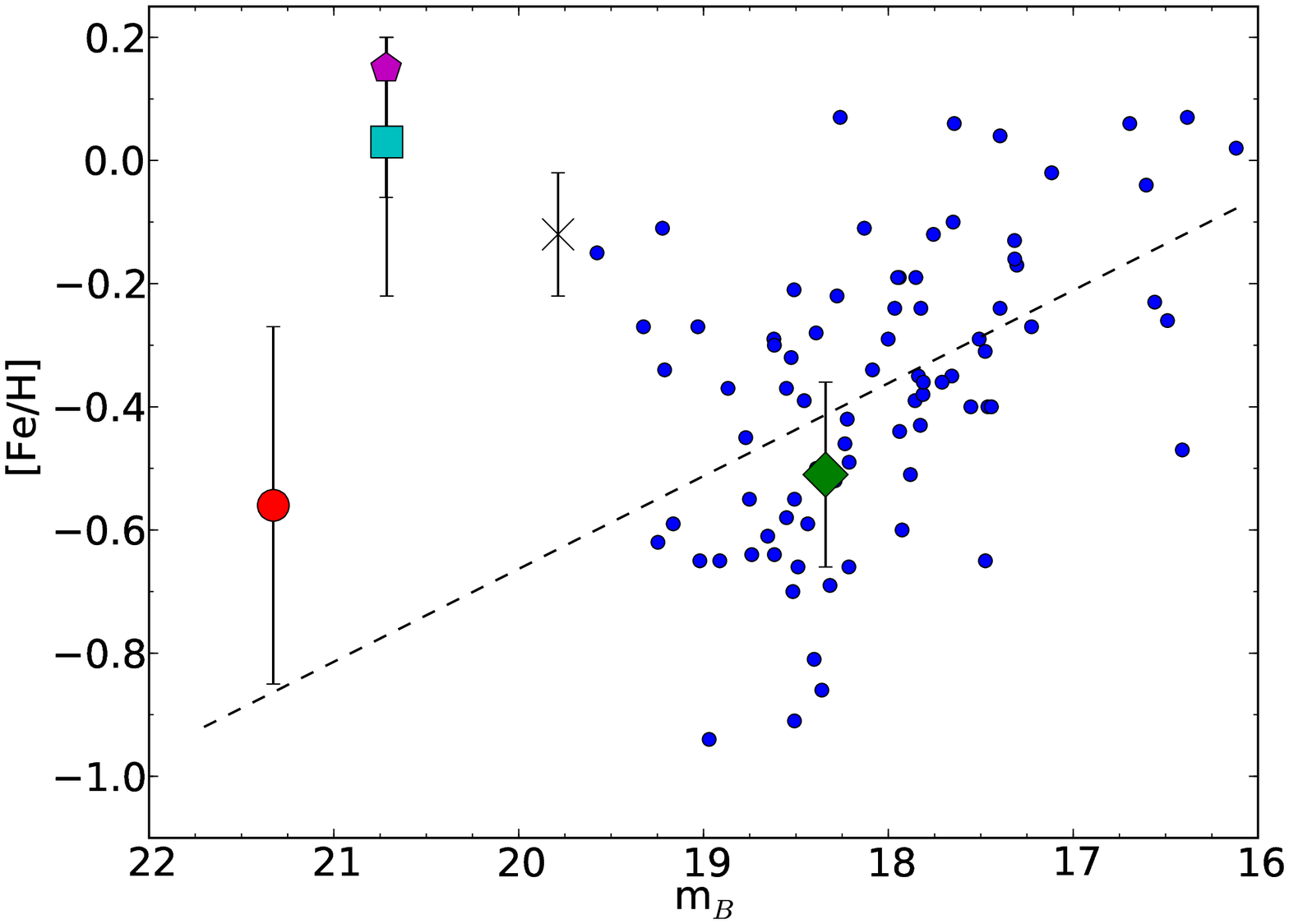}} 
\scalebox{0.28}[0.28]{\includegraphics{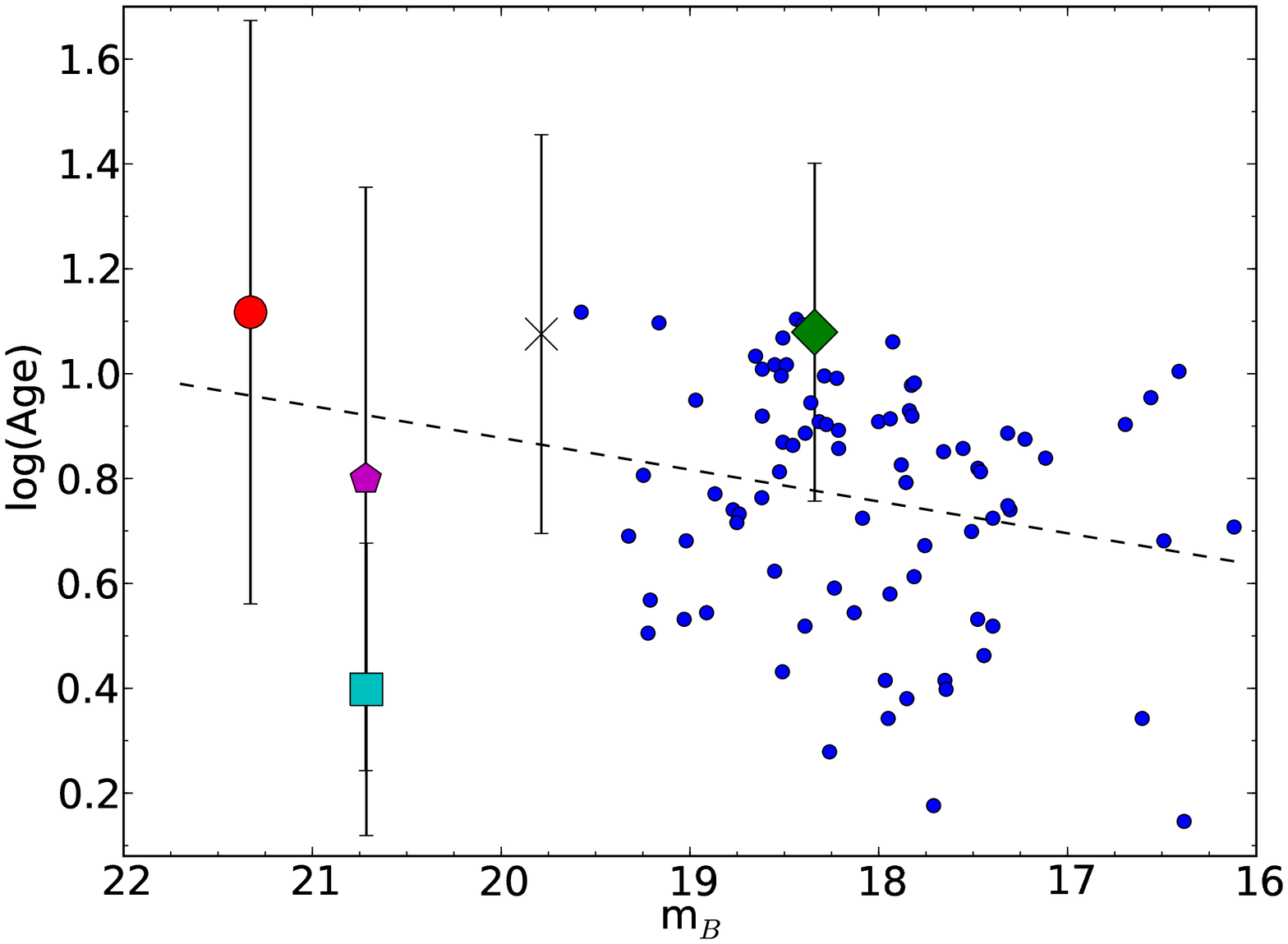}} 
\scalebox{0.28}[0.28]{\includegraphics{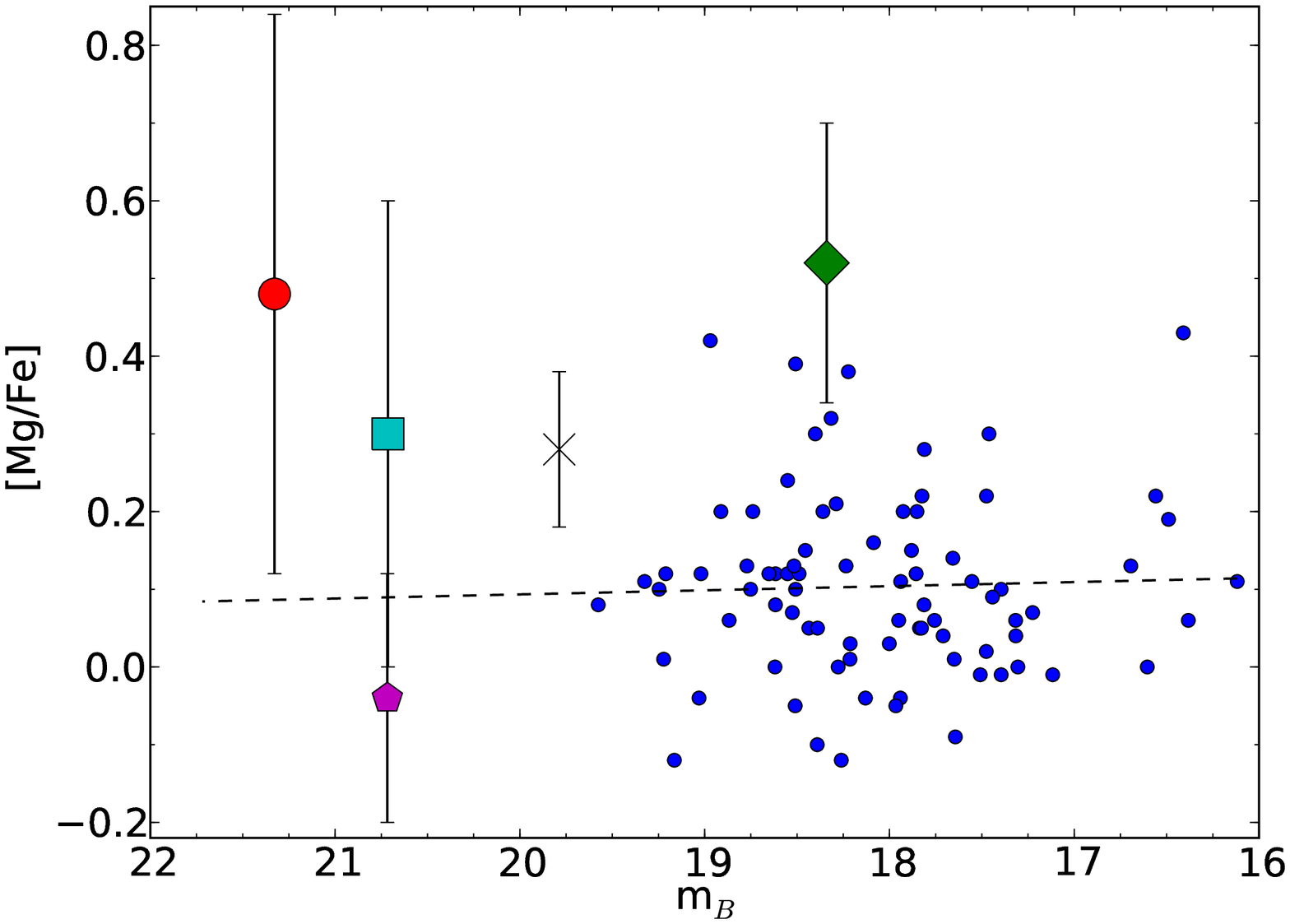}}\\
\caption{Stellar population parameters for the Coma dwarf sample of S09 (blue points) and this work plotted against projected distance from the centre of the cluster (top row) and B band apparent magnitude (bottom row). The black dashed line represents the extrapolation of the relations between magnitude and stellar population parameters identified by S09.}
\label{comadwarfs}
\end{minipage}
\end{figure*} 

Table \ref{agemetmg} displays the stellar population parameters of our sample. In all cases but CcGV1 the default index weighting scheme of $<$Fe$>$ and H$\beta$ for [Fe/H] and age and Mgb for [Mg/Fe] was used. Other combinations were tested but results proved to be stable within errors. The fit of CcGV1 failed with default index weighting because it marginally fell off the H$\beta$--$<$Fe$>$ model grid due to [Fe/H]$\gtrsim$ +0.2. As such we used only Fe5270 to constrain this galaxy's fiducial [Fe/H] and age. Uncertainties are estimated by repeated bootstrapping of the measured indices for each galaxy. We use the measured set of indices, perturbed them randomly by their respective error and ran them through the code. The 1$\sigma$ errors are then taken from the resulting parameter distributions. This somewhat computer intensive approach was chosen, as opposed to using EZ-Ages' built in error scheme, because of situations where the task failed to compute uncertainties on the fiducial [Fe/H], age or both in either positive, negative or both directions. The issue apparently stems from the error bounds being off the model grid and can result in significantly under estimated uncertainties on for instance [Mg/Fe], whose derived error relies on those of the fiducial parameters and the Mgb index. Our method is still not ideal but provides a reasonable solution when trying to suitably assess error propagation. 

Also in Table \ref{agemetmg} we present the inferred stellar M/L ratios for our sample. These are interpolated from Table 33 of \cite{schiavon07} and are based on scaled solar isochrones. The quoted errors are those introduced by the uncertainty on each galaxy's age and [Fe/H] and are also estimated by interpolation. The stellar M/L ratios do not take into account variable abundance patterns in the stellar atmospheres.

Comparing the stellar and dynamical M/L for those galaxies in our sample with both, we see that the three cEs have dark matter fractions of $\sim$ 40 to 70\%. This result is similar to that found by \cite{chilandmamon08} for M59cO but, as one would expect, depends heavily on the choice of the virial coefficient $\beta$. In addition, there is of course an aperture mismatch between that used to derive each of our M/L estimates in the sense that the Hectospec fibres are often smaller than 5$R_{e}$. However, given that all the galaxies have colour profiles which either are flat or get bluer with radius, it is likely that the mean stellar M/L would at best remain constant and likely decrease if the aperture through which it was determined were enlarged.

While it would now be desirable to compare our sample's stellar parameters directly with data for other cEs from the literature, the inherent differences in the models used to derive such parameters makes this a difficult process. Ultimately it is unclear whether the comparison would highlight differences in the models, stellar populations or both and disentangling them is beyond the scope of this work. However the recent work of \citet[hereafter S09]{smith09} employed the same stellar population model as used here to analyse a sample of 79 faint red (g-r $>$ 0.55) dwarf galaxies in the Coma cluster. Their results thus provide a unique comparative sample free of the systematic uncertainties mentioned above.

In Fig. \ref{comadwarfs} we present the stellar population parameters of both data sets against projected radius from the centre of the cluster and B band magnitude. For this we take the centre of the cluster to be RA=12:59:48, Dec=+27:58:50 and transform the SDSS magnitudes of S09 using the methods described previously. Considering the top row of panels first, in the [Fe/H] diagram the compacts are well mixed within the parameter coverage of the normal dwarf galaxies at their respective projected radius. The three bonafide cEs in our sample cover -0.6 $\leq$[Fe/H]$\leq$ -0.1 with the two less extreme compacts, CcGV1 and CcGV9b, having super solar metallicity. In this plot, at least, the compact galaxies are largely indistinguishable from the sample of S09. The same cannot be said for the positions of the three clear cEs in the remaining panels which, despite the sizable errors, are all clustered towards older ages ($\gtrsim$ 12 Gyr) and higher [Mg/Fe] ratios ($\gtrsim$ 0.25) than conventional dwarf galaxies of similar luminosity. This implies that the luminosity weighted stellar content of the cEs was formed rapidly at high redshift, a scenario which is generally associated with more massive galaxies. CcGV1 and CcGV9b on the other hand have comparable ages to the mean of $\sim$ 7$\pm$3 Gyr seen for the S09 sample within 0.5 Mpc of the cluster centre. 

Turning now to the lower row of panels of Fig. \ref{comadwarfs}, in the first diagram we see the relatively strong correlation, with slope $-0.15 \pm 0.03$, between magnitude and [Fe/H] for `normal' red Coma dwarf galaxies identified by S09. The following two panels present substantially weaker relations between magnitude and log(age) (slope $0.06 \pm 0.03$), and magnitude and [Mg/Fe] (slope $-0.01 \pm 0.02$). Unfortunately the magnitude overlap between this work's sample and that of S09 is rather limited making a robust comparison difficult. However, if we make the assumption that the slope of Fig. \ref{cmr} is primarily driven by metallicity, as has been concluded elsewhere \citep{terlevich99,vazdekis01}, we can extrapolate the m$_{B}$--[Fe/H] relation to fainter magnitudes. It is then apparent that two of the Coma compacts, CcGV1 and CcGV9b, have higher metallicities than would be expected even given the significant scatter in the relation. CcGV19a and CcGV19b are also located above the trend, but probably within the extension of the scatter, with only CcGV9a located on it.

\section{Discussion}

As has been previously mentioned, the majority of theories surrounding cE formation rely on a deep potential tidally stripping some type of more massive progenitor galaxy and leaving only its more tightly bound, compact core intact. Logically the core should retain many of the properties of the central regions of the original galaxy, with some dependence perhaps on the severity of the tidal influence, and as such finding a compact object with these particular attributes gives us a direct clue as to its origin. With this in mind we review the findings of this work. 

\begin{enumerate}
\item Six out of seven objects are found to have significantly redder B-I$_{c}$ colours than the sequence defined by normal cluster members. Given how strong colour-magnitude correlations are known to be for cluster early-type galaxies, this clearly implies either a different formation mode to other cluster members or else an evolutionary difference. CcGV9a on the other hand is located within the scatter of the relationship. Crudely extrapolating back to the trend seen in Fig. \ref{cmr} would indicate m$_{B}\sim$ 18 mag or M$_{B}\sim$ -17 mag, the top end of the dwarf regime, as the original magnitudes for the six outliers. Unfortunately given the fact that both E and early-type disk galaxies follow similar if not the same colour-magnitude relation in local galaxy clusters, it is impossible to distinguish likely progenitors from this alone.

\item Fitting PSF convolved point source + S{\'e}rsic, single and double S{\'e}rsic models we find that all of our sample are better represented by two components. Interestingly a similar result has also been found for a number of other cEs studied previously \citep{graham02,mieske05,chilingarian07,chilandmamon08} and recently in the best candidate for a cE found in the Antlia cluster by \cite{smithcastelli08}. 

Three of the sample, CcGV9a, CcGV19a and CcGV19b, are seen to occupy the same region on structural parameter diagrams as currently accepted cE galaxies (see Fig. \ref{structplots}, both panels) and so we conclude they should indeed be classified as such. Under the assumption cEs are of tidal origin, the expected direction of movement on the left panel in Fig. \ref{structplots} of a M$_{B}\sim$ -19 mag, $<\!\mu_{B}\!>_e\sim$ 20 magarcsec$^{-2}$ and R$_{e}\sim$ 1 kpc progenitor whose light profile is stripped from the outside in is likely close to the observed extension of the gE/bulge correlation. The right panel of Fig. \ref{structplots} demonstrates that their inner components begin to fill in the parameter space between Es/bulges and UCDs/nuclei. The other galaxy with a resolved inner component, CcGV9b, is less extreme in terms of surface brightness but is still substantially offset from the general trend seen in the right panel of Fig. \ref{structplots}.
 
All of the remaining three galaxies are well fit by a point source + S{\'e}rsic model with the absolute magnitudes of the point sources comparable to those found for galaxy nuclei in Virgo by \cite{cote06}. Fitting a gaussian to the nuclei luminosity function they obtain $<$M$_{g}$$>$ = -10.74 mag (M$_{B}\sim$ -10.4 mag) with $\sigma$ = 1.47 mag. Additionally they report a median half-light radius of 4.2 pc which would undoubtedly be unresolved at Coma, particularly given their magnitudes.

\item Colour gradients are detected in four of the sample in the sense that their profiles become redder towards the centre which, assuming there is little to no dust in these galaxies, is evidence for radial variations in their stellar populations parameters. CcGV1, CcGV18 and CcGV19b, are found to have little to no colour gradient but this does not rule out age or metallicity variations with radius. In general, significant changes in colour profile slope appear closely associated with the $R_{e}$ of the resolved cores of CcGV9a, CcGV19a and CcGV9b and an upturn is also noted for CcGV12.

\item The velocity dispersions obtained for our sample are found to be greater than those of normal galaxies over the luminosity range in which the two populations overlap.

Across $-12 <$ M$_{B} < -17$ mag there is a notable correlation between magnitude and velocity dispersion for compact galaxies from both the literature and this work. Excluding CcGV9b due to its significant deviation from the trend and considering ground based observations only, we obtain $x = 3.7$ when fitting $L \propto \sigma^{x}$ across this range. It would be difficult to assess the uncertainty on $x$ due to the inhomogeneity of the data but the similarity to that found previously for Es and bulges, $x \approx 4$ (Faber \& Jackson 1976), as opposed to dEs, $x \approx 2$ \citep{matkovic05}, is noteworthy. Additionally, the offset nature of this relation is perhaps one of the strongest pieces of evidence for the progenitors of cEs being more massive galaxies which have undergone some form of stellar mass removal. It is worth pointing out that this would imply taking an early-type galaxy, using some method to dim it by 1-3 magnitudes (or equivalently, assuming a constant stellar M/L, removing between 60\% and 95\% of its stellar mass) and it still remain intact. This issue was highlighted by \cite{bekki01} who performed hydrodynamical simulations of a satellite galaxy interacting with M31. They concluded that the scarcity of cEs can be put down to the small range of initial orbital parameters that allow, in their study, a spiral galaxy to be transformed via galaxy "threshing" into a cE and that once formed the lifetimes of cEs would be short (less than a few Gyr) before being full accreted. Their results also estimate a $\sim$ 60\% mass loss for the disk component of the satellite but only a $\sim$ 20\% mass loss for the bulge. Such a reduction in the bulge stellar content would be insufficient to produce the observed position of many cEs seen in Fig. \ref{fj}. Another prediction of their model is the triggering of a nuclear starburst in the cE remenant of the satellite galaxy. The three cEs in Coma show no evidence for this based on their colour profile slopes and particularly mean luminosity weighted ages within the Hectospec fibres.

Alternatively, \cite{moore96} report that cEs are a rare class of elliptical galaxy that can form alongside normal dEs via the galaxy-galaxy harassment of bulge-less spiral galaxies. However, explaining the two component nature of many cEs would seem difficult in this scenario.

As a third possibility, some cE galaxies could be extreme examples of
``hyper-compact stellar systems'' (HCSSs), defined as clouds of stars
bound to supermassive black holes that have been ejected from the
centres of their host galaxies (Merritt, Schnittman and Komossa 2008).
Effective radii and measured velocity dispersions of HCSSs are predicted
to be
\begin{subequations}
\begin{eqnarray}
R_e &\approx& 50\ {\rm pc} \left(\frac{M_\bullet}{10^9M_\odot}\right)
\left(\frac{V_k}{500\ {\rm km\ s}^{-1}}\right)^{-2}, \\
\sigma &\approx& 100\ {\rm km\ s} \left(\frac{V_k}{500\ {\rm km\ 
s}^{-1}}\right)
\end{eqnarray}
\end{subequations}
where $V_k$ is the kick velocity and $M_\bullet$ is the black  hole mass.
Explaining the Coma cE galaxies in terms of HCSSs would require
large ($\gtrsim 10^9 M_\odot$) black holes and kicks below galactic
escape velocities ($V_{k} \lesssim 500$ km s$^{-1}$).

\item Employing up to date SSP models and fitting code we have constrained the luminosity weighted age, metallicity and [Mg/Fe] ratio of our sample within the $\sim$ 700 pc diameter of the Hectospec fibres at Coma. The three cEs all have old ages ($\gtrsim$ 12 Gyr), intermediate metallicities (-0.6 $<$ [Fe/H] $<$ -0.1) and high [Mg/Fe] ($\gtrsim$ 0.25). Such parameters are often associated with more massive galaxies and particularly places any in-situ formation scenario, that is to say not the transformation of some progenitor, in doubt given the expected lifetimes of cEs. Of the two other Coma compacts with sufficient S/N spectra, both have young to intermediate ages and solar or super-solar metallicities albeit with large error bounds. CcGV1 is found to have solar [Mg/Fe] while CcGV9b has [Mg/Fe] = 0.3 which is unusually high for its age, although again there is the caveat of a large uncertainty.

We find that four out of five galaxies have [Fe/H] either higher (CcGV1 and 9b) or at the upper limit (CcGV19a and 19b) of that expected for their luminosity. As such we can again extrapolate back and note that a 1-3 magnitude shift would be required to bring them back in line with the bulk of normal galaxies. CcGV9a's position on the extreme of the trend in Fig. \ref{cmr} is likely generated by a combination of its old age and metallicity characterised by low [Fe/H] and high [Mg/Fe].

While somewhat unwise to directly compare stellar population parameters derived from different models, it is at least interesting to note those associated with the previously known cE population. A recent study by \cite{rose05} found M32 to have an age and metallicity variation from 4 Gyr and [Fe/H] = 0.0 in the nucleus to 7 Gyr and {[Fe/H] = -0.25} at 1$R_{e}$. \cite{sanchezblaz06b} find NGC 4486B to have age 9.5 Gyr and [Fe/H] = 0.4 and NGC 5846A to have age 10.7 Gyr and [Fe/H] = 0.08. Finally, \cite{chilingarian07} and \cite{chilandmamon08} obtained ages and metallicities for A496cE and M59cO of 16.4 Gyr and [Fe/H] = $-0.04$ and 9.3 Gyr and $-0.03$ respectively. Essentially, the trend appears to be toward intermediate to high ages and metallicities which is in agreement to the parameters found for our cEs. In terms of [Mg/Fe], A496cE and M59cO have super solar values ($\sim 0.2$) while M32 has sub solar, likely a result of its relatively young luminosity weighted age. 

Of course one major issue here, aside from model dependence, is differences in instrumental setup and aperture size given the strong gradients found in M32, at least. Also M32 has significantly better spatial resolution than the other galaxies which makes suitable comparisons difficult. However, we note that \cite{davidge08} conclude from their high resolution K band spectroscopic observations of M32, NGC 4486B and NGC 5846A that the former has the chemical maturity of a spiral galaxy and is less Fe deficient than the latter pair.

\item For those galaxies in our sample which have both a measured dynamical and stellar M/L we estimate that they have dark matter mass fractions $\sim$ 40\% to 90\%. Our colour profiles indicate this result should be robust to the aperture mismatch between the Hectospec fibre (radius $\sim$ 1 to 5$R_{e}$) and the radius used to derive our dynamical mass estimates (5$R_{e}$). However, particularly in the case of CcGV1, the degree to which this figure may be effected by the inherent uncertainties in deriving dynamical M/L for stellar systems which are probably not virialised is unclear.

\end{enumerate}

\subsection{Progenitors}

Combining the above conclusions, the present work supports, to varying degrees, the tidal disruption and stripping scenario as the formation mode of compact galaxies, CcGV12 and CcGV9b being the weakest contributors owing to their poorly constrained velocity dispersions. Assuming this is the correct model, progenitor morphology is nevertheless difficult to assess.

\subsubsection{CcGV9a}

Of the clear cEs, CcGV9a seems to have at least two defining similarities to those of typical disk galaxies in the presence of a weak bar and possible outer ring. The outer component's light profile, with $n = 1.64$, might perhaps then represent the threshed remains of its disk. As such this galaxy is our best candidate to fit a similar formation mechanism to that suggested by \cite{bekki01} and have an early-type disk galaxy as its progenitor. Furthermore, based on the significant fraction of light contained within its outer component and relatively strong signal of its original morphology, we speculate that this galaxy may still be in the early stages of being transformed into a cE. On the other hand, how such a strong colour gradient is generated over such a small spatial scale in a class of galaxy which conventionally has rather shallow gradients is unclear. In fact the observed colour gradient is in the opposite sense to that predicted by the model, and a luminosity weighted age of $\sim$ 12 Gyr and [Fe/H] = $-0.5$ within the Hectospec fibre largely rule out a stripping induced starburst in the centre.

\subsubsection{CcGV19a}

The likely progenitor of CcGV19a is even less well defined. It has a number of properties in common with CcGV9a yet an inner $R_{e}$ = 76 pc would seem rather small for a normal bulge and would have to be the result of it shrinking during the tidal stripping phase. Alternatively, the inner component of this galaxy is only just larger than the upper limit found for elliptical galaxy nuclei ($R_{e}$ = 62 pc) in the Virgo cluster by \cite{cote06}. Ultimately there is no strong pointer either way except perhaps the assumption that S0-Sa galaxies may possess stronger colour gradients than elliptical galaxies.

\subsubsection{CcGV19b}

Our best assessment for the final definite cE, CcGV19b, would be that it is the stripped remains of a M$_{B}$$\sim$ -17 mag dE,N given its compact inner component, flat colour profile, velocity dispersion and stellar population parameters. Furthermore, this galaxy has properties which are close, albeit more luminous and more massive, to those of UCDs in Fornax and Virgo \citep{mieske06,evstigneeva07b}, e.g. UCD 3 in Fornax \citep{phillipps01} which has a compact core and relatively large outer halo, and so we speculate this galaxy may be an example of a UCD progenitor.

\subsubsection{CcGV1, CcGV12, and CcGV18}

All three galaxies from our sample which have unresolved nuclei and outer components with $n$ = 1-1.9, more than likely originated from nucleated elliptical galaxies with initial M$_{B}$$\sim$ -17 mag. In addition, CcGV1 certainly has stellar population parameters which indicate substantial gas processing and only relatively recent star formation truncation which in turn may imply recent entry to the core of the cluster (see S09).

While strictly speaking not sufficiently extreme in terms of surface brightness to be classified as cEs or rather M32-like, these three galaxies are all more spatially compact than conventional dwarf galaxies at their luminosity and have higher velocity dispersions. It would appear likely they originated as the result of either less severe tidal disruption, different progenitor properties, or both, relative to cEs.

\subsubsection{CcGV9b}

CcGV9b is an odd ball even within this sample. Extrapolating back the position of its inner component on Fig. \ref{structplots} points to the dE region of the plot, assuming it experiences similar scenarios to the rest of the sample, and it has a velocity dispersion, albeit poorly constrained, which compliments this conclusion. Yet its inner component is also seemingly too large, $R_{e}$ = 145 pc, to be a classic nucleus and the galaxy's colour profile is strongly sloped. Additionally, this galaxy has the youngest luminosity weighted age of the sample. At best we can comment that this galaxy must have only recently been introduced to the inner regions of the cluster but cannot robustly draw any further conclusions from our data alone.

\section{Summary}

In this paper we have reported the discovery of seven compact galaxies in the Coma cluster that begin to fill in the parameter space between UCDs and cEs. Three were initially identified from HST/ACS imaging and MMT/Hectospec spectroscopy with a further four candidates being eyeball classified as high probability cluster members with support from their colours, size and resolved appearance in the ACS imaging. Subsequent spectroscopy confirmed all four candidates to be cluster members.

All seven compact galaxies are found to be well fit by a two component light profile and have structural parameters that significantly deviate from those expected for galaxies at their luminosity or size. The measured velocity dispersions for our sample exceed those of conventional dEs galaxies and the three extreme enough to be classified as cEs are found to have old, intermediate metallicity stellar populations with a substantial Fe deficiency relative to solar abundance.

Our best explanation is that the compact galaxies in this work originated as more massive systems that were subject to some form of tidal mass removal and that they, particularly the cEs, represent the rare, extreme outcome of such galaxy-galaxy processes. We also speculate that the progenitors of our sample come from at least two distinct early-type morphologies, both of which have suffered similar nurturing by the dense cluster environment.

In light of the success reported here which employed relatively simple criteria, we propose that further searches should be mounted in other massive clusters to continue to study this extreme end of galaxy formation and evolution. 

\section*{Acknowledgements}
We thank Nelson Caldwell and Dan Fabricant for their support of the Hectospec observations and Richard Cool for providing HSRED and for advice on its use. We also thank Chien Peng for advice on using Galfit and Mark Taylor for writing {\it Tool for Operations on Catalogues And Tables} (http://www.star.bris.ac.uk/$\sim$mbt/topcat/) and providing support on its use. Finally, we thank our referee for the helpful comments and suggestions which have improved this paper.

JP acknowledges support from the UK Science and Technology Facilities Council. SP and AH acknowledge the support of a grant from The Leverhulme Trust to the University of Bristol. R.O.M. acknowledges the support of NSF grant AST-0607866. DM was supported by grant NNX07AH15G from NASA and by grants AST-0821141 and AST-0807810 from the NSF. RJS is supported by STFC rolling grant PP/C501568/1 Extragalactic Astronomy and Cosmology at Durham 2005-2010. DC is supported by STFC rolling grant PP/E001149/1 Astrophysics Research at LJMU. PE was supported by DFG Priority Program 1177. Support for HST Program GO-10861 was provided by NASA through a grant from the Space Telescope Science Institute.

Based on observations with the NASA/ESA {\it Hubble Space Telescope} obtained at the Space Telescope
Science Institute, which is operated by the association of Universities for Research in Astronomy, Inc., under
NASA contract NAS 5-26555. These observations are associated with program GO10861. Observations reported here were obtained at the MMT Observatory, a joint facility of the Smithsonian Institution and the University of Arizona

\end{document}